\documentclass[a4paper,11pt]{article}
\pdfoutput=1
\usepackage{jheppub}

\usepackage[utf8]{inputenc}

\usepackage{tabularx}
\makeatletter
\def\hlinewd#1{%
\noalign{\ifnum0=`}\fi\hrule \@height #1 %
\futurelet\reserved@a\@xhline}
\makeatother

\usepackage{multirow}
\usepackage{amssymb,amsmath,bm}
\usepackage{xcolor}
\usepackage{slashed}
\usepackage{graphicx}
\usepackage[utf8]{inputenc}
\usepackage{amsfonts}
\usepackage{lscape}
\usepackage{amsthm}
\usepackage{booktabs}
\usepackage{array}
\usepackage{rotating}
\usepackage{float}
\usepackage{verbatim}
\usepackage{mathrsfs}

\usepackage[colorlinks=true,urlcolor=blue,anchorcolor=blue,citecolor=blue,filecolor=blue,linkcolor=blue,menucolor=blue,pagecolor=blue]{hyperref}

\usepackage[small,bf]{caption}
\newcommand{\GeV}{{\, \rm GeV}}

\newcommand{\eps}{\epsilon}

\newcommand{\be}{\begin{equation}}
\newcommand{\ee}{\end{equation}}

\newcommand{\ONR}[1]{\mathcal{O}_{#1}^{\mathrm{NR}}}

\newcommand{\PlotsFolder}{Plots}

\newlength{\Oldarrayrulewidth}

\begin{document}

\title{Faint Light from Dark Matter: Classifying and Constraining Dark Matter-Photon Effective Operators}

\author[a]{Bradley J. Kavanagh,}
\author[b, c]{Paolo Panci}
\author[b]{and Robert Ziegler}

\affiliation[a]{GRAPPA, Institute of Physics, University of Amsterdam, 1098 XH Amsterdam, The Netherlands}

\affiliation[b]{CERN, Theory Division, CH-1211 Geneva 23, Switzerland}

\affiliation[c]{Laboratori Nazionali del Gran Sasso (INFN-LNGS), Via G. Acitelli, 22, 67100 Assergi, Italy}

\emailAdd{b.j.kavanagh@uva.nl}
\emailAdd{paolo.panci@cern.ch}
\emailAdd{robert.ziegler@cern.ch}

\preprint{CERN-TH-2018-200}	

\abstract{Even if Dark Matter (DM) is neutral under electromagnetism, it can still interact with the Standard Model (SM) via photon exchange from higher-dimensional operators. Here we classify the general effective operators coupling DM to photons, distinguishing between Dirac/Majorana fermion and complex/real scalar DM. We provide model-independent constraints on these operators from direct and indirect detection. We also constrain various DM-lepton operators, which induce DM-photon interactions via RG running or which typically arise in sensible UV-completions. This provides a simple way to quickly assess constraints on any DM model that interacts  mainly via photon exchange or couples to SM leptons. }

\maketitle

\clearpage

\section{Introduction}

The origin and nature of Dark Matter (DM) remains a key puzzle of modern physics \cite{Jungman:1995df,Bertone:2004pz,Feng:2010gw} and a varied program is underway, searching for particle DM through direct detection \cite{Undagoitia:2015gya}, indirect detection \cite{Gaskins:2016cha} and at colliders \cite{Kahlhoefer:2017dnp}. Though referred to as `Dark', DM may still have couplings to the Standard Model photon. This coupling may be at tree-level, in which case the DM is electrically charged \cite{DeRujula:1989fe}, a possibility which has received much attention as a possible explanation of the EDGES 21cm anomaly \cite{Barkana:2018lgd,Munoz:2018pzp,Fraser:2018acy,Berlin:2018sjs}. Such a renormalisable coupling to photons would be valid up to an arbitrarily high scale, though the DM in this case must be `milli-charged' \cite{Holdom:1985ag,Holdom:1986eq,Abel:2003ue,Batell:2005wa,DelNobile:2015bqo}, with a charge substantially smaller than that of the electron \cite{Davidson:2000hf,McDermott:2010pa}. Alternatively, DM may couple to the photon through higher-dimensional, non-renormalizable operators, which arise at low energy after integrating out heavy Beyond-the-Standard-Model (BSM) degrees of freedom.

Such higher-dimensional DM-photon interactions allow us to constrain the scale of New Physics beyond the Standard Model, which is the aim of the current work. There are a wide range of these interactions and it is not always clear which constraints should be the most relevant for a given interaction or which interaction is most constraining for a given UV model. Our philosophy in this work will therefore be to classify these DM-photon operators and present updated constraints on the ones most relevant for phenomenology. We classify these operators as follows: in terms of their dimension, anticipating that the lowest dimensional operators will typically dominate; in terms of the nature of the DM candidate (Majorana or Dirac fermion, as well as real or complex scalar); and in terms of the CP parity of the interactions. Such a classification allows one to determine which operators (and therefore which bounds) are the most relevant for a given UV model, without detailed calculation. This classification will be a useful tool in constraining New Physics beyond the Standard Model through DM-photon couplings.


Many DM-photon operators have been discussed previously in the literature. These include electric and magnetic dipoles (e.g.~\cite{Pospelov:2000bq, Sigurdson:2004zp, Barger:2010gv, Banks:2010eh,DelNobile:2017fzy}), anapole (e.g.~\cite{Pospelov:2000bq, Ho:2012bg,Gao:2013vfa,DelNobile:2014eta,Alves:2017uls}) and Rayleigh interactions (e.g.~\cite{Weiner:2012cb, Weiner:2012gm, Frandsen:2012db, Latimer:2016kdg}). However, it is useful to update and summarize the current bounds from direct detection (DD) and indirect detection (ID). Moreover, we consider simple UV completions in which DM couples to a heavy particle and a lepton. We provide completely general expressions for the Wilson coefficients, and use the bounds on the effective operators to constrain the parameters of this model, which are relevant in any scenario in which DM couples to charged particles. These so-called `lepton portal' models have been studied previously \cite{Bai:2014osa}; we update the bounds using the latest results from direct, indirect and collider DM searches. We also aim to carefully make the connection between such simple UV models and DM-photon interactions in effective field theories. In some cases, the strongest constraints come from DM-photon operators which are expected to be sub-dominant based on na\"ive dimension-counting. In other cases, the dominant constraints do not come from DM-photon interactions themselves but instead from other operators appearing in sensible UV completions.

The paper is structured as follows. We begin in Sec.~\ref{sec:classification} with a classification of non-renormalisable DM-photon interactions; the operators we focus on are summarized in Table~\ref{Table:Operators}. In Sec.~\ref{sec:Bounds}, we present bounds on the coefficients of these operators from direct and indirect detection experiments. We also present bounds on a number of `UV-related' operators, which may appear in typical UV-complete models. In Sec.~\ref{sec:UVmodels}, we present two simple UV models which give rise to DM-photon interactions at low energy. We also map these UV models onto the non-renormalisable operators presented in Sec.~\ref{sec:classification}, allowing us to compare experimental bounds on such UV models coming from a full calculation and from an effective field theory approach. Finally, in Sec.~\ref{sec:summary}, we summarise our results and present our conclusions.


\section{Classifying DM-photon Interactions}
\label{sec:classification}

Our aim is to classify and study models of Dark Matter (DM) in which possible signals in direct and indirect detection arise through photon exchange.  We begin in Sec.~\ref{sec:photops} by listing the DM-photon interactions which produce signals at the energy scales relevant for direct and indirect searches. These are described by an effective field theory (EFT) that is invariant only under the electromagnetic gauge symmetry, dubbed $\mathrm{EMSM_\chi}$ \cite{DEramo:2014nmf}. 
In general a given DM-photon interaction in  $\mathrm{EMSM_\chi}$ can be induced in two ways, either by a tree-level matching or by  renormalisation group (RG) running.  In Sec.~\ref{sec:running}, we will therefore also identify all operators in $\mathrm{EMSM}_\chi$  that $i)$ radiatively induce DM-photon interactions and $ii)$ are constrained dominantly by photon exchange at low energies. Finally in Sec.~\ref{sec:UVrelated}, we identify a number of additional operators which do not lead to DM-photon interactions but which are expected to appear in any UV-complete model due to SM gauge invariance. In some cases, these `UV-related' operators may lead to stronger bounds that those from photon exchange.

\subsection{Photon Operators}
\label{sec:photops}
Here, we list all possible operators that couple DM to photons up to dimension 7. We organise the operators according to the nature of the DM particle. We first discuss ``Majorana fermion" operators ($\chi$), which are allowed in the case of Majorana DM. We then discuss ``Dirac fermion" operators ($\psi$), which are additional operators which vanish for a Majorana fermion. Similarly, we distinguish between ``real scalar" operators ($R$) and ``complex scalar" operators ($S$), which vanish for a real scalar. All operators can only be induced at loop level, so we include suitable loop factors of $e/(16\pi^2)$ or $e^2/(16\pi^2)$ in the Lagrangians. The UV scale $\Lambda$ denotes the scale of generic heavy physics that has been integrated out, above which the EFT is expected to break down.

\paragraph{Majorana Fermion:}

For Majorana DM, only operators with scalar, pseudo-scalar and axial-vector bilinears do not  vanish identically. The Lagrangian that describes the interactions with photons is given by
 \begin{equation}
 \label{MajoranaL}
 \begin{split}
 {\cal L}_{\rm Majorana }^\chi  & =    \frac{ e}{16 \pi^2 } \left[   \frac{\mathcal C_{\chi 5 \chi \partial F}}{\Lambda^2}  \,  \underbrace{ \frac{1}{2} \overline{\chi }  \gamma^{\mu}\gamma^5 \chi  \cdot \partial^\nu F_{\mu \nu}}_{{\cal O}_{\chi 5 \chi \partial F}} \right]  \\
 & +  \frac{ e^2}{16 \pi^2 } \left[ \frac{\mathcal C_{\chi  \chi FF}}{\Lambda^3} \underbrace{ \frac{1}{2} \overline{\chi}  \chi \cdot F^{\mu \nu} F_{\mu \nu}}_{{\cal O}_{\chi  \chi FF} } +   \frac{\mathcal C_{\chi 5 \chi F \tilde{F}}}{ \Lambda^3}  \, \underbrace{\frac{i}{2} \overline{\chi} \gamma^5 \chi \cdot F^{\mu \nu} \tilde{F}_{\mu \nu}}_{{\cal O}_{\chi 5 \chi F \tilde{F}}} \right]   \\
& +  \frac{ e^2}{16 \pi^2 } \left[ \frac{\mathcal C_{\chi  \chi F \tilde{F}}}{\Lambda^3} \underbrace{ \frac{1}{2} \overline{\chi}  \chi \cdot F^{\mu \nu} \tilde{F}_{\mu \nu}}_{{\cal O}_{\chi  \chi F \tilde{F}} } +   \frac{ \mathcal C_{\chi 5 \chi F F}}{ \Lambda^3}  \, \underbrace{\frac{i}{2} \overline{\chi} \gamma^5 \chi \cdot F^{\mu \nu} F_{\mu \nu}}_{{\cal O}_{\chi 5 \chi F F}} \right] \,,
\end{split}
\end{equation}
where $\tilde{F}_{\mu \nu}  = \frac{1}{2} \eps_{\mu \nu \rho \sigma} F^{\rho \sigma}$. 
The leading operators we consider here are labelled with under-braces. The term in the first line is the Anapole moment, which is the only allowed dimension-6 operator and describes the lowest order  interactions of Majorana DM with photons. There are also four dimension-7 Rayleigh operators, which can be divided into CP-even (second line) and CP-odd operators (third line).

\paragraph {Dirac Fermion:}
For Dirac DM, also tensor and vector bilinears are allowed. In addition to the interactions described in Eq.~\eqref{MajoranaL} one can write three other leading order operators. The Lagrangian is 
\begin{equation}
\label{DiracL}
 \begin{split}
 {\cal L}_{\rm Dirac}^\psi  & =  2 {\cal L}_{\rm Majorana }^{\chi \rightarrow \psi}  \\  & +  \frac{ e}{16 \pi^2 } \left[ \frac{\mathcal C_{\psi \psi F}}{2\Lambda} \underbrace{\overline{\psi } \sigma^{\mu \nu} \psi \cdot F_{\mu \nu}}_{{\cal O}_{\psi \psi F} } +   \frac{\mathcal C_{\psi 5 \psi F}}{2 \Lambda}  \, \underbrace{i\, \overline{\psi }  \sigma^{\mu \nu}\gamma^5 \psi \cdot F_{\mu \nu} }_{{\cal O}_{\psi 5 \psi F} } + \frac{\mathcal C_{\psi \psi \partial F}}{\Lambda^2} \underbrace{\overline{\psi } \gamma^\mu \psi \cdot \partial^\nu  F_{\mu \nu}}_{{\cal O}_{\rm \psi \psi \partial F}}  \right] \,,
 \end{split}
\end{equation}
where $\sigma^{\mu \nu} = \frac{i}{2} \left[ \gamma^\mu, \gamma^\nu \right] $.  
The first and the second terms are the dimension-5 magnetic and electric dipoles respectively, while the third one is the dimension-6  fermion charge radius operator.   The factor 2 appearing in the first line has been chosen so that the Feynman rules for Majorana and Dirac fermions are the same.

\paragraph{Real Scalar:}
For real scalars, the lowest order interactions with photons are described by two dimension-6  Rayleigh operators. The Lagrangian is given by
\begin{equation}
\label{RScalarL}
 {\cal L}_{\rm Real Scalar}^R  =   \frac{ e^2}{16 \pi^2 } \left[ \frac{\mathcal C_{R R FF}}{\Lambda^2} \underbrace{ \frac{1}{2} R^2 \cdot F^{\mu \nu} F_{\mu \nu}}_{{\cal O}_{R R F F} } +   \frac{\mathcal C_{R R F \tilde{F}}}{ \Lambda^2}  \, \underbrace{\frac{1}{2} R^2 \cdot F^{\mu \nu} \tilde{F}_{\mu \nu}}_{{\cal O}_{R R F \tilde{F}}} \right] \,.
\end{equation}

\paragraph{Complex Scalar:}
For complex scalars, in addition to the dimension-6 Rayleigh interactions in Eq.~\eqref{RScalarL}, one can directly couple a complex scalar current to the electromagnetic field strength tensor, which is the scalar charge radius.  The Lagrangian reads
\begin{equation}
 \label{CScalarL}
 \begin{split}
{\cal L}_{\rm Complex Scalar}^S & =  2 {\cal L}_{\rm Real Scalar}^{R \rightarrow S}    +  \, \frac{e}{16 \pi^2 }   \frac{ \mathcal C_{\partial S \partial S F}}{\Lambda^2}  \, \underbrace{i \, \partial^\mu S \partial^\nu S^*  \cdot F_{\mu \nu}}_{{\cal O}_{\partial S \partial S F}} \, . \\
 \end{split}
 \end{equation}
As for  fermion DM, the factor of 2 multiplying $\mathcal{L}_\mathrm{RealScalar}$ ensures  the same Feynman rules  for complex and real scalars.

%
%
%

\subsection{Running to the Nuclear Scale}
\label{sec:running}

The Lagrangians in the previous section are defined at the UV scale $\Lambda$, which denotes the scale of generic heavy physics that has been integrated out. From this scale we should run the theory  down to the relevant IR scale $\mu_{\rm IR}$ (e.g.~the nuclear energy scale $\mu_n \sim 1\,\,\mathrm{GeV}$ for direct detection\footnote{In principle, if we have only light leptons running in the loops, the relevant IR scale can be set by the mass of the leptons or the typical momenta involved. However, when we couple to the axial-vector current of leptons, this induces an axial-vector current interaction with quarks, which in turn couple to the pion. In this case, the relevant IR scale is the hadronic mass scale. We therefore fix $\mu_n = 1\,\,\mathrm{GeV}$ (as in Refs.~\cite{Crivellin:2014qxa,DEramo:2014nmf,DEramo:2016gos}) for consistency in this scenario.}). Since all relevant photon operators  can only arise at loop level at the energy scale $\Lambda$, we should also take into account the RG contribution of possible tree-level operators that can mix into the photon operators. Focusing on operators that are constrained dominantly by photon exchange, we restrict to tree-level operators that  do not involve quarks and Higgs fields. Operators with quark fields, even if not valence quarks, will always induce DM-gluon couplings at 1-loop \cite{Hisano:2015bma}. On the other hand operators with Higgs fields will  give rise to DM couplings to valence quarks already at tree-level due to $Z$-boson exchange~\cite{DEramo:2014nmf}. Thus, the only operators that are relevant for our purposes are tree-level operators that involve vector currents of SM leptons. We therefore consider also the following additional operators:
\begin{equation}
\begin{split}
\label{D4l}
& \Delta {\cal L}_{\rm Majorana}^{\ell \ell}   = \frac{\mathcal C_{\chi 5 \chi \ell \ell}}{\Lambda^2} \, \underbrace{\frac{1}{2} \overline{\chi } \gamma^\mu \gamma_5 \chi \cdot \overline{\ell}  \gamma_\mu   \ell}_{ {\cal O}_{\chi 5 \chi \ell \ell} }\, ,    \\
&  \Delta {\cal L}_{\rm Dirac}^{\ell \ell}   =  \frac{\mathcal C_{\psi \psi \ell \ell}}{\Lambda^2} \,  \underbrace{\overline{\psi } \gamma^\mu \psi \cdot \overline{\ell}  \gamma_\mu   \ell}_{\mathcal O_{\psi \psi \ell \ell}}   + \frac{\mathcal C_{\psi 5 \psi \ell \ell}}{\Lambda^2} \,  \underbrace{\overline{\psi } \gamma^\mu \gamma_5 \psi \cdot \overline{\ell}  \gamma_\mu   \ell}_{\mathcal O_{\psi 5 \psi \ell \ell}}  \, ,     \\
&  \Delta {\cal L}_{\rm Complex \, Scalar}^{\ell \ell}   = \frac{ \mathcal C_{S\partial S \ell \ell}}{\Lambda^2} \, \underbrace{i \, S^* \overset{\leftrightarrow}{\partial^\mu} S  \cdot \overline{\ell} \, \gamma_\mu  \ell }_{{\cal O}_{S\partial S \ell \ell} } \, . 
\end{split}
\end{equation}
Note that to avoid strong constraints from lepton flavor-violation, we will assume a coupling only to a single DM lepton ($\ell = e,\,\mu,\,\tau$) at a time. 

These operators induce fermion charge radius, anapole and scalar charge radius operators through RG evolution, according to
\begin{equation}
\begin{split}
\label{eq:LoopMatch}
{\cal C}_{\psi \psi \partial F} |_{\mu_{\rm IR}}  & = {\cal C}_{\psi \psi \partial F}  |_{\Lambda} +  \mathcal K(\Lambda) \, {\cal C}_{\psi \psi \ell \ell} |_{\Lambda}   \, , \\
{\cal C}_{\psi 5 \psi \partial F} |_{\mu_{\rm IR}} & = {\cal C}_{\psi 5 \psi \partial F}   |_{\Lambda}  + \mathcal K(\Lambda) \, {\cal C}_{\psi 5 \psi \ell \ell}  |_{\Lambda}  \, ,  \\
{\cal C}_{\chi 5 \chi \partial F} |_{\mu_{\rm IR}}  & = {\cal C}_{\chi 5 \chi \partial F } |_{\Lambda}   +   \mathcal K(\Lambda) \, {\cal C}_{\chi 5 \chi  \ell \ell } |_{\Lambda}  \, ,  \\
{\cal C}_{\partial S \partial S F} |_{\mu_{\rm IR}}  & = {\cal C}_{\partial S \partial S F}  |_{\Lambda}  + \mathcal K(\Lambda)  \, {\cal C}_{S \partial S  \ell \ell }  |_{\Lambda}  \, , 
\end{split}
\end{equation}
where $\mathcal K(\Lambda) \approx (4/3) \, \log(\Lambda^2/\mu_{\rm IR}^2)$. The precise value is calculated using the \textsc{runDM} code \cite{runDM}, which takes into account the running of couplings between $\Lambda$ and the nucleon level~\cite{Crivellin:2014qxa,DEramo:2014nmf,DEramo:2016gos}. We have checked that there is a very good numerical agreement between \textsc{runDM} and the analytic expression for $\mathcal K(\Lambda)$ given above\footnote{Running effects also play a role in the calculation of the direct detection signal for Rayleigh interactions ($\propto F^{\mu\nu}F_{\mu\nu},\,F^{\mu\nu}\tilde{F}_{\mu\nu}$), as we discuss in more detail in Appendix~\ref{sec:DD_cross_sections}.}.

Since the operators in the UV can only arise at loop-level themselves, the logarithmically enhanced contribution from the tree-level operators with leptons typically dominates. In the next section we will demonstrate that this enhancement of the radiatively-induced operators can be significant, particularly in the case of Majorana fermion DM, and lead to stronger direct detection bounds than on the fermion charge radius, anapole and scalar charge radius operators themselves. 
 
 We note that in general loops which couple DM to SM quarks through the insertion of \textit{two} DM-photon effective operators are expected to be further suppressed compared to the single-insertion interactions we consider here. Naively, one would expect these to therefore be negligible in an EFT analysis. However, in some cases, such loops can give rise to interactions with a larger rate (for example, being spin-independent or velocity-independent) than the tree-level EFT operators, partially compensating for the suppression by additional powers of the cut-off scale $\Lambda$. This effect was discussed in detail in Ref.~\cite{Haisch:2013uaa}. For example, two insertions of the dimension-6 anapole operator $\mathcal{O}_{\psi 5 \psi \partial F}$ leads to a scalar-scalar coupling between DM and SM quarks, which scales as $\Lambda^{-4}$. This does indeed lead to spin-independent, velocity-unsuppressed direct detection constraints but these may dominate over the tree-level anapole constraints only for light DM ($m_\chi \lesssim \mathcal{O}(\text{few GeV})$), where the momentum-suppression of the tree-level anapole substantially weakens constraints. We focus here on DM masses at the scale of 10 GeV and above and we therefore neglect this type of loop interaction for the remainder of this work.

\begin{table}[th!]
\centering
\footnotesize
\def\arraystretch{1.8}
\begin{tabular}{|c|c||c|c|c||}
\hlinewd{1.3pt}
& CP & Dim.~5 
 & Dim.~6  
 &  Dim.~7  \\
\hlinewd{1.3pt}
\multirow{ 4}{*}{\shortstack{{\bf Dirac} \\{\bf Fermion} $\psi$ }}  & \multirow{ 2}{*}{{\bf +}} & \multirow{ 2}{*}{${\cal O}_{\psi \psi F} $}  & ${\cal O}_{\psi \psi \ell \ell}$ $\longrightarrow$ (${\cal O}_{\psi \psi \partial F}$)  & (${\cal O}_{\psi  \psi F F} $)   \\
&  & &      ${\cal O}_{\psi 5 \psi \ell \ell}$ $\longrightarrow$ $({\cal O}_{\psi 5 \psi \partial F}) $ & (${\cal O}_{\psi 5 \psi F \tilde{F}} $)  \\
\cline{2-5}
 & \multirow{ 2}{*}{{\bf --}} &   \multirow{ 2}{*}{$ {\cal O}_{\psi 5 \psi F} $ }  &   & (${\cal O}_{\psi  \psi F \tilde{F}} $)  \\
& &      & & (${\cal O}_{\psi  5 \psi F F} $)   \\

\hlinewd{1.3pt}
\multirow{ 4}{*}{\shortstack{{\bf Majorana}\\ {\bf Fermion} $\chi$  }}  &  \multirow{ 2}{*}{{\bf +}}  &  & & ${\cal O}_{\chi \chi FF} $    \\
 & & &     ${\cal O}_{\chi 5 \chi \ell \ell} \longrightarrow ({\cal O}_{\chi 5 \chi \partial F})$  & ${\cal O}_{\chi 5 \chi F \tilde{F}} $  \\
\cline{2-5}
& \multirow{ 2}{*}{{\bf --}}  &   &  & (${\cal O}_{\chi  \chi  F \tilde{F}} $)    \\
& &  &   & ${\cal O}_{\chi 5 \chi  F F} $ \\

\hlinewd{1.3pt}
\multirow{ 2}{*}{\shortstack{{\bf Complex}\\{\bf Scalar} $S$}}  &   {\bf +}  &  & (${\cal O}_{ S  S F F} $) , ${\cal O}_{S \partial S \ell \ell} \longrightarrow  ({\cal O}_{\partial S \partial S F} $) & \\
\cline{2-5}
&   {\bf --}  &   & (${\cal O}_{ S  S F \tilde{F}} $) &   \\
\hlinewd{1.3pt}
\multirow{ 2}{*}{\shortstack{{\bf Real}\\{\bf Scalar }$R$}}  &   {\bf +} &  & ${\cal O}_{ R  R F F } $   & \\
\cline{2-5}
  &  {\bf --} &   & ${\cal O}_{ R  R F \tilde{F}} $ &  \\
\hlinewd{1.3pt}
\end{tabular}

\caption{Relevant Dark Matter-photon operators at the UV scale  $\Lambda$. Operators in round brackets give a subleading contribution to signals in direct and indirect searches, compared to operators with the same CP quantum numbers, or are dominantly generated in the IR by other operators via RG evolution, denoted by $\longrightarrow$. In subsequent sections, we set limits only on those operators which are not in brackets (except for ${\cal O}_{\chi 5 \chi \partial F}$ and ${\cal O}_{ \partial S \partial S F}$, whose limits we will show for illustrative purposes). \label{Table:Operators} }
\end{table}

\subsection{Summary}

We collect the relevant operators we have discussed so far in Table~\ref{Table:Operators}. We arrange the operators according to the type of DM particle and, for a given DM particle, we distinguish between those operators which are CP-even (top row) and CP-odd (bottom row). We also order the operators according to their dimension (Dim.~5, Dim.~6 or Dim.~7). We focus on operators that lead to the strongest constraints, contributing to signals in either direct detection or indirect detection. Therefore, when presenting bounds in Sec.~\ref{sec:Bounds}, we will neglect operators that have the same CP quantum numbers as the dominant operators but give weaker constraints.
That is, for two operators with the same dimension and CP properties, we present bounds only on the more strongly constrained of the two; the subdominant operator we denote with round brackets in Table~\ref{Table:Operators}.  In particular, this concerns the fermion charge radius, the anapole and the scalar charge radius operators, which at low energy are dominantly generated from the tree-level operators via RG evolution, denoted by $\longrightarrow$ in the table. In addition, as we discuss in more detail later, some of the Rayleigh operators give sub-leading constraints (and in some cases give no constraints at all).


\subsection{UV-related Operators}
\label{sec:UVrelated}
Up to now we have taken into account only operators that are dominantly constrained by photon exchange at low energies. However, in a UV-complete model SM gauge invariance implies that these operators are necessarily accompanied by additional operators that are potentially more strongly constrained than the operators in Table~\ref{Table:Operators}. In this section we briefly comment on this issue and discuss the possible operators arising in UV-complete models respecting eletroweak gauge invariance, which we therefore refer to as ``UV-related" operators. 

First of all,  due to electroweak symmetry it is clear that all photon operators will be accompanied by operators involving the $Z$-boson. However, such operators typically give effects in DD and ID which are subleading compared to the corresponding photon operators, since they are suppressed by the $Z$-boson mass appearing the propagator\footnote{In direct detection this is a general statement, while there may be exceptions for indirect detection. In particular, in cases where the DM-photon coupling gives a gamma-ray line signature, the corresponding coupling with the $Z$-boson will produce a gamma-ray continuum extending down to energies below the DM mass. In such cases, it may be possible to constrain heavier DM (at a given gamma-ray energy) in the case of DM-$Z$ interactions. See also Ref.~\cite{Weiner:2012cb}.}. We will therefore not explicitly consider the bounds on these kinds operators in the following. 

The other class of operators that typically arise in UV models are tree-level operators that couple the DM field to lepton currents. For example, the operator ${\cal O}_{\psi \psi \ell \ell } = \overline{\psi} \gamma^\mu \psi \overline{\ell} \gamma_\mu \ell$ generically comes together with the operator $\overline{\psi} \gamma^\mu \psi \overline{\ell} \gamma_\mu \gamma_5 \ell$, since only left-handed and right-handed leptons  have definite SM quantum numbers. In order to be complete, we will consider all possible tree-level operators with a similar structure to the ones in  Eq.~\eqref{D4l}, i.e.~for fermion DM all dimension six operators with two DM fermions and two leptons, and for scalar DM all dimension six  operators with two DM scalars, two leptons and one derivative. According to this approach we define 
  \begin{align}
\label{D4l5}
 &\Delta {\cal L}_{\rm Majorana}^{\ell 5 \ell}     = \frac{\mathcal C_{\chi 5 \chi \ell 5 \ell}}{\Lambda^2} \, \underbrace{\frac{1}{2} \overline{\chi } \gamma^\mu \gamma_5 \chi \cdot \overline{\ell}  \gamma_\mu \gamma_5   \ell}_{ {\cal O}_{\chi 5 \chi \ell 5 \ell} }  \, , \nonumber \\
& \Delta {\cal L}_{\rm Dirac}^{\ell 5 \ell}    =  \frac{\mathcal C_{\psi \psi \ell 5 \ell}}{\Lambda^2} \,  \underbrace{\overline{\psi } \gamma^\mu \psi \cdot \overline{\ell}  \gamma_\mu \gamma_5  \ell}_{\mathcal O_{\psi \psi \ell 5\ell}}   + \frac{\mathcal C_{\psi 5 \psi \ell 5 \ell}}{\Lambda^2} \,  \underbrace{\overline{\psi } \gamma^\mu \gamma_5 \psi \cdot \overline{\ell}  \gamma_\mu \gamma_5  \ell}_{\mathcal O_{\psi 5 \psi \ell 5 \ell}} \, ,     \\
&  \Delta {\cal L}_{\rm Real \, Scalar}^{\ell 5 \ell}     =  \frac{ \mathcal C_{RR \ell 5 \partial  \ell}}{\Lambda^2} \, \underbrace{ \frac{1}{2} R^2  \cdot \overline{\ell} \,  \gamma_5 \slashed{D} \ell }_{{\cal O}_{RR \ell 5 \partial \ell} } + \frac{ \mathcal C_{RR \ell  \partial  \ell}}{\Lambda^2} \, \underbrace{\frac{i}{2}  R^2 \cdot \overline{\ell} \,  \slashed{D} \ell }_{{\cal O}_{RR \ell  \partial \ell} } \, , \nonumber \\
&   \Delta {\cal L}_{\rm Complex \, Scalar}^{\ell 5 \ell}     = \frac{ \mathcal C_{S\partial S \ell 5 \ell}}{\Lambda^2} \, \underbrace{i\, S^* \overset{\leftrightarrow}{\partial^\mu} S  \cdot \overline{\ell} \, \gamma_\mu \gamma_5 \ell }_{{\cal O}_{S\partial S \ell 5 \ell} }  + \frac{ \mathcal C_{SS \ell 5 \partial  \ell}}{\Lambda^2} \, \underbrace{ S^* S  \cdot \overline{\ell} \,  \gamma_5 \slashed{D} \ell }_{{\cal O}_{S S \ell 5 \partial \ell} } + \frac{ \mathcal C_{SS \ell  \partial  \ell}}{\Lambda^2} \, \underbrace{i \,  S^* S  \cdot \overline{\ell} \,  \slashed{D} \ell }_{{\cal O}_{S S \ell  \partial \ell} } \nonumber \, . 
  \end{align} 
Using the EOM for the lepton, $\slashed{D} \ell = - i m_\ell \ell $,  the scalar operators involving derivatives can be re-written as
\begin{equation}
\begin{split}
&{\cal O}_{R R \ell  5 \partial \ell}   = - \frac{i m_\ell}{2} \, R^2 \cdot \overline{\ell} \gamma_5 \ell \, ,  
\qquad\qquad \,
{\cal O}_{R R \ell  \partial \ell}   = \frac{m_\ell}{2} \, R^2 \cdot \overline{\ell}  \ell \, ,  \\
&{\cal O}_{S S \ell  5 \partial \ell}  = - i m_\ell \, S^* S \cdot \overline{\ell} \gamma_5 \ell \, ,  
\qquad\qquad
{\cal O}_{S S \ell  \partial \ell}  =  m_\ell \, S^* S \cdot \overline{\ell}  \ell \, .
\end{split}
\end{equation} 
Finally we add the four-fermion operators with scalar currents, which must involve additional fermion masses if the UV completion does not contain additional sources of chiral symmetry breaking\footnote{Photon operators are induced via mixing only by operators with vector lepton currents, which respect the chiral symmetry $\ell \to e^{i \alpha \gamma_5} \ell$. This symmetry is always broken by the lepton mass, and if that is the only source of the breaking, all scalar and pseudo-scalar currents must be proportional to the lepton mass. A similar conclusion holds for the DM  fermion  currents.}. Under this assumption these operators arise only at dimension-8, and are given by 
\begin{align}
\label{dim8}
& \Delta {\cal L}_{\rm Majorana}^{\ell  \ell-8}  =  \frac{\mathcal C_{\chi \chi \ell  \ell-8}}{\Lambda^4} \,  \underbrace{ m_\ell m_\chi  \, \frac{1}{2} \overline{\chi }  \chi \cdot \overline{\ell}   \ell}_{\mathcal O_{\chi \chi \ell \ell-8}}   +  \frac{\mathcal C_{\chi \chi \ell 5 \ell-8}}{\Lambda^4} \,  \underbrace{ i m_\ell m_\chi  \, \frac{1}{2} \overline{\chi }  \chi \cdot \overline{\ell}  \gamma_5 \ell}_{\mathcal O_{\chi \chi \ell 5 \ell-8}} \, , \nonumber    \\
& \qquad\qquad\quad +  \frac{\mathcal C_{\chi 5 \chi \ell  \ell-8}}{\Lambda^4} \,  \underbrace{i m_\ell m_\chi  \, \frac{1}{2}  \overline{\chi } \gamma_5 \chi \cdot \overline{\ell}   \ell}_{\mathcal O_{\chi 5 \chi \ell  \ell-8}} +  \frac{\mathcal C_{\chi 5 \chi \ell 5 \ell-8}}{\Lambda^4} \,  \underbrace{ m_\ell m_\chi  \, \frac{1}{2} \overline{\chi } \gamma_5 \chi \cdot \overline{\ell}  \gamma_5 \ell}_{\mathcal O_{\chi 5 \chi \ell 5 \ell-8}} \, , \nonumber \\
& \Delta {\cal L}_{\rm Dirac}^{\ell  \ell-8}    = 2 \Delta {\cal L}_{\rm Majorana}^{\ell  \ell-8}  (\chi \to \psi) \, .
\end{align}
Note that all of the operators in this section are CP-even, apart from those which contain one pure axial bilinear (e.g.~$\overline { \psi } \psi \cdot \overline { \ell } \gamma _ { 5 } \ell$).

In the next section we will show that the bounds on these additional UV-related operators are rarely stronger than the ones in Table~\ref{Table:Operators} when considering DD (where their contributions are loop-suppressed). In ID, however, they may be relevant as they induce tree-level annihilation of the DM into SM leptons.


\section{Bounds on EFT Operators}
\label{sec:Bounds}

We now present bounds from direct and indirect detection on the DM-photon operators described in the previous section.  We also briefly discuss bounds on 4-fermion effective operators from Bhabha scattering at LEP-2.

\paragraph{Direct Detection:}
For direct detection, we show current bounds from the full LUX exposure (``LUX (2016)") \cite{Akerib:2016vxi}, the one tonne-year XENON1T exposure (``XENON1T (2018)") \cite{Aprile:2018dbl} and the projected sensitivity of the LZ experiment (``LZ-projected") \cite{Akerib:2015cja,Mount:2017qzi}. Similar constraints for high mass DM come from PANDAX-II \cite{Cui:2017nnn}. Here, we will focus on DM at the GeV-scale and above, so we do not consider a number of constraints which are relevant only at low mass \cite{Hehn:2016nll,Agnese:2017jvy,Petricca:2017zdp}. Details of direct detection cross sections and bounds are given in Appendix~\ref{app:DirectDetection}. In all cases, we calculate an approximate single bin Poisson limit using the approach of Ref.~\cite{DelNobile:2013sia} to translate the bounds between different interactions. Details of reproducing the bounds from LUX can be found in Appendix~B.2 of Ref.~\cite{Kavanagh:2016pyr} and details of calculating the LZ projections are given in Appendix~D of Ref.~\cite{DEramo:2016gos}. The corresponding calculation for XENON1T is given here in Appendix~\ref{app:Xenon1T}.

\paragraph{Indirect Detection:}
Bounds from indirect detection are presented for the FERMI search for spectral lines using 5.8 years of data and an isothermal DM density profile (``FERMI 5.8 yrs (2013)") \cite{Ackermann:2015lka} and the H.E.S.S.-I search for high-energy gamma ray lines using 112 hours of data and an Einasto DM profile\footnote{For conservative choices of the DM density profile (e.g. an isothermal sphere), the limits from H.E.S.S. become very weak (see for example the right panel of Fig.~5 in Ref.~\cite{Lefranc:2016fgn}). } (``H.E.S.S.-I 112h (2013)") \cite{Abramowski:2013ax}. For operators which allow DM annihilation directly into charged leptons (such as ${\cal O}_{\psi \psi \ell \ell}$), we also show conservative bounds from FERMI observations of dwarf Spheroidal galaxies (``FERMI dSphs (2016)") \cite{Ackermann:2015zua}. In this case, FERMI searched for a continuum spectrum of gamma rays (rather than a line) and we differentiate between limits for different final states: $e$, $\mu$ or $\tau$\footnote{Loop-induced annihilation into quarks is discussed in Ref.~\cite{DEramo:2017zqw}. We have checked explicitly that these bounds are subdominant, apart from when annihilation occurs very close to the $Z$-pole ($m_\mathrm{DM} \sim m_Z/2$) in which case loop-induced annihilation may be competitive with tree-level annihilation to charged leptons.}. When the DM couples to a single $F_{\mu \nu}$, we set limits by summing over all fermion final states, weighting by the photon-mediated cross section, using the publicly available likelihood tools (in the Supplemental Material to Ref.~\cite{Ackermann:2015zua}, available at \href{https://www-glast.stanford.edu/pub\_data/1048/}{https://www-glast.stanford.edu/pub\_data/1048/}). We also show conservative limits from the FERMI Galactic Halo search using an isothermal DM density profile (``FERMI GH (2013)") \cite{Ackermann:2012rg}. 
 While this constraint is typically slightly weaker than the FERMI dSphs search for $s$-wave annihilation, in some cases we study operators which lead to $p$-wave annihilation, $\langle \sigma v \rangle \propto v^2$. Given the larger velocities in the Galactic Halo compared to those expected in dwarfs, the Fermi GH search may come to dominate in these cases. Note that limits for $e^\pm$ final states are not reported in Ref.~\cite{Ackermann:2012rg}. However, the $\mu^\pm$ can be taken as a conservative estimate of the $e^\pm$ bound; in the latter case the gamma-ray flux is expected to also include a stronger inverse compton component (see e.g.~\cite{Cirelli:2009vg,Cirelli:2009dv}). Apart from the bounds obtained by the H.E.S.S.-I search  for high-energy gamma ray lines, we do not consider any other limits from Cherenkov telescope arrays because they strongly depend on assumptions about the DM density profile.  Annihilation cross sections for the operators we consider in this work are given in Appendix~\ref{app:IndirectDetection}.


\paragraph{Colliders:}
The only relevant constraint from colliders comes from measurements of Bhabha scattering at LEP-2\footnote{LHC bounds for the anapole operator are given in~Ref.~\cite{Gao:2013vfa} probing energy scales below $\Lambda \sim 20\,\mathrm{GeV}$. These bounds will be improved by a factor of 3 at the High Luminosity LHC \cite{Alves:2017uls}. For other studies of projected collider bounds on EFT operators, see e.g.~\cite{Fichet:2016clq}.}, giving bounds on 4-fermion operators involving electrons
\begin{align}
{\cal L} & = \frac{1}{v^2} \left[ \frac{c_{VV}}{2} (\overline{e} \gamma^\mu e) \cdot (\overline{e} \gamma_\mu e)  +  \frac{c_{AA}}{2} (\overline{e} \gamma^\mu \gamma_5 e) \cdot (\overline{e} \gamma_\mu \gamma_5 e) \right] \, ,
\end{align}
where $v= 246$~GeV.
These operators can in turn be induced at one-loop with 2 insertions of DM-DM-$e$-$e$ operators, giving at leading order in $m_{\rm DM}^2/\Lambda^2$
\begin{align}
c_{VV} & = - \frac{v^2}{8 \pi^2 \Lambda^2} \left( {\cal C}_{\psi \psi \ell \ell}^2 + {\cal C}_{\psi 5 \psi \ell \ell}^2 +  \frac{1}{2}  {\cal C}_{S \partial S \ell \ell}^2\right)   \, ,  \\
c_{AA} & = - \frac{v^2}{8 \pi^2 \Lambda^2} \left( {\cal C}_{\psi \psi \ell 5 \ell}^2      + {\cal C}_{\psi 5 \psi \ell 5 \ell}^2   + \frac{1}{2} {\cal C}_{S \partial S \ell 5 \ell}^2   \right) \, , 
\label{VVAAmatching}
\end{align}
where we have used the UV scale $\Lambda$  to cut off (quadratically divergent) loops in Euclidean momentum space. As discussed in the previous section, more realistic UV models involve chiral combinations of operators, i.e. have correlated operator coefficients
\begin{align}
{\cal C}_{\psi \psi \ell  5 \ell}  & = {\cal C}_{\psi \psi \ell   \ell} \equiv \frac{1}{2} \, {\cal C}_{\psi \psi \ell  R \ell} \, ,    
& 
{\cal C}_{\psi 5 \psi \ell  5 \ell}  & = {\cal C}_{\psi 5 \psi \ell   \ell} \equiv \frac{1}{2}  \, {\cal C}_{\psi 5 \psi \ell  R \ell} \, ,    \nonumber \\
- {\cal C}_{\psi \psi \ell  5 \ell}  & =  {\cal C}_{\psi \psi \ell   \ell} \equiv \frac{1}{2}  \, {\cal C}_{\psi \psi \ell  L \ell} \, ,    
& 
- {\cal C}_{\psi 5 \psi \ell  5 \ell}  & = {\cal C}_{\psi 5 \psi \ell   \ell} \equiv \frac{1}{2}  \, {\cal C}_{\psi 5 \psi \ell  L \ell} \, ,    \nonumber \\
{\cal C}_{S \partial S \ell  5 \ell}  & = {\cal C}_{S \partial S \ell   \ell} \equiv \frac{1}{2}  \, {\cal C}_{S \partial S \ell  R \ell} \, ,    
& 
- {\cal C}_{S \partial S \ell  5 \ell}  & = {\cal C}_{S \partial S \ell   \ell} \equiv \frac{1}{2}  \, {\cal C}_{S \partial S \ell  L \ell} \, .    
\end{align}
These induce chiral operators involving only electrons 
\begin{align}
{\cal L} & = \frac{1}{v^2} \left[ \frac{c_{LL}}{2} (\overline{e} \gamma^\mu P_L e) \cdot (\overline{e} \gamma_\mu P_L e)  +  \frac{c_{RR}}{2} (\overline{e} \gamma^\mu P_R  e) \cdot (\overline{e} \gamma_\mu  P_R e) \right] \, ,
\end{align}
with coefficients given by, analogously to Eq.~\eqref{VVAAmatching}, 
\begin{align}
c_{LL} & = - \frac{v^2}{8 \pi^2 \Lambda^2} \left( {\cal C}_{\psi \psi \ell L \ell}^2 + {\cal C}_{\psi 5 \psi \ell L \ell}^2 +  \frac{1}{2}  {\cal C}_{S \partial S \ell L \ell}^2\right)   \, ,  \\
c_{RR} & = - \frac{v^2}{8 \pi^2 \Lambda^2} \left( {\cal C}_{\psi \psi \ell R \ell}^2      + {\cal C}_{\psi 5 \psi \ell R \ell}^2   + \frac{1}{2} {\cal C}_{S \partial S \ell R \ell}^2   \right) \, .
\label{RRmatching}
\end{align}
For the low-energy bounds on $c_{VV}, c_{AA},  c_{LL},  c_{RR}$ we use Refs.~\cite{Falkowski:2015krw, Falkowski:2017pss}, which have re-analyzed the LEP data in Ref.~\cite{LEP:2003aa} to derive the constraints (assuming the presence of one operator at a time) 
\begin{align}
c_{VV} & = \left( 1.5 \pm 0.8 \right) \cdot 10^{-3} \, , & c_{LL} & = \left( 8.0 \pm 2.8 \right) \cdot 10^{-3} \, , \nonumber \\
c_{AA} & = \left( 2.7 \pm 2.2 \right) \cdot 10^{-3} \, , & c_{RR} & = \left( 3.8 \pm 2.8 \right) \cdot 10^{-3} \, . 
\end{align}
The resulting upper bounds on the tree-level DM-electron operator coefficients are shown in Table~\ref{LEP}.  These numbers should be taken with a grain of salt, since they arise from  quadratically divergent loops, which are sensitive to the specific UV completion. Nevertheless the above procedure gives indicative bounds, which will reproduce the results of the particular UV completions that we consider in the next section. Note that these bounds are quite stringent, as a result of the preference of LEP-2 data for $\emph{positive}$ coefficients of 4-electron operators that already disfavor the SM, while the new loop contributions are always $\emph{negative}$ (in particular for LL operators, where at 95\% CL the coefficient is still positive, so we decide to set a bound on 99\% CL). We do not show these bounds explicitly in the rest of Sec.~\ref{sec:Bounds} (though we will in Sec.~\ref{sec:UVmodels}). However, we remind the reader that these rather stringent bounds apply for all DM masses in the scenario where DM couplings to electrons at tree-level.

\begin{table}[t!]
\centering
\begin{tabular}{|c||c|c|c|c|c|c|}
\hline
 Coefficient & $\Lambda/{\cal C}_{\psi \psi ee}$ & $\Lambda/{\cal C}_{\psi 5 \psi ee}$ & $\Lambda/{\cal C}_{\psi \psi e5 e}$ & $\Lambda/{\cal C}_{\psi 5 \psi e 5 e}$ & $\Lambda/{\cal C}_{S \partial S e e}$ & $\Lambda/{\cal C}_{S \partial S e 5 e}$ \\
\hline
Bound & $2.8 \, {\rm TeV}$& $2.8 \, {\rm TeV}$ & $0.67 \, {\rm TeV}$ & $0.67 \, {\rm TeV}$ & $2.0 \, {\rm TeV}$ & $0.47 \, {\rm TeV}$ \\
\hline
\hline
Coefficient & $\Lambda/{\cal C}_{\psi \psi e L e}$ & $\Lambda/{\cal C}_{\psi 5 \psi e L e}$ & $\Lambda/{\cal C}_{\psi \psi e R e}$ & $\Lambda/{\cal C}_{\psi 5 \psi e R e}$ & $\Lambda/{\cal C}_{S \partial S e L e}$ & $\Lambda/{\cal C}_{S \partial S e R e}$ \\
\hline
Bound & $1.4 \, {\rm TeV}^*$& $1.4 \, {\rm TeV}^*$ & $0.65 \, {\rm TeV}$ & $0.65 \, {\rm TeV}$ & $0.98 \, {\rm TeV}^*$ & $0.46 \, {\rm TeV}$ \\
 \hline
\end{tabular}
\caption{\label{LEP} Lower 95\% CL bounds on dimension 6 operators  involving electrons from LEP-2. Bounds with $*$ are 99\% CL,  and all numbers are rather indicative,  see text for details. }
\end{table}

\paragraph{Relic density:} We assume throughout this work that the new particle we consider makes up all of the DM. It is therefore necessary for the particle to be produced with the correct relic abundance, matching the observed cosmological abundance of DM (  $\Omega _ \mathrm{DM} h ^ { 2 } = 0.1206 \pm 0.0021$  \cite{Aghanim:2018eyx}). 

One of the standard ways to achieve this is through the so-called thermal {\it freeze-out} mechanism. A crucial assumption for the determination of the thermal annihilation cross section is given by the energy budget of the Universe at the time of DM production. In the scenario in which the Universe was dominated by radiation at freeze-out, the observed cosmological  abundance of DM is reproduced if $\langle \sigma v \rangle \simeq 3\times 10^{-26}$ cm$^3$/s \cite{Bertone:2004pz}.

Within this scenario we can map out the parameter space in the plane ($M_\mathrm{DM}$, $\Lambda/\mathcal C^\frac{1}{(d-4)}$) where freeze-out is viable (for a given operator of dimension $d$). For example, for the dipole interactions $\mathcal{O}_{\psi \psi F}$ and $\mathcal{O}_{\psi 5 \psi F}$ the DM is always overproduced (see e.g.~\cite{DelNobile:2012tx}) because the energy scale of the effective operator allowed by direct detection experiments is large and therefore the annihilation cross section is too small. To give another example, for the 4-fermion operators (such as $\mathcal{O}_{\psi\psi\ell\ell}$) one can have regions where the relic abundance is correctly reproduced depending on the DM mass and Lorentz structures (see e.g.~\cite{DEramo:2017zqw}). 

However, it is important to bear in mind that we do not have any direct information regarding the expansion rate and energy budget of the Universe at temperature higher than the BBN scale ($\sim 1$ MeV). If one allows for a different thermal history, the canonical thermal value of the annihilation cross section can dramatically change (see e.g.~\cite{Chung:1998rq,Giudice:2000ex,DEramo:2017gpl,Hamdan:2017psw,Binder:2017rgn}). In addition, there are a number of alternatives to the standard thermal freeze-out scenario by which the correct relic density can be produced. These include mechanisms such as {\it freeze-in} \cite{Hall:2009bx}, asymmetric DM \cite{Petraki:2013wwa}, forbidden DM \cite{DAgnolo:2015ujb} and many others. As a consequence, we do not show the regions of parameter space where the relic density is reproduced because such regions would be dramatically model-dependent.


\paragraph{Perturbative Unitarity:}
Additional constraints on EFT operator coefficients can be obtained from requiring perturbative unitarity. These are relevant only for dimension-six tree-level operators, and we have estimated the bounds from perturbative unitarity by imposing that the real part of the Born amplitude for a 2-2 scattering process is smaller than 1/2. 
Neglecting possible additional order-one factors, unitarity is  violated if 
\begin{equation}
\sqrt s \geq  \sqrt{4\pi} \Lambda_{\rm eff}\,,
\end{equation}
where $s$ is the center-of-mass energy of the process and $\Lambda_{\rm eff}$ is the effective suppression scale of the dimension-six operator including dimensionless couplings. 
For non-relativistic DM particles one roughly has $\sqrt{s} \sim m_{\rm DM}$ and therefore perturbative unitarity is violated if $m_{\rm DM} \geq  \sqrt{4\pi}\, \Lambda_{\rm eff}$. Note that this limit holds for both direct and indirect detection, as the center-of-mass energy of the system is always larger than $m_{\rm DM}$.

\subsection{Dirac Fermion Dark Matter}

\begin{figure}[!t]
\centering

\includegraphics[width= 0.48 \textwidth]{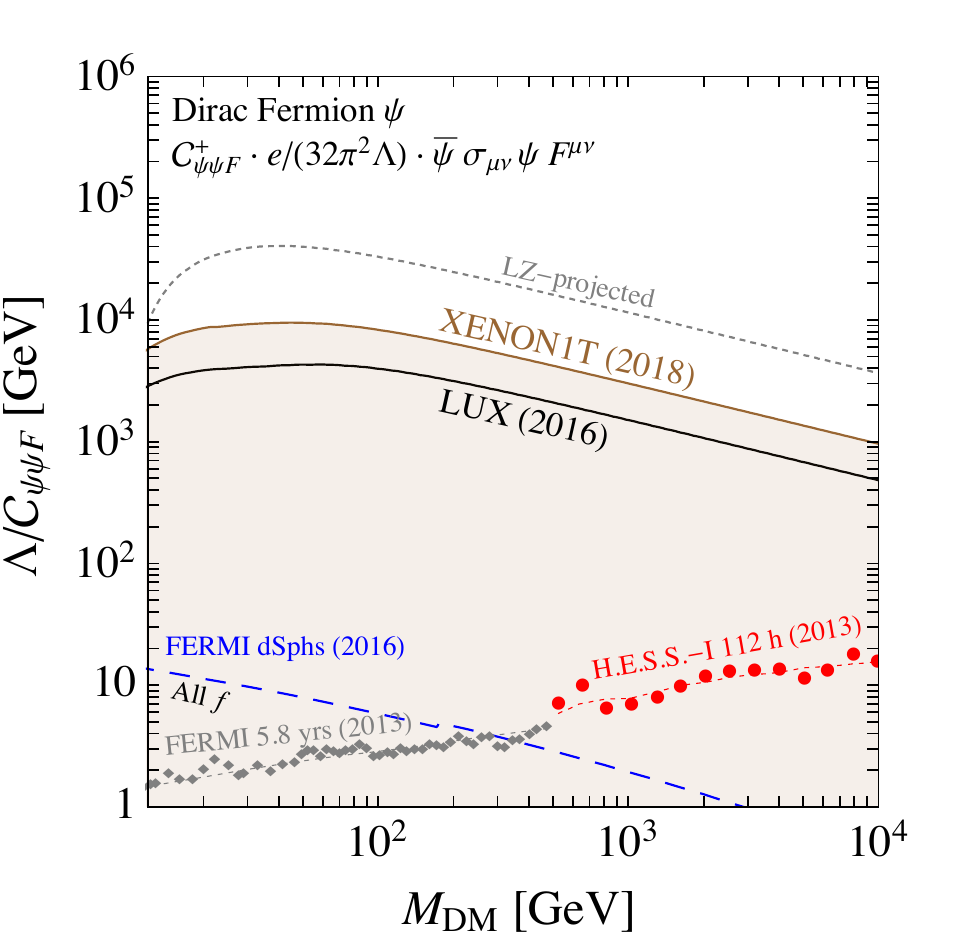} \
\includegraphics[width= 0.48 \textwidth]{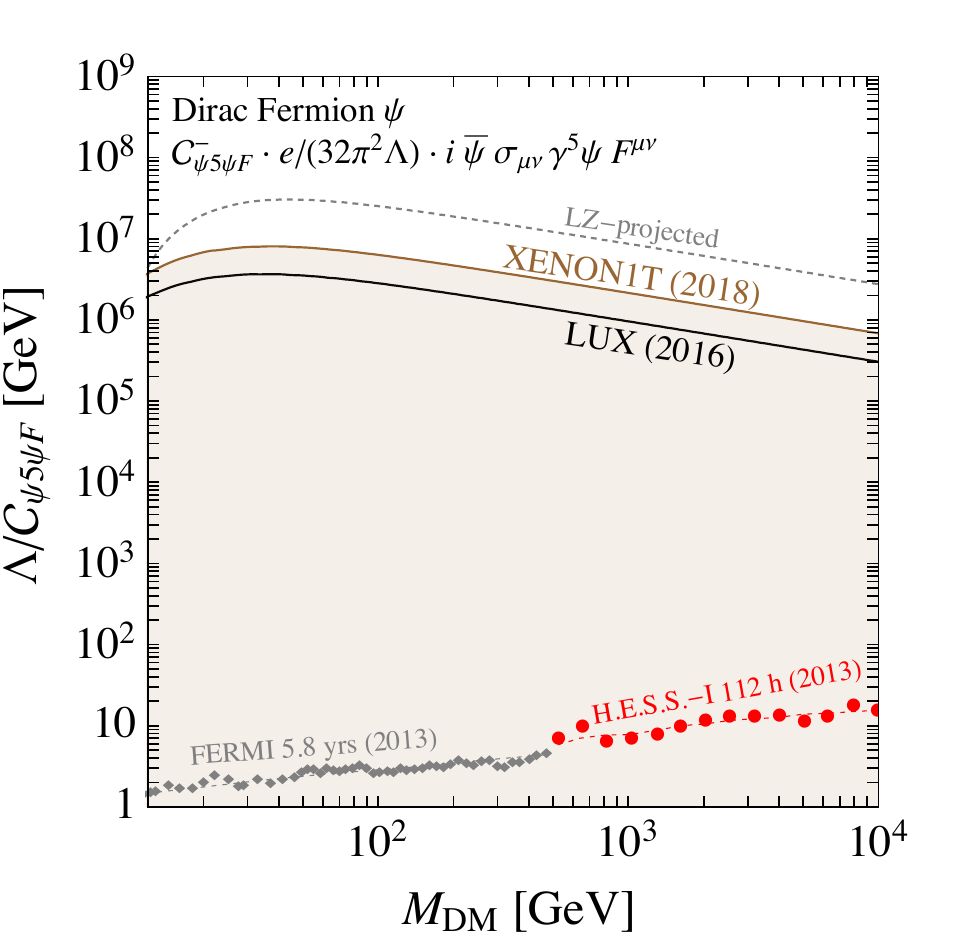} \\

\includegraphics[width= 0.485 \textwidth]{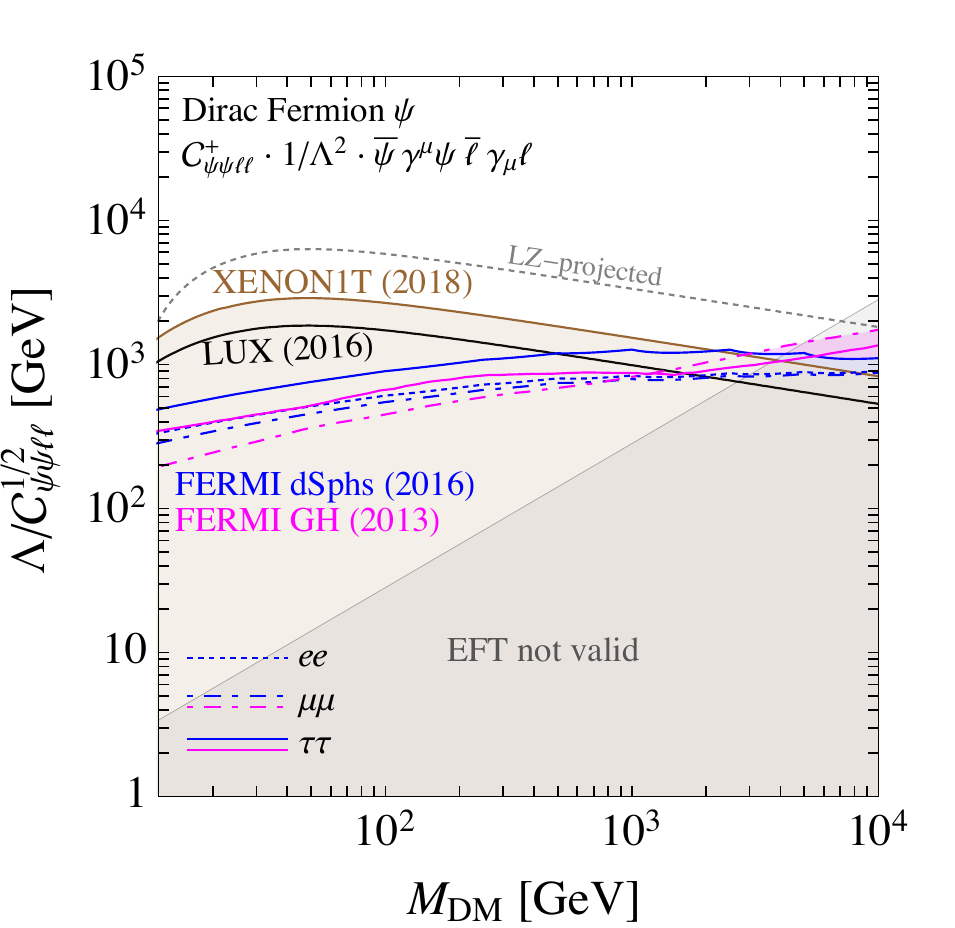} \
\includegraphics[width= 0.485 \textwidth]{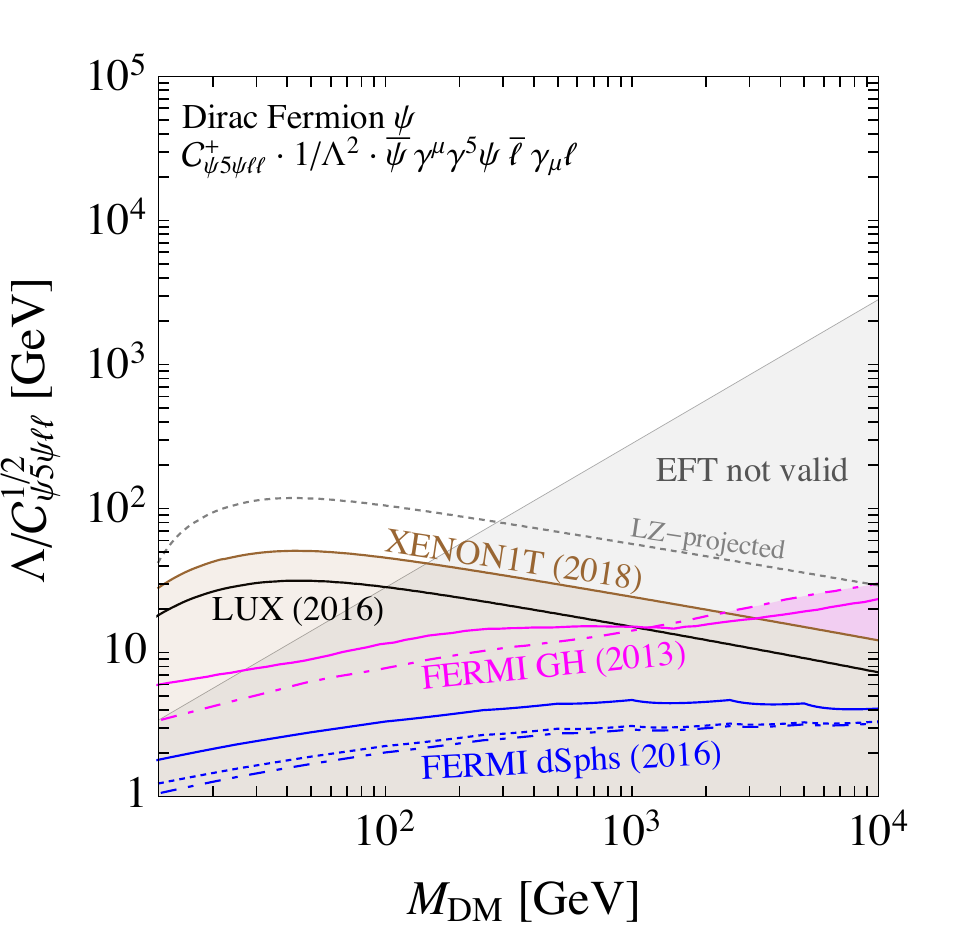} \\

\caption{Constraints from direct and indirect detection on high-scale operators which give rise to DM-photon couplings at low energy, in the case of Dirac fermion DM $\psi$.  The relevant operators are discussed in detail in Sec.~\ref{sec:photops} and Sec.~\ref{sec:running}, and we have indicated the CP parity $\pm$ in the operator coefficient. In each panel we show constraints on a different operator, specified in the upper left. More details of the bounds are given at the start of Sec.~\ref{sec:Bounds}. Note the different vertical scales in the top two panels. \label{fig:EFT_Constraints_Dirac}}
\end{figure}

In Fig.~\ref{fig:EFT_Constraints_Dirac}, we show constraints on the 4 most relevant high-scale operators which give rise to low energy interactions between photons and Dirac fermion Dark Matter, $\psi$. 

In the top left panel of Fig.~\ref{fig:EFT_Constraints_Dirac}, we show constraints on the  magnetic dipole operator, $\mathcal{O}_{\psi\psi F}$. This operator gives rise to a coherently enhanced direct detection cross section through the coupling of the DM magnetic dipole to the nuclear charge, in addition to a spin dependent dipole-dipole interaction. These interactions have both a long-range contribution and a contact contribution and lead to constraints on the EFT cut-off at the level of $\Lambda/\mathcal{C}_{\psi\psi F} \gtrsim 8000\,\,\mathrm{GeV}$ for $m_\mathrm{DM} \sim 50\,\,\mathrm{GeV}$. The cross section for Dark Matter annihilation into 2 photons scales as $\Lambda^{-4}$, meaning that indirect detection bounds are suppressed relative to those from direct detection (where $\sigma \sim \Lambda^{-2}$)\footnote{We note that the limits from gamma-ray line searches (grey and red points) do not strictly apply when they are weaker than limits from gamma-ray continuum searches (blue and magenta dashed lines). In such cases, the line signal would be weaker than the continuum signal and so no bound can be obtained on annihilation into a line.}.

In the top right panel of Fig.~\ref{fig:EFT_Constraints_Dirac}, we show constraints on the electric dipole operator, $\mathcal{O}_{\psi 5 \psi F}$. This too gives rise to coherently-enhanced spin-independent scattering with the nucleus, though in this case the interaction is purely long range and therefore enhanced at low recoil energies. This results in strong constraints from direct detection, $\Lambda/\mathcal{C}_{\psi 5 \psi F} \gtrsim 7 \times 10^6\,\,\mathrm{GeV}$. Constraints on direct annihilation into gamma rays are the same as in the case of the magnetic dipole operator and therefore substantially weaker than direct detection constraints. Annihilation into fermions is $p$-wave suppressed, so that constraints from dwarf Spheroidals are negligible.

In the bottom left panel of Fig.~\ref{fig:EFT_Constraints_Dirac}, we show constraints on the 4-fermion interaction $\mathcal{O}_{\psi\psi \ell \ell}$, which induces the fermion charge radius operator ${\cal O}_{\rm \psi \psi \partial F} $ at low energy. This gives rise to the standard (coherently-enhanced) spin-independent interaction, though coupling only to the proton content of the nucleus. Despite being a dimension-6 operator, the 4-fermion interaction gives constraints which are not much weaker than the magnetic dipole, $\Lambda/\mathcal{C}_{\psi\psi \ell \ell}^{1/2} \gtrsim 3000\,\,\mathrm{GeV}$. As discussed in Sec.~\ref{sec:running}, this is because ${\cal O}_{\rm \psi \psi \partial F}$ is induced through RG evolution, giving an enhancement  of $\sim \log(\Lambda/\mu)$ relative to the magnetic dipole (induced only at the loop level at the scale $\Lambda$). Finally, the dominant annihilation channel of $\mathcal{O}_{\psi\psi \ell \ell}$ is directly into charged leptons, leading to competitive constraints from FERMI searches for diffuse gamma rays in both dwarf Spheroidal galaxies and the Milky Way Galactic halo.

In the bottom right panel of Fig.~\ref{fig:EFT_Constraints_Dirac}, we show constraints on the 4-fermion interaction, $\mathcal{O}_{\psi 5 \psi \ell \ell}$. In this case, RG evolution induces the anapole operator $\mathcal{O}_{\psi 5 \psi \partial F}$. The resulting interaction with nuclei is velocity suppressed and therefore gives much weaker constraints than the other operators (with the same CP quantum number) we have considered in this section, $\Lambda/\mathcal{C}_{\psi 5 \psi \ell \ell}^{1/2} \gtrsim 50 \,\,\mathrm{GeV}$. As in the case of $\mathcal{O}_{\psi \psi \ell \ell}$, the dominant constraint in indirect detection comes from annihilation directly into charged leptons, though in this case the cross section is $p$-wave suppressed, leading again to rather weak constraints. Constraints from the FERMI Galactic halo search dominate over that in dwarf Spheroidals by a factor of a few, owing to the larger DM velocities expected in the Milky Way halo. Note that bounds from perturbative unitarity dominate at DM masses above 100 GeV, but the stronger constraints from the operators on the left panel with the same CP quantum numbers render this restriction rather irrelevant.   
\subsection{Majorana Fermion Dark Matter}

\begin{figure}[!t]
\centering

\includegraphics[width= 0.48 \textwidth]{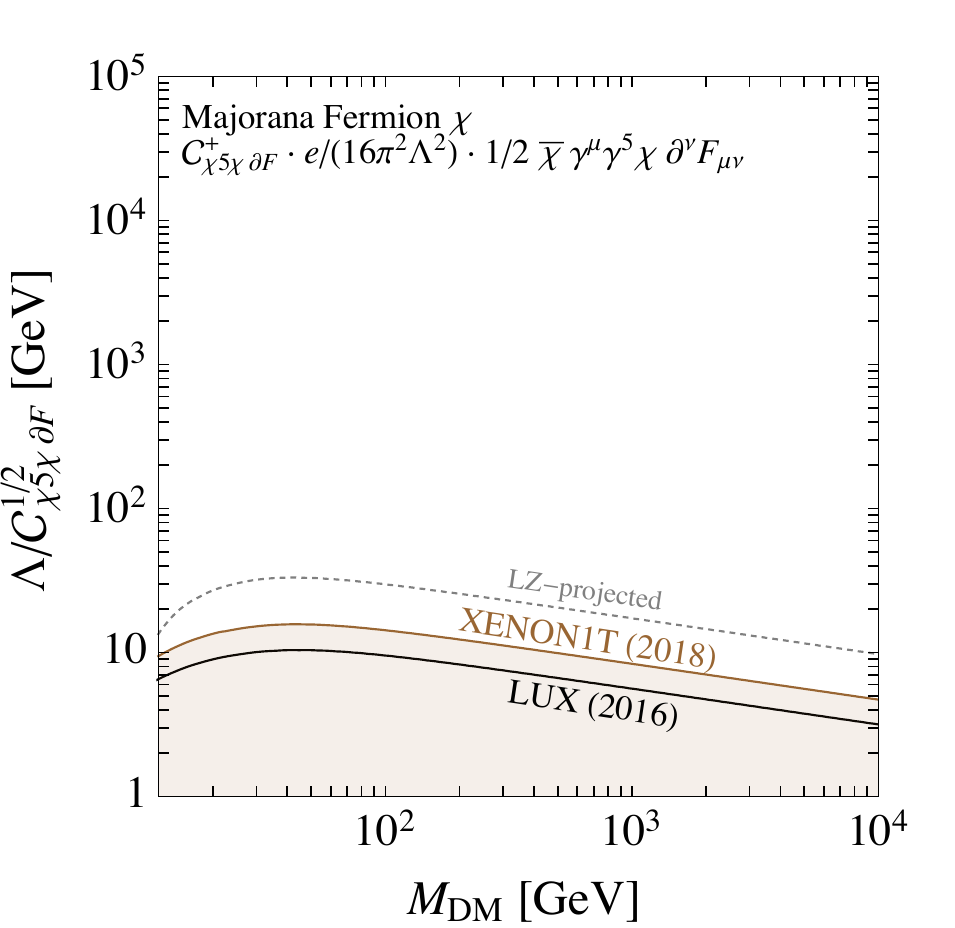} \
\includegraphics[width= 0.485 \textwidth]{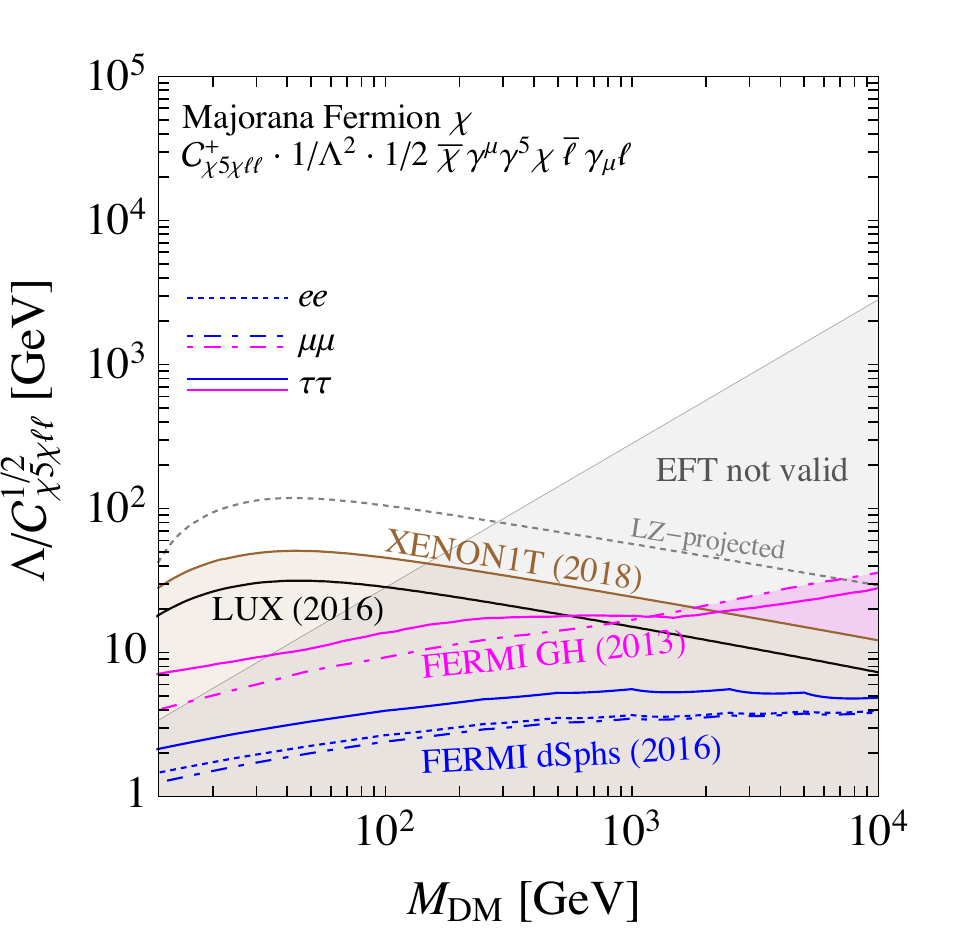}

\includegraphics[width= 0.48 \textwidth]{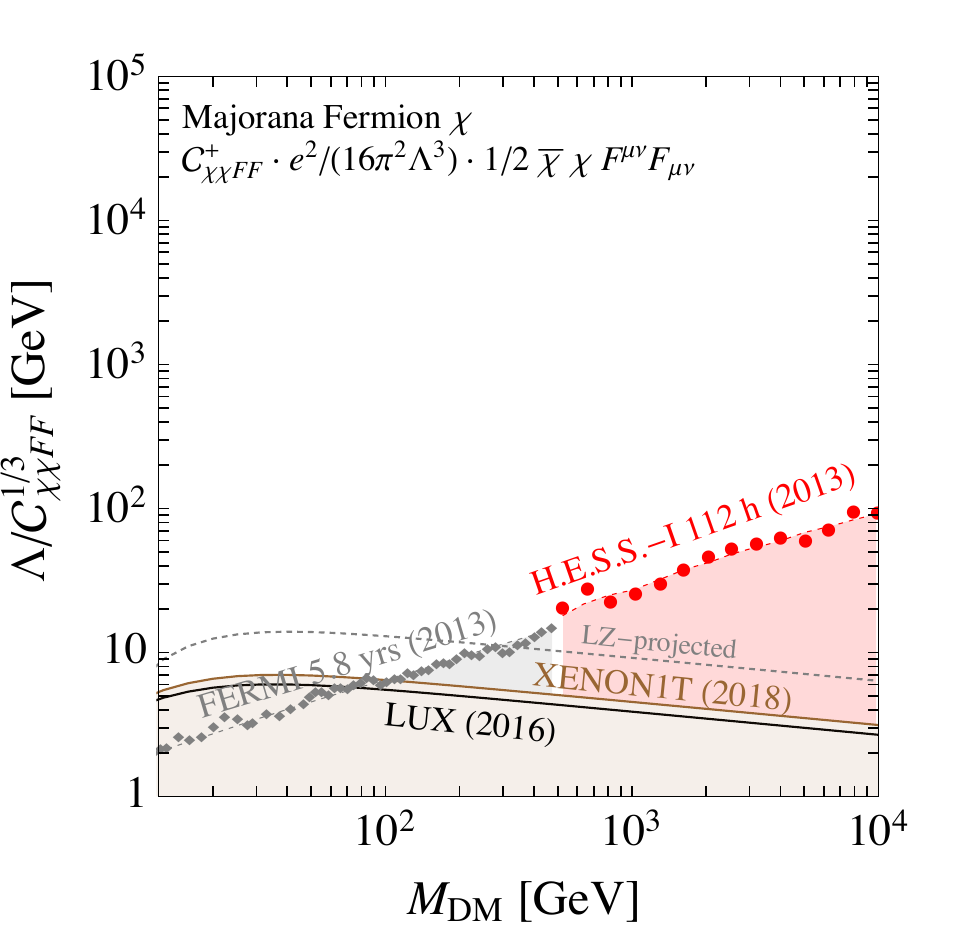} \
\includegraphics[width= 0.485 \textwidth]{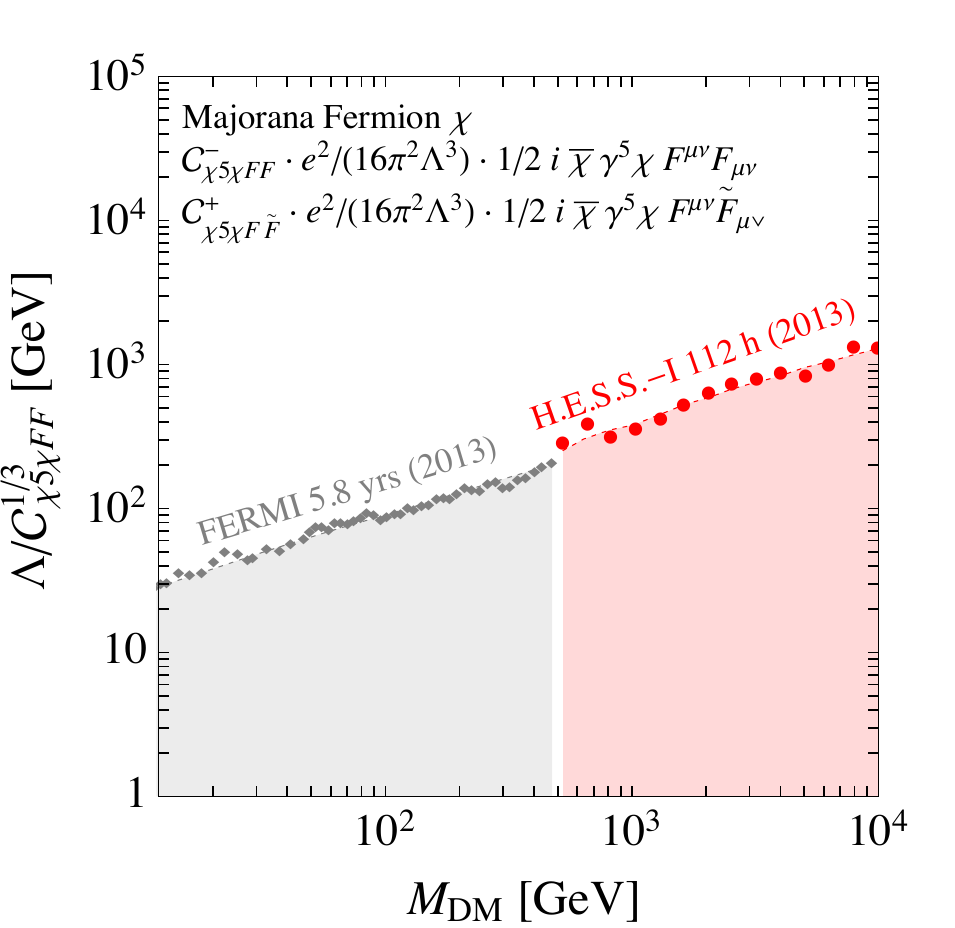} \\

\caption{Same as Fig.~\ref{fig:EFT_Constraints_Dirac} but for Majorana Fermion Dark Matter $\chi$. Note that bounds on the operators $\mathcal{O}_ {\chi 5 \chi FF}$ and $\mathcal{O}_ {\chi 5 \chi F\tilde{F}}$ (both shown in the bottom right panel) are identical.\label{fig:EFT_Constraints_Majorana}}
\end{figure}

In Fig.~\ref{fig:EFT_Constraints_Majorana}, we show constraints on the 4 most relevant high-scale operators which give rise to low energy interactions between photons and Majorana fermion Dark Matter, $\chi$. In contrast with the Dirac fermion case, the magnetic and electric dipole operators are forbidden, so we now include the Rayleigh interactions ($\propto F_{\mu\nu}F^{\mu\nu},\,F_{\mu\nu}\tilde{F}^{\mu\nu}$). We also include the anapole operator and the 4-fermion operator which can induce it through Standard Model loops.

In the top left panel of Fig.~\ref{fig:EFT_Constraints_Majorana}, we show constraints on the anapole operator $\mathcal{O}_{\chi 5 \chi \partial F}$. In the top right panel, we show constraints on the 4-fermion operator $\mathcal{O}_{\chi 5 \chi \ell \ell}$, which induces $\mathcal{O}_{\chi 5 \chi \partial F}$ at the low scale through RG evolution. Constraints on the anapole are weaker than those on the 4-fermion operator, $\Lambda/\mathcal{C}_{\chi 5 \chi \partial F}^{1/2} \gtrsim 10 \,\,\mathrm{GeV}$ compared to $\Lambda/\mathcal{C}_{\chi 5 \chi \ell \ell}^{1/2} \gtrsim 50 \,\,\mathrm{GeV}$ (corresponding to a difference in the rate of a factor of $\sim600$). However, as in the Dirac case the bounds from perturbative unitarity dominate the constraints for DM masses above 100 GeV.  


In the bottom left panel of Fig.~\ref{fig:EFT_Constraints_Majorana}, we show constraints on the Rayleigh operator $\mathcal{O}_{\chi\chi F F}$. As detailed in Appendix \ref{app:DirectDetection} (and originally in Refs.~\cite{Weiner:2012cb,Frandsen:2012db}), this interaction leads to coherent nuclear scattering, with a cross section scaling as $Z^4/\Lambda^6$ for a nucleus of charge $Z$. However, the Rayleigh operator is dimension-7 and the scaling with $\Lambda$ dominates (along with the loop-suppression prefactor), resulting in constraints at the level of $\Lambda/\mathcal{C}_{\chi \chi F F}^{1/3} \gtrsim 4 \,\mathrm{GeV}$. The cross section for annihilation into 2 photons also scales as $\Lambda^{-6}$, as well as being $p$-wave suppressed. Even so, above DM masses around 100 GeV, indirect detection constraints begin to dominate over direct searches, as the annihilation cross section is enhanced as $m_\psi^4$.

In the bottom right panel of Fig.~\ref{fig:EFT_Constraints_Majorana}, we show constraints on the Rayleigh operator $\mathcal{O}_{\chi5 \chi F \tilde{F}}$. In this case, the direct detection signal is negligible \cite{Frandsen:2012db} as the cross section is momentum-suppressed, while the annihilation cross section is $s$-wave, leading to much stronger constraints than in the case of $\mathcal{O}_{\chi\chi F F}$. The cut-off scale is constrained to be $\Lambda/\mathcal{C}_{\chi 5 \chi F \tilde{F}}^{1/3} \gtrsim 100 \,\,\mathrm{GeV}$ at $m_\mathrm{DM} = 100 \,\,\mathrm{GeV}$, again with the cross section growing rapidly with $m_\psi$. Note that constraints on the Rayleigh operator $\mathcal{O}_{\chi 5 \chi F F}$ are identical to those on $\mathcal{O}_{\chi5 \chi F \tilde{F}}$ (shown in the bottom right panel of Fig.~\ref{fig:EFT_Constraints_Majorana}); the two share the same momentum-suppressed direct detection cross section and $s$-wave annihilation cross section. As we note in Sec.~\ref{sec:DD_cross_sections}, direct and indirect constraints on the remaining Majorana Rayleigh operator $\mathcal { O } _ { \chi \chi  F \tilde { F }}$ are negligible. 

We note here that the Rayleigh operators $\mathcal{O}_{\chi 5 \chi F F}$ and $\mathcal{O}_{\chi5 \chi F \tilde{F}}$ are more strongly constrained than the (induced) anapole operator, due to the $s$-wave annihilation cross section and scaling with the DM mass. Of course, the relative importance of the different operators depends on the exact UV completion, which we explore later in Sec.~\ref{sec:UVmodels}.

\subsection{Scalar Dark Matter}
\label{sec:ScalarEFTBounds}
\begin{figure}[!t]
 \centering
 
 \includegraphics[width= 0.4925 \textwidth]{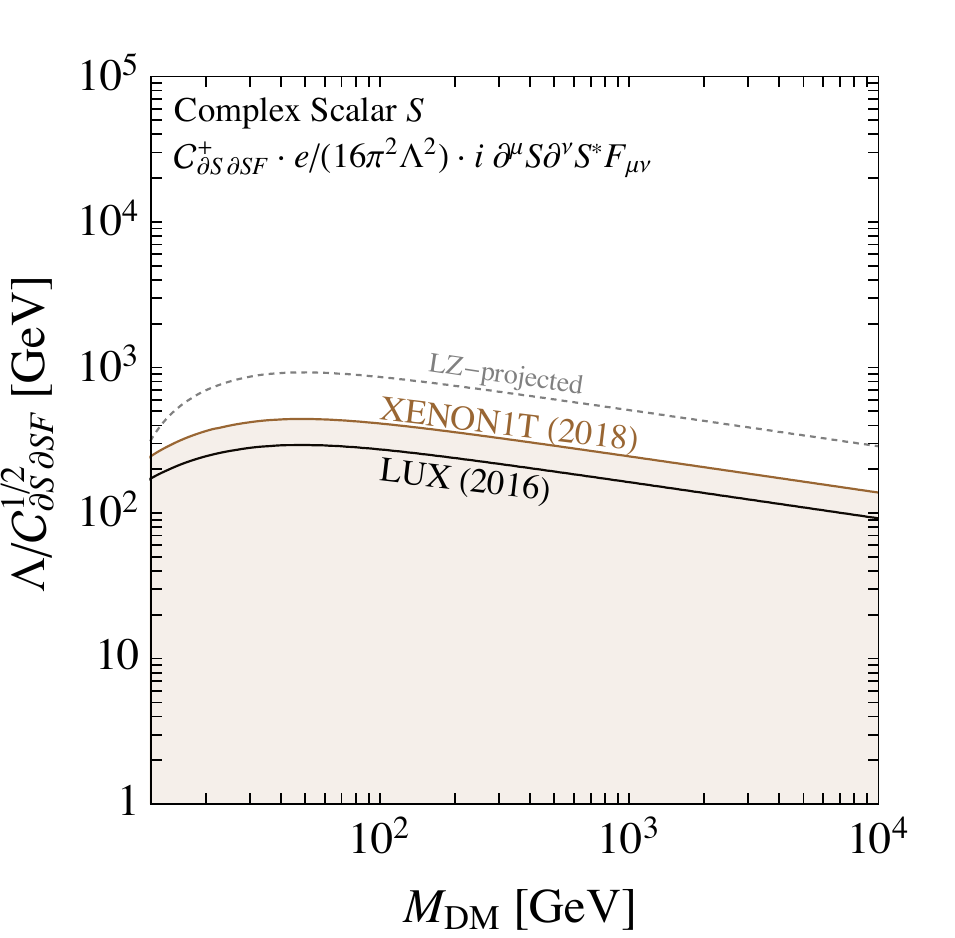} \
\includegraphics[width= 0.4925 \textwidth]{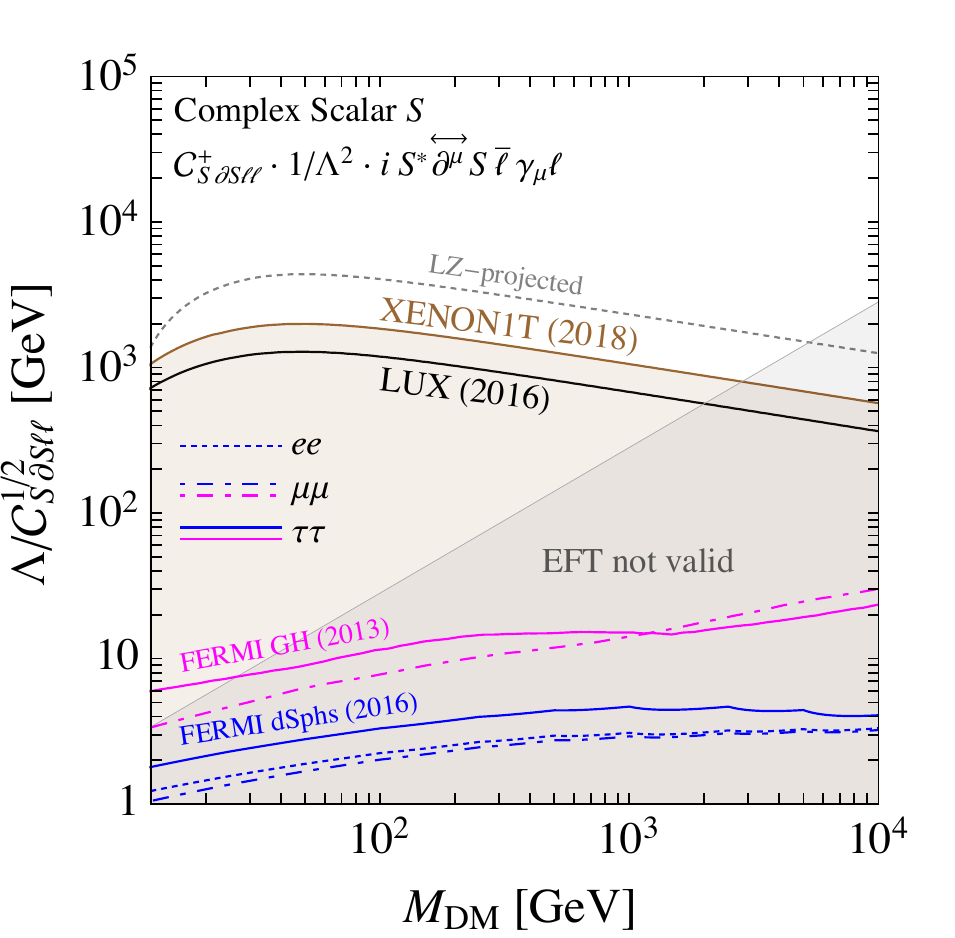}
 
\includegraphics[width= 0.4925 \textwidth]{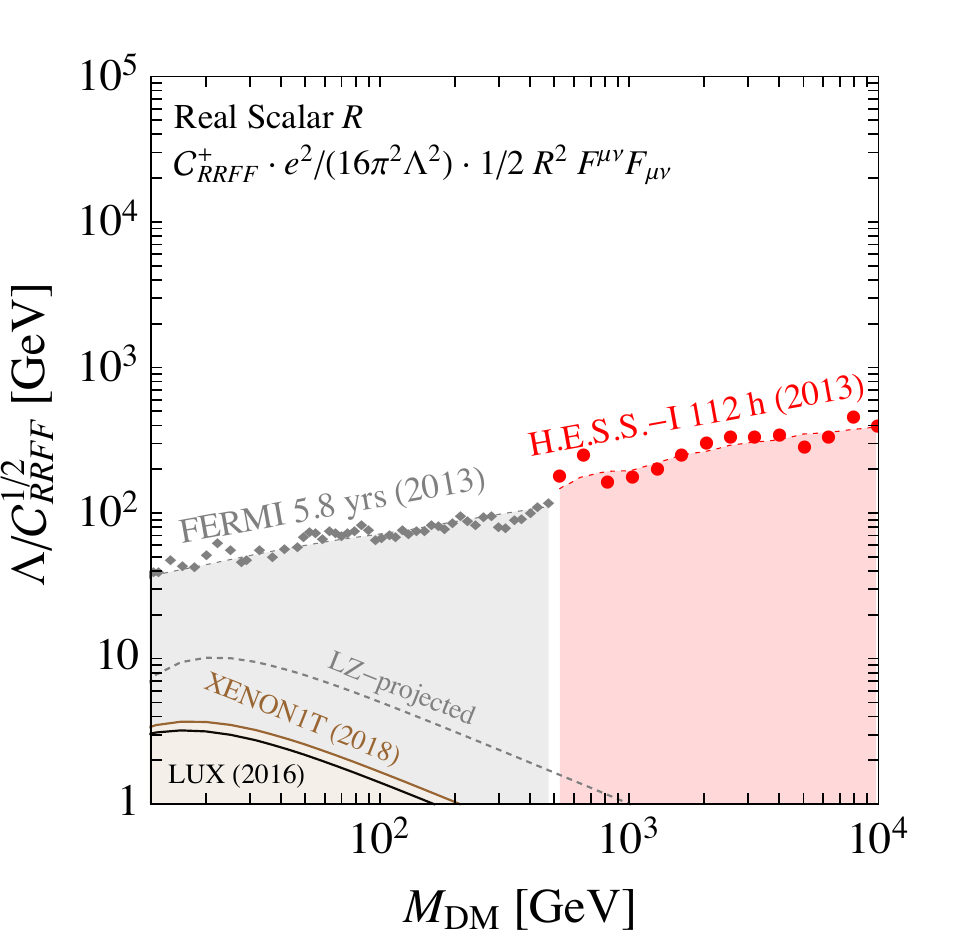} \
\includegraphics[width= 0.4925 \textwidth]{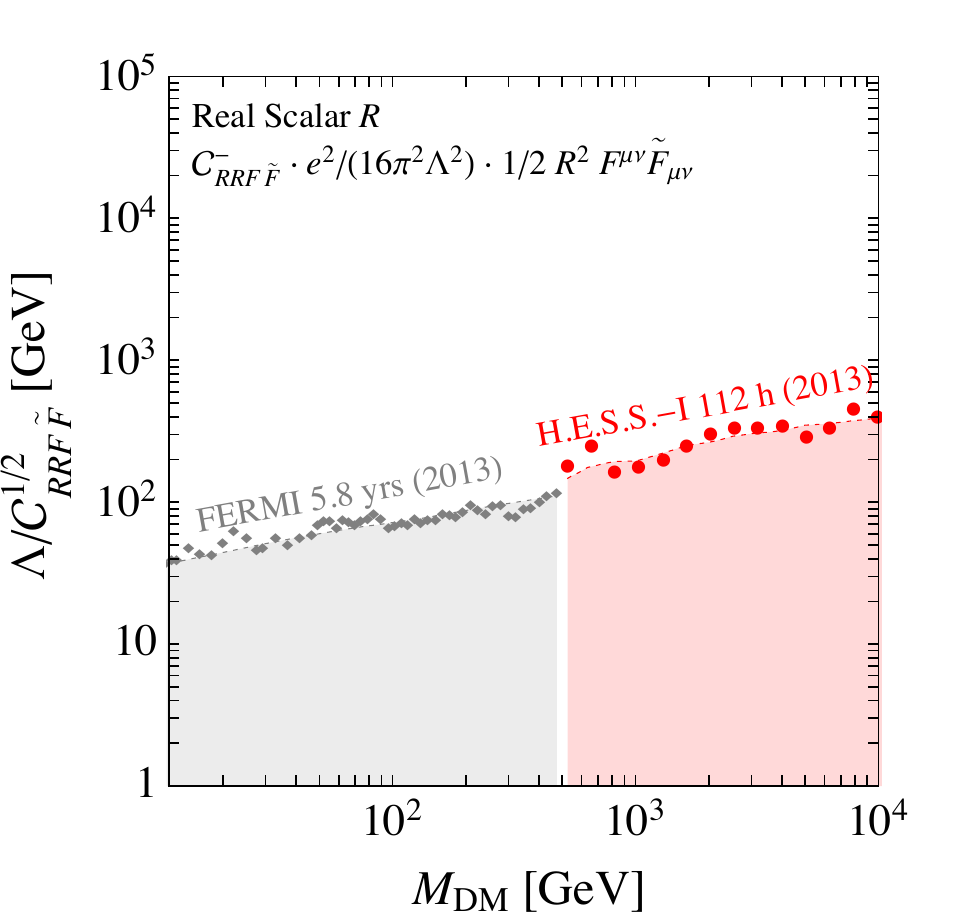} \\

\caption{Same as Fig.~\ref{fig:EFT_Constraints_Dirac} but for Scalar Dark Matter. In the top row, we consider operators for complex scalar DM $S$ and in the bottom row we consider real scalar DM $R$.\label{fig:EFT_Constraints_Scalar}}
\end{figure}

In Fig.~\ref{fig:EFT_Constraints_Scalar}, we show constraints on the most  revelant high-scale operators which give rise to low energy interactions between photons and complex scalar Dark Matter $S$ (top row) or real scalar Dark Matter $R$ (bottom row). 

In the top left panel of Fig.~\ref{fig:EFT_Constraints_Scalar}, we show constraints on the scalar charge radius operator, $\mathcal { O } _ { \partial S \partial S F }$. In direct detection, this gives rise to a short range coherently-enhanced coupling to the protons in the nucleus (up to $\mathcal{O}(1)$ factors, the cross section is the same as for the fermion charge radius). The resulting constraint is $\Lambda/\mathcal{C}_{ \partial S \partial S F}^{1/2} \gtrsim 400\,\,\mathrm{GeV}$ for complex scalar DM with a mass around 50 GeV. For indirect detection, the charge radius operator couples only to electromagnetic currents, so annihilations into photons is not possible at tree-level. This means that searches for gamma ray lines are not constraining.

In the top right panel of Fig.~\ref{fig:EFT_Constraints_Scalar}, we show constraints on the complex scalar coupling to Standard Model leptons, $\mathcal{O}_{S\partial S \ell \ell}$, which induces the scalar charge radius operator radiatively. As we saw in previous sections (and in particular in the top row of Fig.~\ref{fig:EFT_Constraints_Majorana}), the logarithmic enhancement to the scalar charge radius operator coming from the RG evolution dominates over the contribution from the tree-level $\mathcal { O } _ { \partial S \partial S F }$ defined at the high scale. The resulting constraint is $\Lambda/\mathcal{C}_{S\partial S \ell \ell} \gtrsim 2000 \,\,\mathrm{GeV}$, stronger than the constraint on $\mathcal { O } _ { \partial S \partial S F }$. Indirect detection constraints come from annihilation directly into charged leptons, although the cross section is $p$-wave suppressed, leading to much weaker constraints than those coming from direct detection. 

In the bottom panels of Fig.~\ref{fig:EFT_Constraints_Scalar}, we show constraints on the dimension-6 real scalar Rayleigh operators, $\mathcal{O}_{RRFF}$ and $\mathcal{O}_{RRF\tilde{F}}$. The resulting constraints follow a similar pattern as for the Majorana fermion Rayleigh operators (bottom row of Fig.~\ref{fig:EFT_Constraints_Majorana}). While the scalar Rayleigh operators are dimension-6 (rather than dimension-7 as in the case of the Majorana fermion), direct detection constraints on $\mathcal{O}_{RRBB}$ are still very weak. They also weaken rapidly with increasing $m_\mathrm{DM}$, as the cross section scales as $\sigma_{RRFF} \sim m_\mathrm{DM}^{-2} \Lambda^{-4}$. For both operators $\mathcal{O}_{RRFF}$ and $\mathcal{O}_{RRF\tilde{F}}$, the annihilation cross section is $s$-wave and scales as $m_\mathrm{DM}^2 \Lambda^{-4}$, meaning that constraints from indirect detection are similar  for the two operators and  dominate over direct detection in both cases.

\subsection{UV-related Operators}
Finally we discuss the bounds on the UV-related tree-level operators given in Eqs.~\eqref{D4l5} and \eqref{dim8}. In particular, we highlight cases where bounds on UV-related operators are stronger than bounds on DM-photon operators with the same CP properties. In such cases, realistic UV models may be more strongly constrained than would be expected from considering only the DM-photon interactions which they give rise to.

 \begin{figure}[!t]
 \centering
 
 \includegraphics[width= 0.4925 \textwidth]{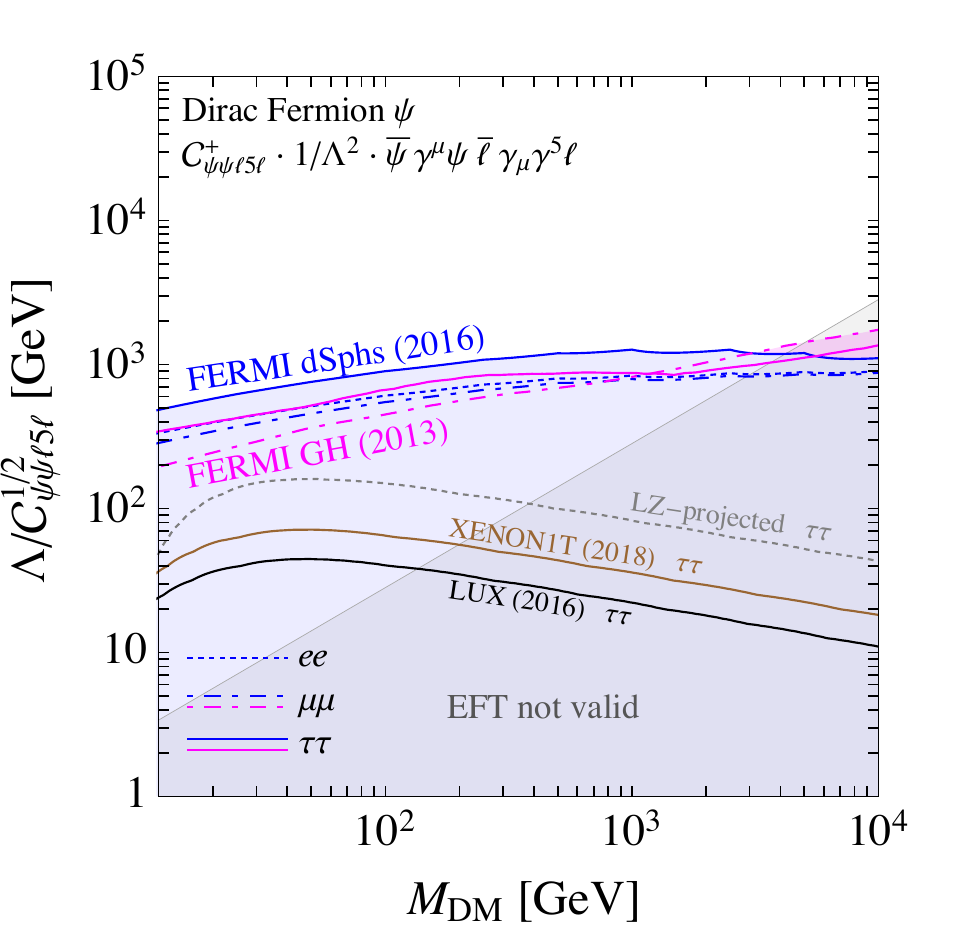} \
 \includegraphics[width= 0.4925 \textwidth]{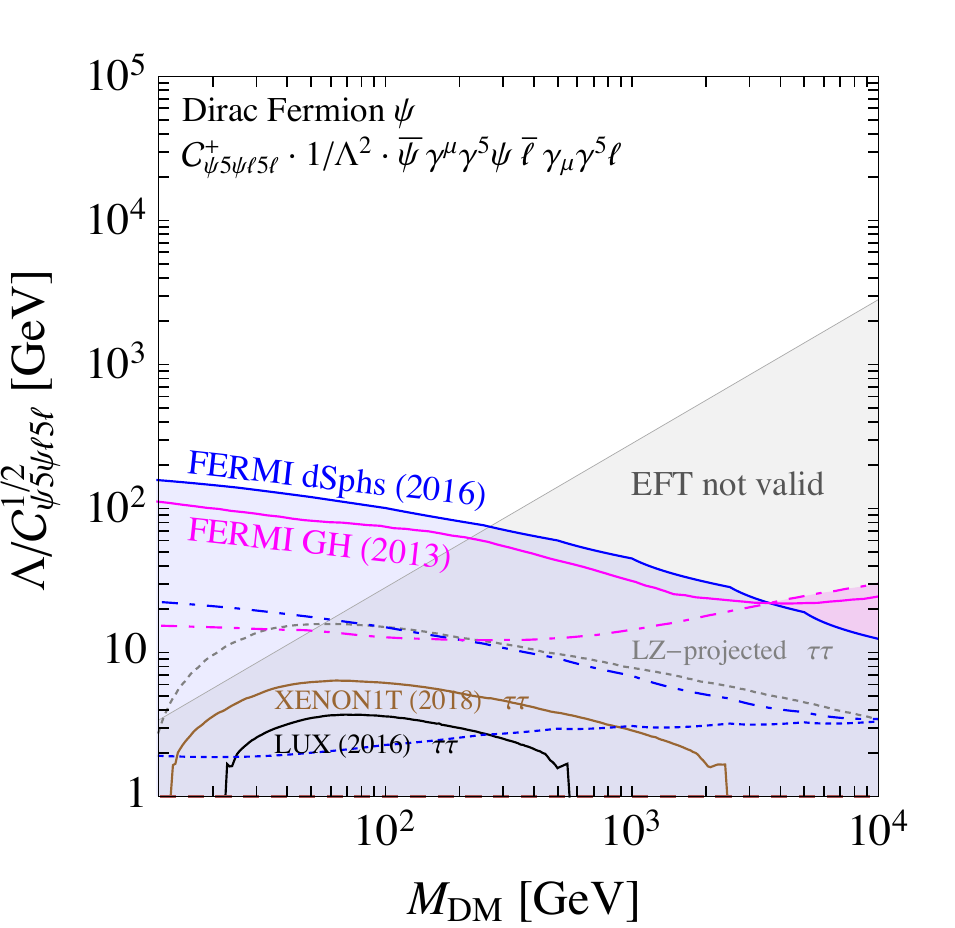} \\
 
  \includegraphics[width= 0.4925 \textwidth]{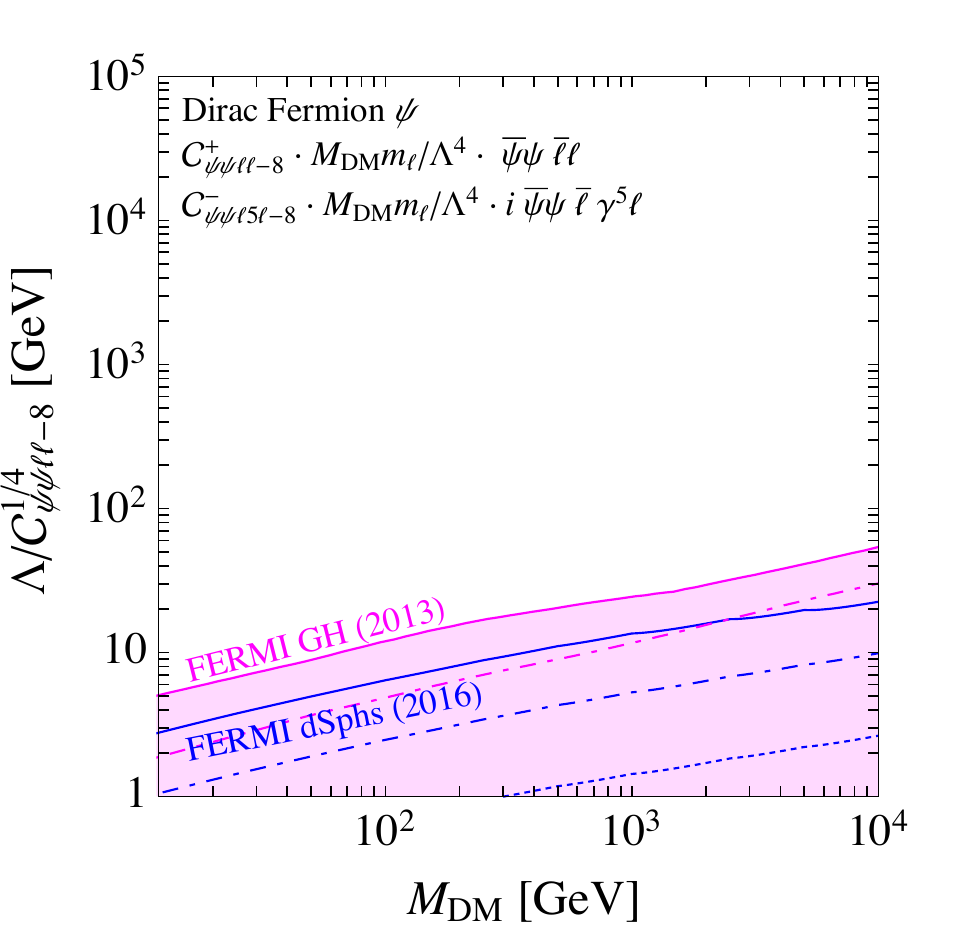} \
 \includegraphics[width= 0.4925 \textwidth]{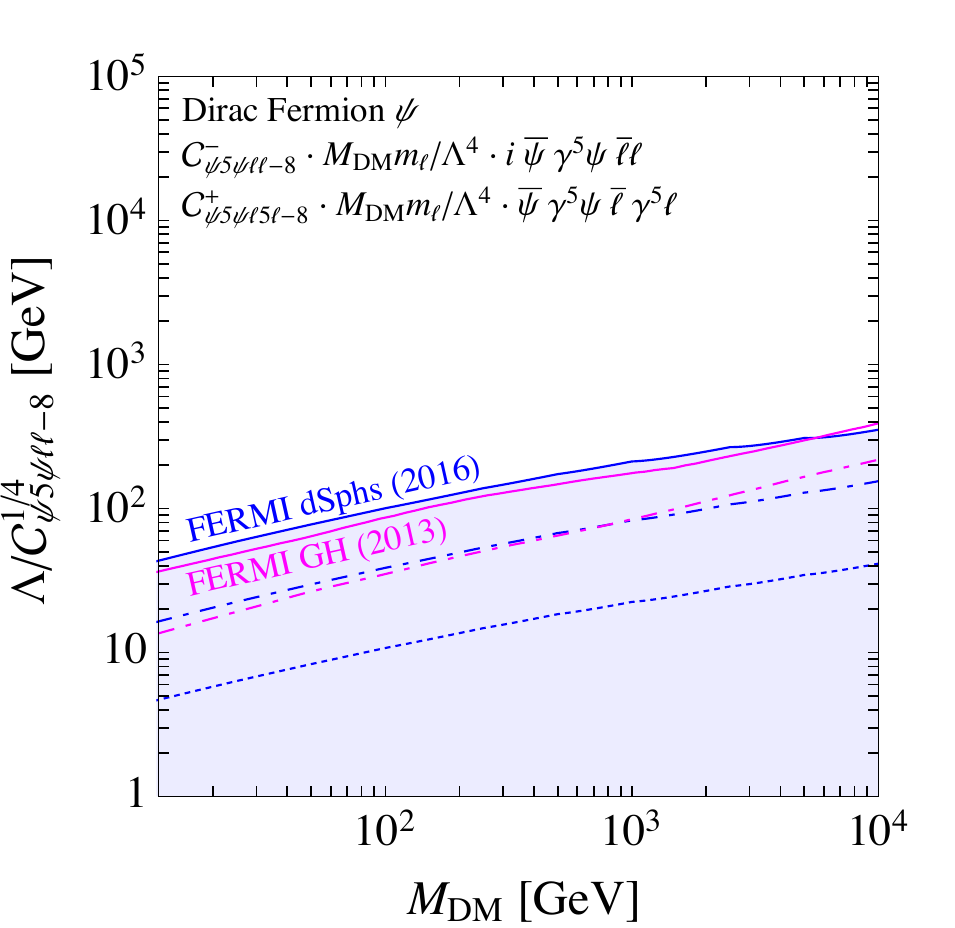}

\caption{Constraints from direct and indirect detection on UV-related operators for Dirac Fermion DM $\psi$. In the top row, we show the dimension-6 operators with axial-vector lepton currents. In the bottom row, we show the dimension-8 scalar operators. The UV-related operators are discussed in detail in Sec.~\ref{sec:UVrelated}. These constraints should be compared with the corresponding constraints from DM-photon operators shown in Fig.~\ref{fig:EFT_Constraints_Dirac}.\label{fig:UVrelated_Constraints_Dirac}}
\end{figure}

In Fig.~\ref{fig:UVrelated_Constraints_Dirac}, we show bounds on dimension-7 (top row) and dimension-8 (bottom row) UV-related operators for Dirac Fermion DM $\psi$.  In the top left panel, we show constraints on the operator $\mathcal{O}_{\psi \psi \ell 5 \ell}$. These come predominantly from annihilation into charged leptons at tree-level, with constraints from dwarf Spheroidals and the Galactic halo being comparable ($\Lambda/\mathcal{C}_{\psi \psi \ell 5 \ell}^{1/2} \gtrsim 200 - 1000\,\,\mathrm{GeV}$). Direct detection constraints arise only at loop-level, driven by the charged lepton Yukawa \cite{DEramo:2017zqw}. For coupling to $e$ and $\mu$, DD constraints are negligible, while for coupling to the $\tau$ constraints may extend up to $\mathcal{O}(50 \,\mathrm{GeV})$. We see from Fig.~\ref{fig:UVrelated_Constraints_Dirac} that this operator $\mathcal{O}_{\psi\psi \ell 5 \ell}$ is the most strongly constrained of the Dirac fermion UV-related operators. Even so, these constraints are much weaker than the corresponding constraints from the magnetic dipole operator $\mathcal{O}_{\psi\psi F}$ (CP-even) or the electric dipole operator $\mathcal{O}_{\psi 5 \psi F}$ (CP-odd), presented in the top row of Fig.~\ref{fig:EFT_Constraints_Dirac}. For Dirac DM then, we expect that the DM-photon EFT should capture the most relevant constraints, even including additional operators which may appear in the UV.

 \begin{figure}[!t]
 \centering

 \includegraphics[width= 0.4925 \textwidth]{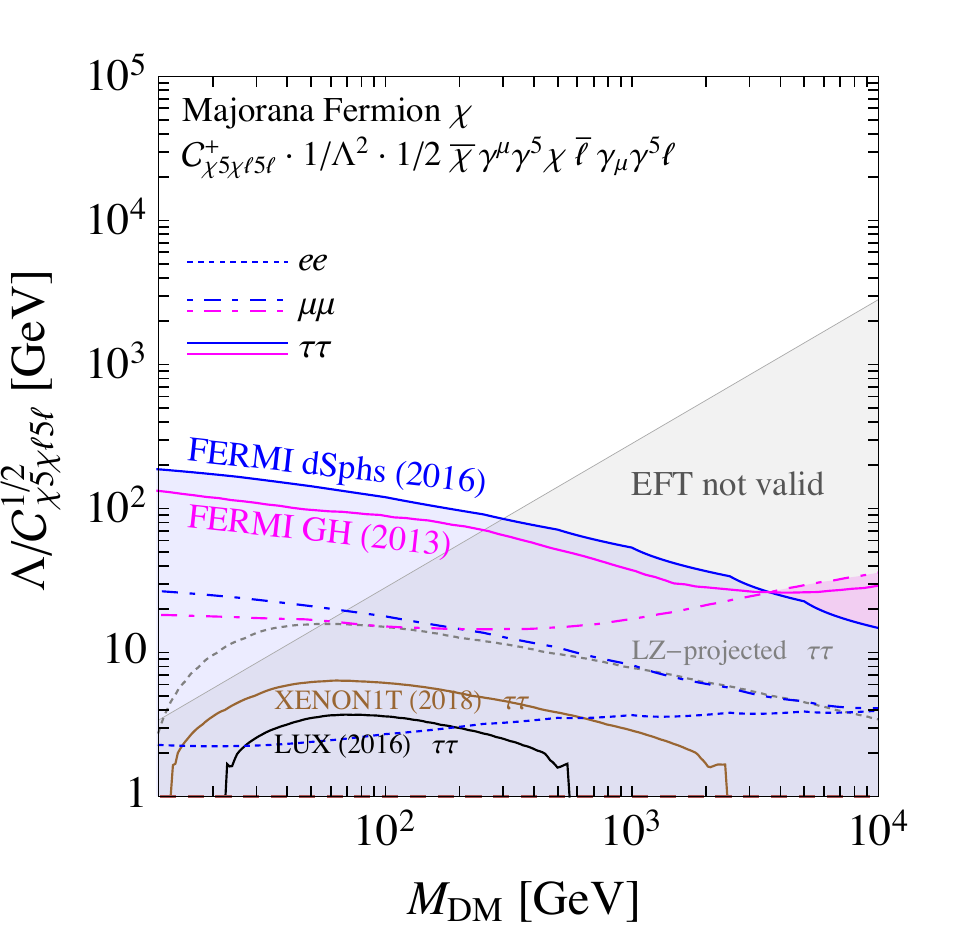} \\
 
  \includegraphics[width= 0.4925 \textwidth]{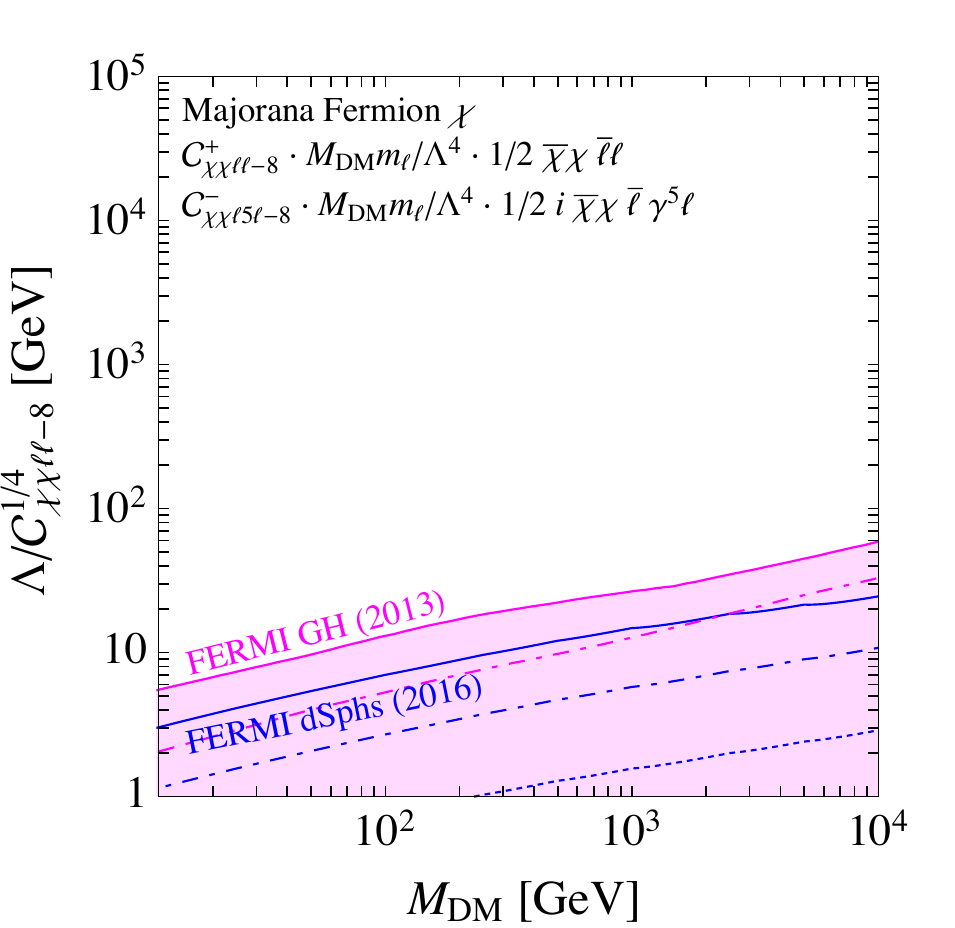} \
 \includegraphics[width= 0.4925 \textwidth]{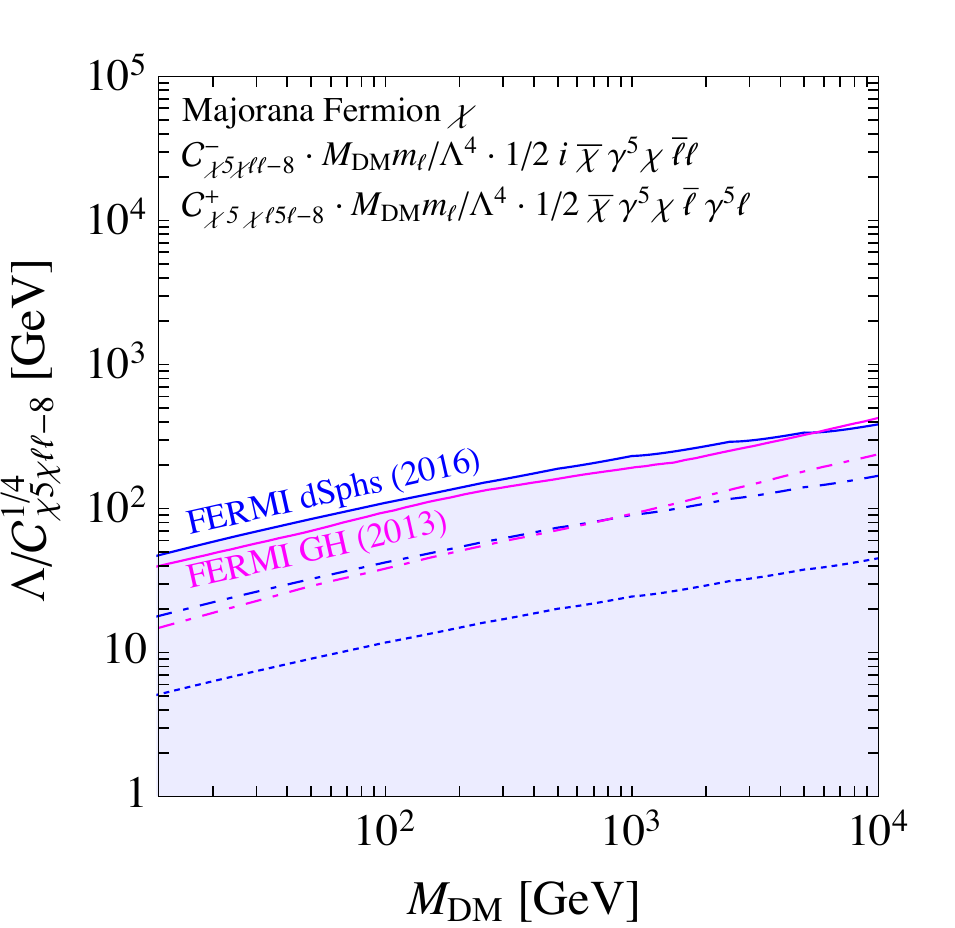} 

\caption{Same as Fig.~\ref{fig:UVrelated_Constraints_Dirac} but for the UV-related operators for Majorana Fermion $\chi$. These constraints should be compared with the corresponding constraints from DM-photon operators shown in Fig.~\ref{fig:EFT_Constraints_Majorana}. \label{fig:UVrelated_Constraints_Majorana}}
\end{figure}

In Fig.~\ref{fig:UVrelated_Constraints_Majorana}, we show constraints on the UV-related operators for Majorana Fermion DM $\chi$. Considering the operator $\mathcal{O}_{\chi 5 \chi \ell 5 \ell}$ (top), we see that DD constraints are again loop-suppressed and non-zero only in the case of coupling to $\tau$ leptons. In ID, the annihilation cross section has both an $s$-wave contribution (proportional to $m_\ell^2$) and a $p$-wave contribution (proportional to $m_\chi^2$). For large DM masses, the FERMI constraints are similar to those from $\mathcal{O}_{\chi 5 \chi \ell \ell}$ (shown in Fig.~\ref{fig:EFT_Constraints_Majorana}), arising from the $p$-wave contribution (with the limits being almost universal for $e$, $\mu$ and $\tau$ for $m_\chi > 2 \,\mathrm{TeV}$). Instead, for lighter DM, the $s$-wave contribution becomes dominant. Constraints for different charged leptons then diverge, substantially strengthening in the case of coupling to the heavier $\tau$. These ID constraints may be competitive with the DD constraints arising from the operator $\mathcal{O}_{\chi 5 \chi \ell \ell}$ (which induces the anapole operator at low energy). In contrast to the Dirac case, the strongly-constrained electric and magnetic dipoles operators are forbidden for Majorana DM so constraints from the UV-related $\mathcal{O}_{\chi 5 \chi \ell 5 \ell}$ may be as relevant as the strongest constraints from DM-photon interactions, together with bounds from perturbative unitarity which dominate above DM masses of around 300 GeV.

In the bottom row of Fig.~\ref{fig:UVrelated_Constraints_Majorana}, we show constraints on the dimension-8 UV-related operators. These couple DM (pseudo-)scalar currents to lepton (pseudo-)scalar currents and do not give rise to DD signals. For operators coupling to the scalar DM current $\overline{\chi}\chi$ (bottom left panel), the annihilation cross section is $p$-wave suppressed, leading to rather weak constraints, at the level of $\Lambda/\mathcal{C}_{\chi\chi \ell (5)\ell-8}^{1/4} \gtrsim 10\,\mathrm{GeV}$, depending on the lepton. Instead, annihilation through the pseudo-scalar DM current $\overline{\chi} \gamma^5\chi$ (bottom right panel) is $s$-wave, leading to constraints on $\Lambda$ up to $100\,\mathrm{GeV}$ for couplings to the $\tau$. Again, such constraints may be competitive with constraints from DD and ID shown in Fig.~\ref{fig:EFT_Constraints_Majorana}, coming from the operator $\mathcal{O}_{\chi 5 \chi \ell \ell}$ (CP-even) or the Rayleigh operators $\mathcal{O}_{\chi 5 \chi F \tilde{F}}$ (CP-even) and $\mathcal{O}_{\chi 5 \chi F F}$ (CP-odd). This suggests that the UV-related operators may be particularly relevant in the case of Majorana DM (though typically only for coupling to the $\tau$).

 \begin{figure}[!t]
 \centering

 \includegraphics[width= 0.4925 \textwidth]{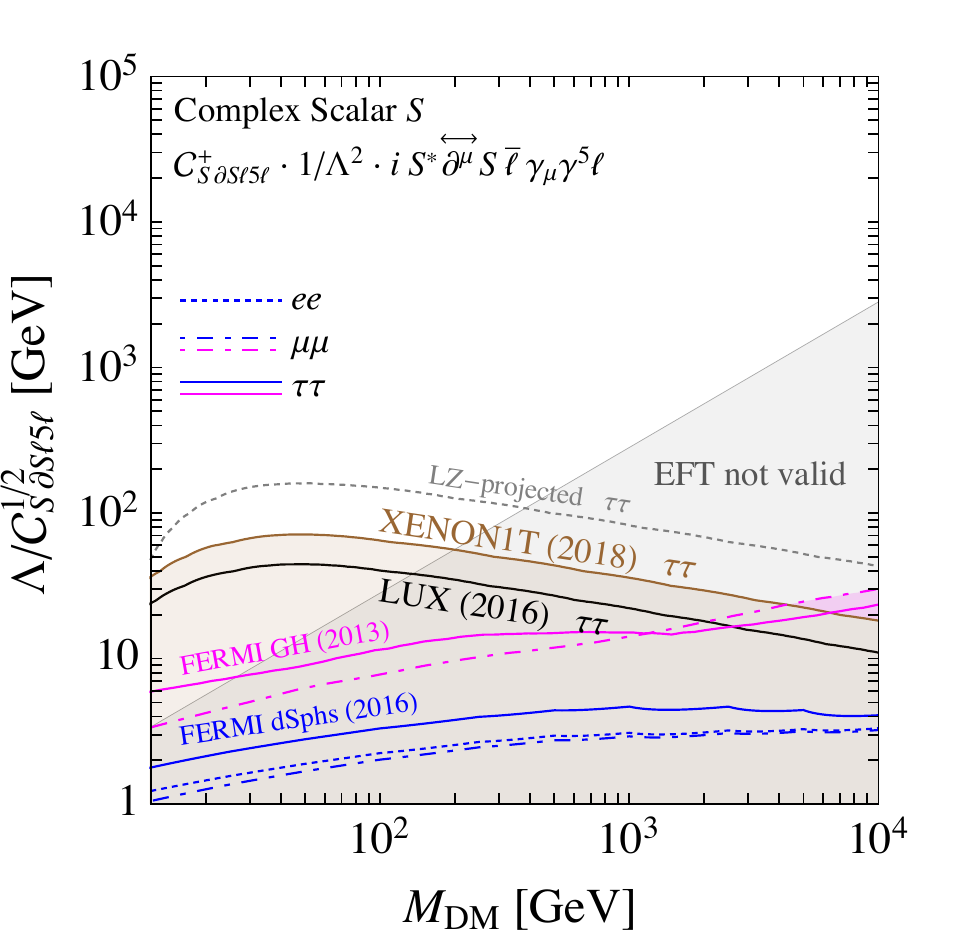} \
  \includegraphics[width= 0.4925 \textwidth]{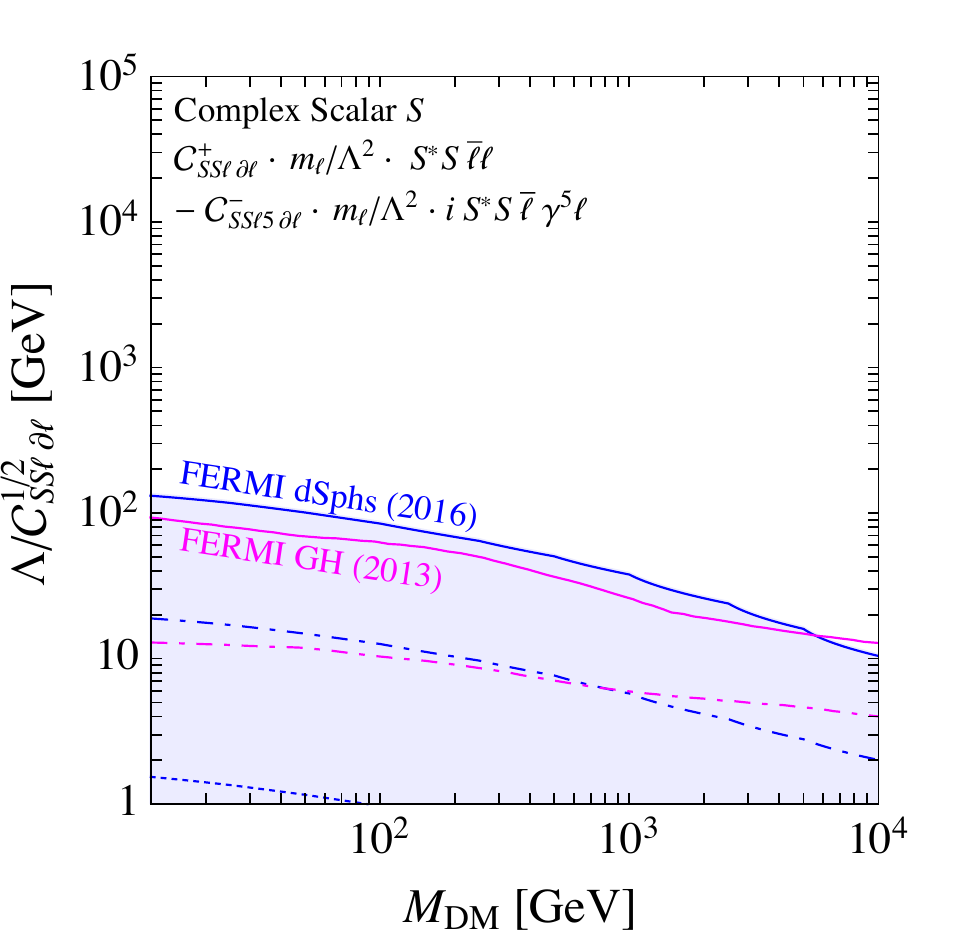} \\
 
 \includegraphics[width= 0.4925 \textwidth]{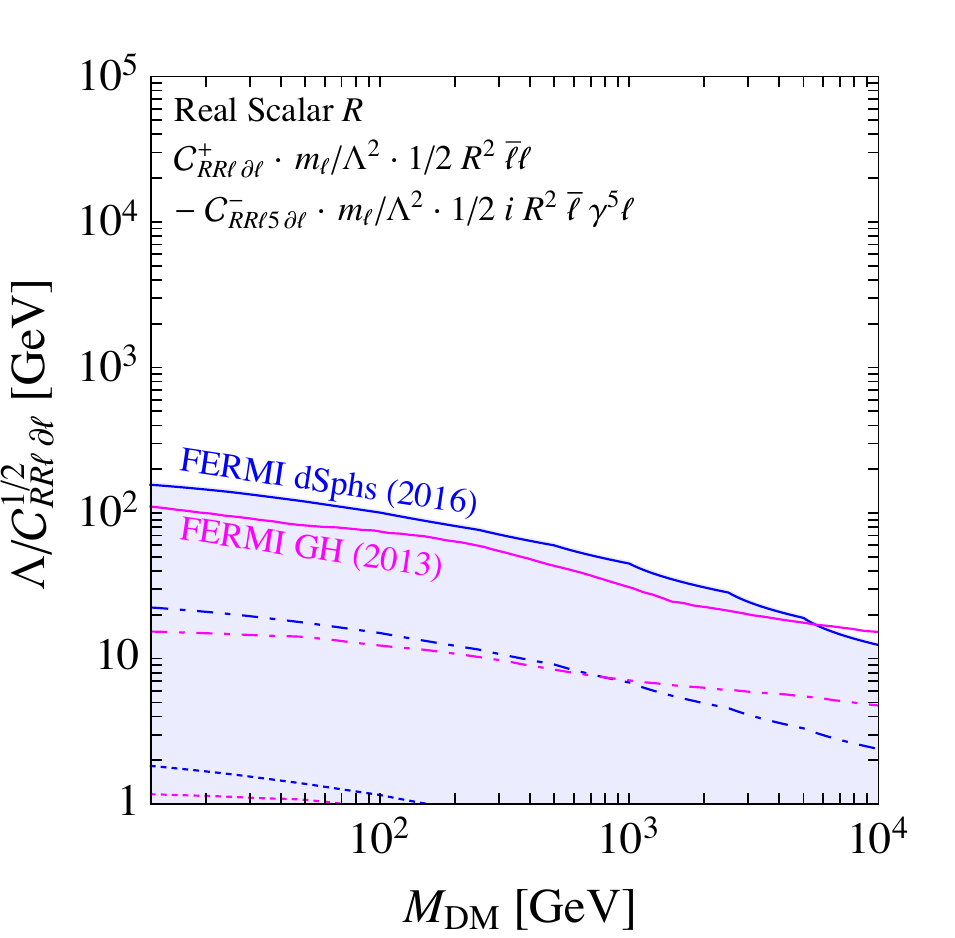}

\caption{Same as Fig.~\ref{fig:UVrelated_Constraints_Dirac} but for the UV-related operators for complex (top) and real (bottom) Scalar DM. These constraints should be compared with the corresponding constraints from DM-photon operators shown in Fig.~\ref{fig:EFT_Constraints_Scalar}.\label{fig:UVrelated_Constraints_Scalar}}
\end{figure}

In Fig.~\ref{fig:UVrelated_Constraints_Scalar}, we show constraints on UV-related operators for complex (top row) and real (bottom) scalar DM. For the operator $\mathcal { O } _ { S \partial S \ell 5 \ell }$ (top left), the dominant constraints come from direct detection (which is loop-suppressed  and only relevant for coupling to the $\tau$, as in the case of fermionic DM); in this case the annihilation cross section is $p$-wave suppressed. For the remaining operators, the bounds arise from $s$-wave annihilation into leptons at tree-level, with negligible DD constraints. For the case of complex scalar DM, these limits are always weaker than limits from the (induced) scalar charge radius (top row of Fig.~\ref{fig:EFT_Constraints_Scalar}): the UV-related operators can safely be ignored and the EFT describing DM-photon interactions should capture all the relevant effects. For the real scalar, these bounds may be competitive with constraints from the Rayleigh operators $\mathcal{O}_{RRFF}$ and $\mathcal{O}_{RRF\tilde{F}}$ (bottom row of Fig.~\ref{fig:EFT_Constraints_Scalar}), at low DM mass and for coupling to the $\tau$ only. The annihilation cross section for the dimension-6 scalar operators scales as $m_\ell^2$, meaning that constraints for coupling to $e$ and $\mu$ are much weaker and can typically be neglected.


\section{Example UV Completions}
\label{sec:UVmodels}

In this section we consider two simple examples of UV models that induce DM-photons interactions at one-loop by coupling DM to SM leptons. This mainly serves as a illustration of the usefulness of the EFT analysis for deriving constraints from DD and ID, which we compare to bounds obtained from a fixed-order calculation. This allows us to discuss the region of UV parameter space where the EFT is valid and identify the regime of EFT breakdown, in particular in DD. We complete this section by considering also collider constraints, where Bhabha scattering at LEP turns out to provide an important bound on the UV completion with electrons.   

Note that the two explicit `Lepton Portal' scenarios we consider here were also studied in detail in e.g.~Ref.~\cite{Bai:2014osa} (see also Refs.~\cite{Agrawal:2011ze, Giacchino:2013bta, Garny:2015wea, Baker:2018uox}). More complex UV completions, involving richer dark sectors, are of course possible (see e.g~Refs.~\cite{Pierce:2014spa,Baker:2018uox,Herrero-Garcia:2018koq}) but we restrict ourself here to the `minimal' scenario of DM coupling to light SM leptons.




%
\subsection{Fermion DM: DD and ID Constraints}
\label{sec:FermionUV}
We add to the SM a singlet fermion $\chi$ that is either Dirac or Majorana and a complex scalar $S$ with hypercharge $-1$. The relevant part of the Lagrangian is given by 
\begin{align}
{\cal L} & = \left( \lambda_i \, S  \overline{\chi} P_R  \ell_i  + {\rm h.c.} \right) - M_{\rm DM} \overline{\chi} \chi  - M_S^2 |S|^2 \, , 
\end{align} 
where $\ell_i$ denotes a SM lepton. In order to avoid constraints from lepton flavor-violating processes, we only consider the case of a single non-vanishing $\lambda_i$. We can then always make $\lambda_i$ real through a phase redefinition of $S$, so this theory conserves CP. We take $M_{\rm DM} \ll M_S$, so $\chi$ is a DM candidate that is stable because of a $Z_2$ symmetry. 

\begin{figure}[!t]
 \centering
\includegraphics[width= 0.315 \textwidth]{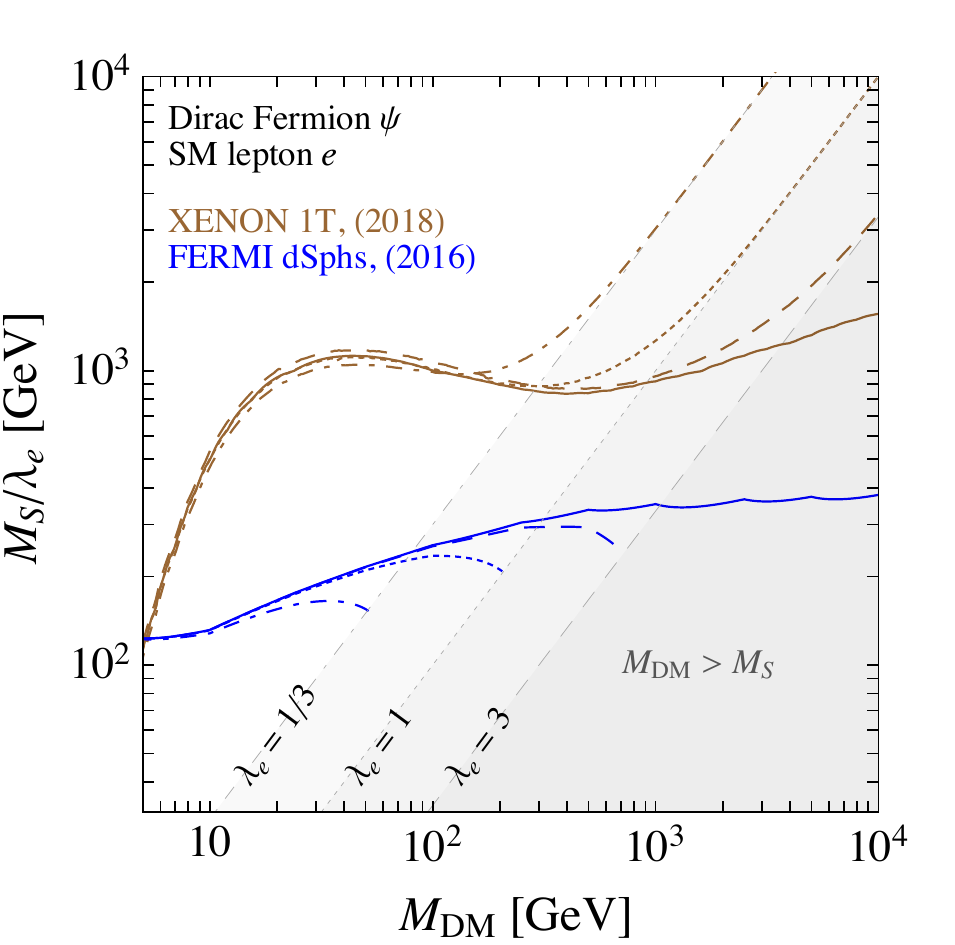} \
\includegraphics[width= 0.315 \textwidth]{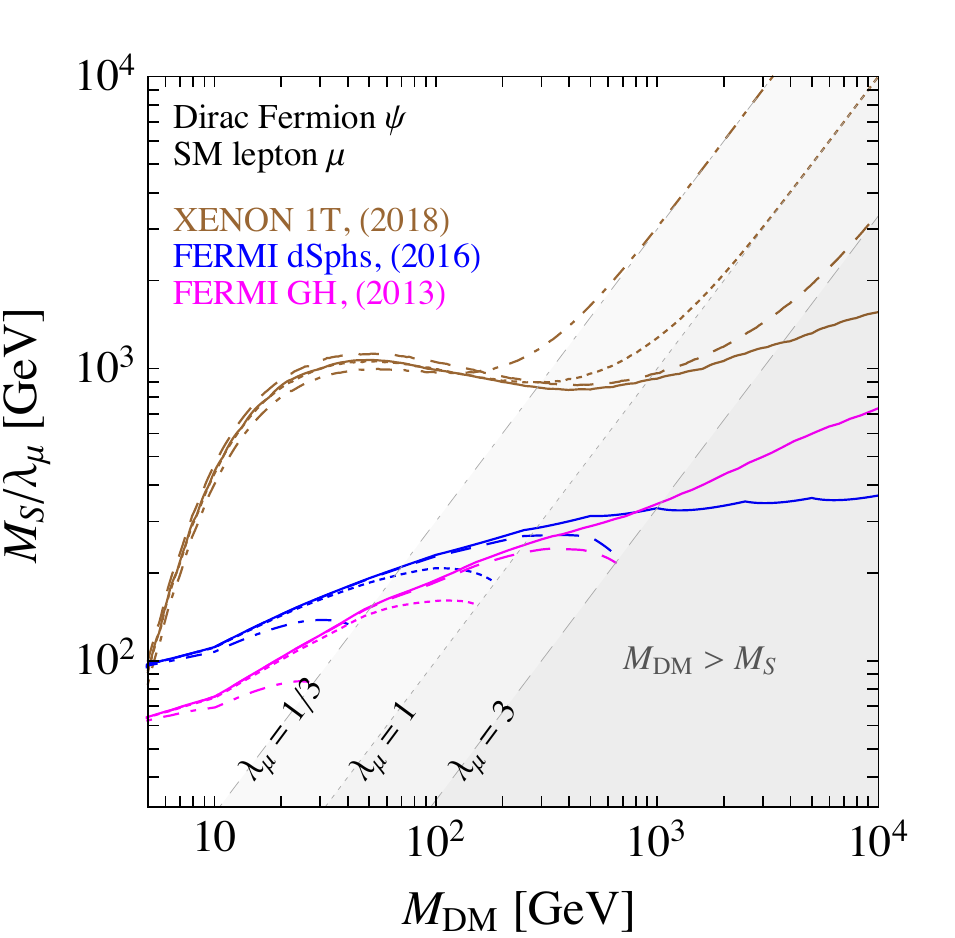} \
\includegraphics[width= 0.315 \textwidth]{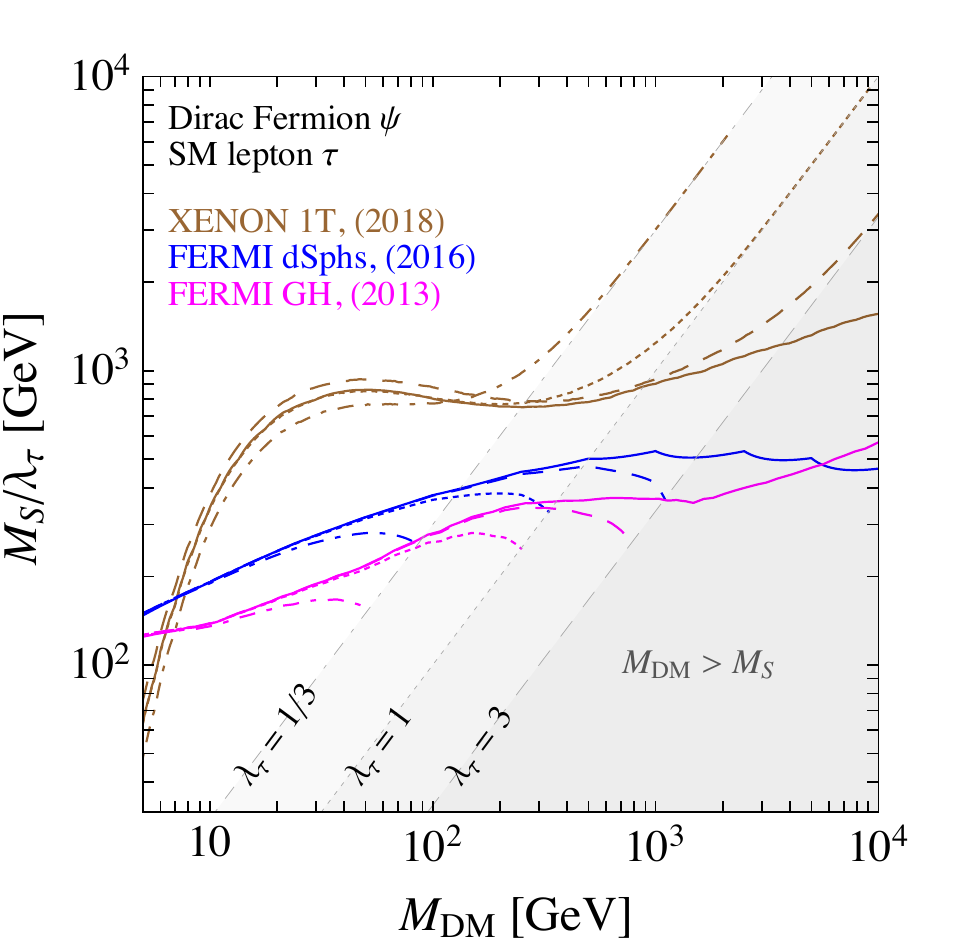} \\

\includegraphics[width= 0.315 \textwidth]{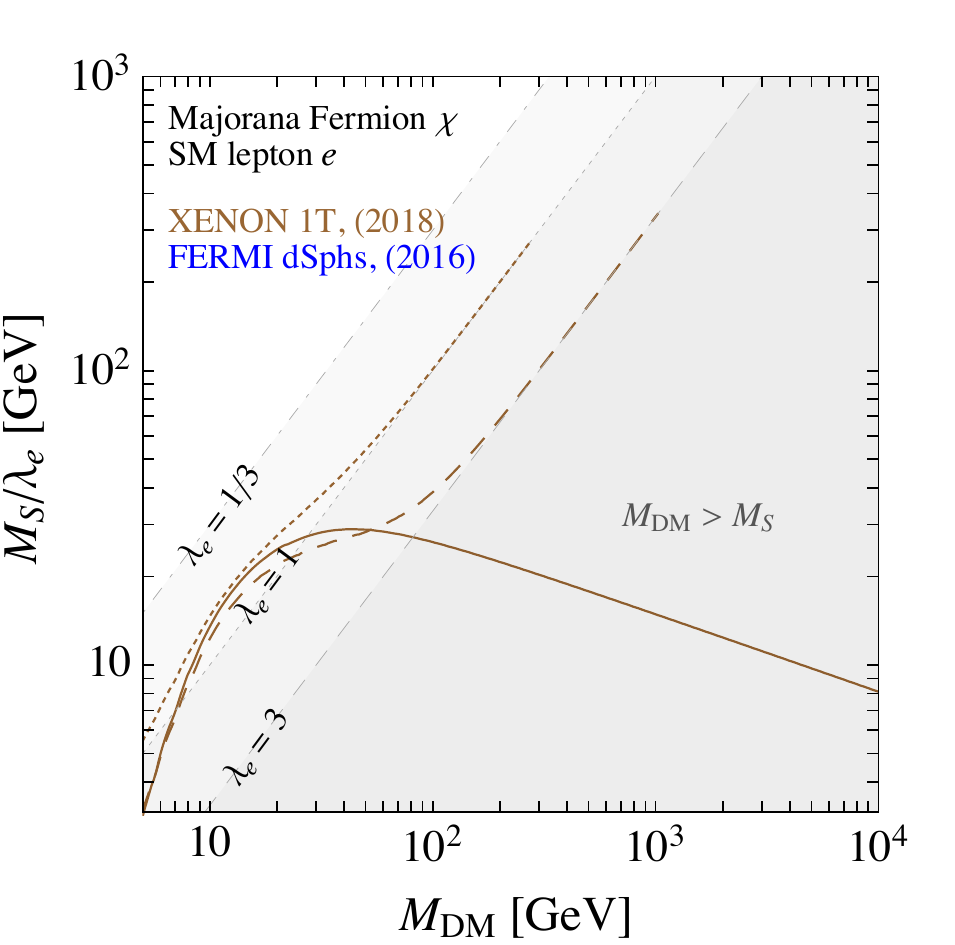} \
\includegraphics[width= 0.315 \textwidth]{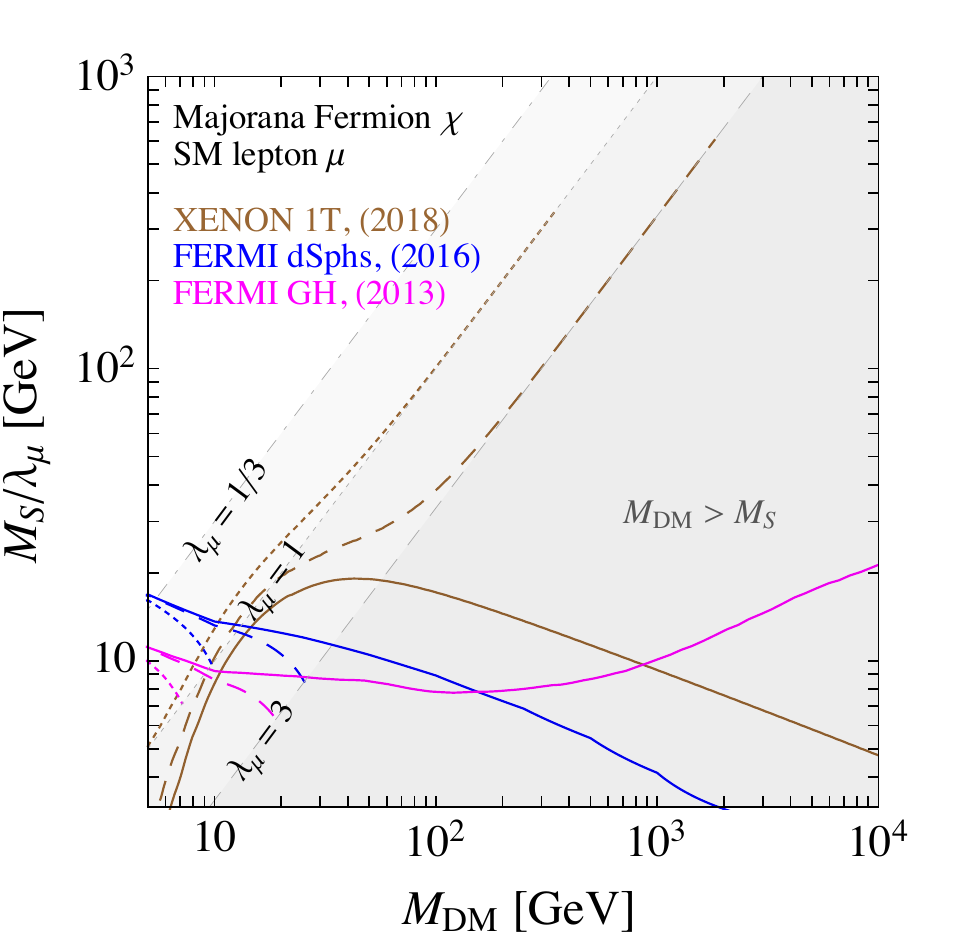} \
\includegraphics[width= 0.315 \textwidth]{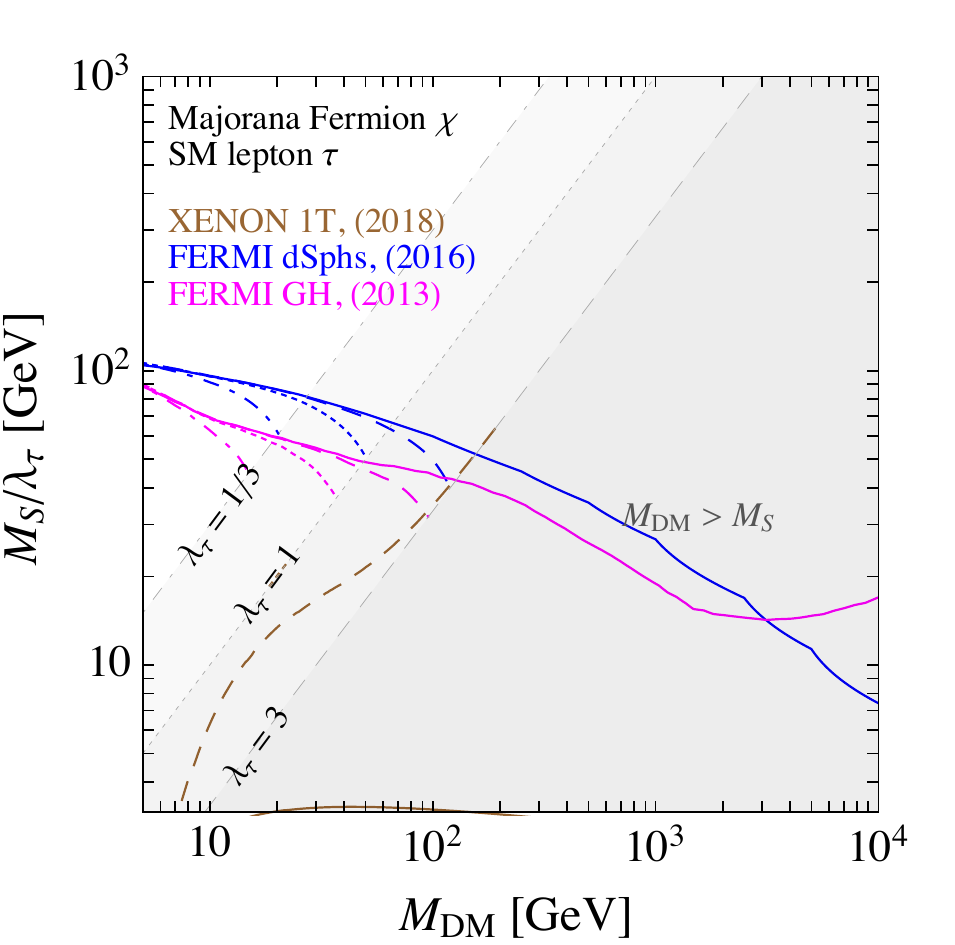} 

\caption{\label{fig:EFT_vs_FUVC} Constraints on the UV models for Dirac (upper panels) and Majorana DM (lower panels), for couplings to $e, \mu$ and $\tau$. In grey is shown the excluded region of parameter space where $M_{\rm DM} > m_S$, which depends on $\lambda_i$ due to the choice of the y-axis. Regions below the brown lines are excluded by DD constraints from XENON1T, while the blue (magenta) lines denote the bounds from ID, coming from FERMI observations of dwarf Spheroidal galaxies (Galactic Halo). Constraints shown as solid lines arise from considering just the effective operators, and only depend on the combination $m_S/\lambda_i$. The full fixed order calculation also depends on $\lambda_i$ and $m_S$ separately, and we show the resulting constraints for $\lambda_i = 1/3$ (dot-dashed), $\lambda_i = 1$ (dotted) and $\lambda_i = 3$ (dashed).}
\end{figure}

We now match this theory to the relevant (dominant) effective operators in Table~\ref{Table:Operators}, by integrating out the heavy scalar and expanding in powers of external momenta or equivalently $M_{\rm DM}$. As a result of CP invariance, we have $C_{\psi 5 \psi F} = 0$, while for the non-vanishing operators the parametric dependence on $\lambda_i, M_{\rm DM}, M_S$ can be inferred from dimensional analysis, and only the numerical factors have to be calculated. Below we show only the results for the most relevant operator coefficients\footnote{We do not consider the Rayleigh operator coefficients, since they arise only at dimension 8.}. For the Dirac case, one obtains 
\begin{equation}
\begin{split}
&\frac{C_{\psi \psi F}}{\Lambda}  =   \frac{ \lambda_i^2 M_{\rm DM}}{4 M_S^2} \, , \quad \,\,\,
\frac{C_{\psi 5 \psi F}}{\Lambda}  = 0 \, ,  \qquad \quad \,\,\,\,\,
\frac{C_{\psi \psi \partial F}}{\Lambda^2}  = - \frac{ \lambda_i^2}{4 M_S^2} \, , \quad \, 
\frac{C_{\psi 5 \psi \partial F}}{\Lambda}  = \frac{\lambda_i^2}{4 M_S^2} \, ,  \\
&\frac{C_{\psi \psi \ell \ell}}{\Lambda^2}  = - \frac{\lambda_i^2}{8 M_S^2} \, , \quad \,\,
\frac{C_{\psi  \psi \ell 5 \ell}}{\Lambda^2}  = - \frac{\lambda_i^2}{8 M_S^2}  \, ,  \quad \,
\frac{C_{\psi  5 \psi \ell  \ell}}{\Lambda^2}  = \frac{\lambda_i^2}{8 M_S^2}  \, ,  \quad \,\,\,\,\,
\frac{C_{\psi  5 \psi \ell 5 \ell}}{\Lambda^2}  = \frac{\lambda_i^2}{8 M_S^2} \, ,
\label{FDMmatchingD}
\end{split}
\end{equation}
while in the Majorana case one finds 
\begin{align}
 \frac{C_{\chi 5 \chi \partial F}}{\Lambda} & = \frac{\lambda_i^2}{4 M_S^2} \, , & \frac{C_{\chi  5 \chi \ell  \ell}}{\Lambda^2} & = \frac{\lambda_i^2}{4 M_S^2}  \, , & 
\frac{C_{\chi  5 \chi \ell 5 \ell}}{\Lambda^2} & = \frac{\lambda_i^2}{4 M_S^2} \, .
\label{FDMmatchingM}
\end{align}
Using these coefficients, one can derive the allowed regions in the UV parameter space $M_{\rm DM} - M_S/\lambda_i$, just as we derived bounds on the individual effective operators from Sec.~\ref{sec:Bounds}. These are shown as solid lines in the upper (Dirac) and lower (Majorana) panels of Fig.~\ref{fig:EFT_vs_FUVC}, for all three SM leptons.

It is instructive to compare these bounds, which have been obtained from matching to the EFT at leading order in $M_{\rm DM}/M_S$ and  $m_{\ell_i}/M_S$, to the constraints obtained from a fixed order calculation in the full UV theory. This gives the result for the Dirac fermion DM-quark amplitude  
\begin{align}
 {\cal A} = \frac{e^2  Q_q}{16 \pi^2} \overline{u}_\chi (p^\prime) \left[ \gamma^\mu \left( C_\gamma  + C_{\gamma5} \gamma_5 \right) 
 + C_\sigma i  \sigma^{\mu \nu} \frac{q_\nu}{q^2} \right] u_\chi (p) \cdot  \overline{u}_q (k^\prime) \gamma_\mu u_q (k) \, ,
 \label{FDMauv}
\end{align}
where $p$ ($k$) is the momentum of the incoming DM fermion $\chi$ (incoming quark $q = u,d$), $p^\prime$ ($k^\prime$) is the momentum of the outgoing DM fermion $\chi$ (outgoing quark $q$) and $q = p - p^\prime = k^\prime - k$. Furthermore $Q_q$ denotes the electric charges of the SM quarks, and the functions $C_\alpha (\lambda_i, m_{\ell_i}, q, M_{\rm DM}, M_S)$ can be found in Appendix~\ref{appfull}. The case of Majorana DM is given by the Dirac DM result with the replacements $C_\sigma \to 0, C_\gamma \to 0, C_{\gamma 5} \to 2 \, C_{\gamma 5}$.

From these expressions one can recover the EFT coefficients by taking the limit of small lepton masses, small DM masses, and small momentum transfer, $m_{\ell_i}^2, M_{\rm DM}^2, q^2 \ll M_S^2$. In this limit some diagrams acquire an IR divergence, which is cut-off by $m_{\ell_i}^2$ or $q^2$, depending on which is larger. For $m_{\ell_i}^2 > q^2$ one finds at leading order in $M_{\rm DM}^2/M_S^2$ 
\begin{align}
C_\gamma  & = - C_{\gamma 5} = \frac{\lambda_i^2}{12 M_S^2} \left( 3 + 2 \log \frac{m_{\ell_i}^2}{M_S^2} \right) \, , & C_\sigma  & =  \frac{\lambda_i^2 M_{\rm DM}}{4 M_S^2} \, .
\label{matching}
\end{align}
 One can check that these expressions yield the EFT result in Eq.~\eqref{FDMmatchingD} and Eq.~\eqref{FDMmatchingM}, apart from the logarithm that is reproduced in the EFT by running down the four-fermion operators, and indeed  recovers Eq.~\eqref{eq:LoopMatch}.
 
 The resulting bounds on the UV parameter space from DD (XENON1T) are shown in the upper (Dirac) and lower (Majorana) panels of Fig.~\ref{fig:EFT_vs_FUVC}, for all three SM leptons. In contrast to the EFT calculation, these bounds depend not only on the combination $\lambda_i^2/M_S^2$ but also on the ratio $M_{\rm DM}^2/M_S^2$. We have therefore chosen to present the bounds for three cases of the UV coupling $\lambda_i =  \{ 1/3, 1, 3\}$, which are denoted by brown dot-dashed, dotted and dashed lines, respectively. 
 
Before we discuss the results, we also need to consider the bound from ID  due to the direct annihiliation into leptons in the UV theory. The annihilation cross-sections into leptons $\ell_i$ are given by:
\begin{align}
\sigma(\overline{\chi}\chi \to \ell_i \overline{\ell}_i) \, v   & = \frac{ \lambda_i^4 }{32 \pi M_S^2} \frac{r_F^2}{(1+r_F^2)^2} \, , \\
\sigma(\chi \chi \to \ell_i \overline{\ell}_i) \, v  & = \frac{\lambda_i^4}{ 32 \pi M_S^2} \frac{1}{(1+r_F^2)^2}  \left( \frac{m_{\ell_i}^2}{M_S^2} + \frac{2 v^2 r_F^2 (1 + r_F^4)}{3 (1+r_F^2)^2} \right)   \, , 
\end{align}
for the case of Dirac/Majorana fermion, respectively, with $v$ denoting the relative velocity and $r_F = M_{\rm DM}/M_S < 1$. This  implies a dependence on three independent parameters $M_{\rm DM}, M_S$ and $\lambda_i$,  therefore we again present the bounds from ID (FERMI) in Fig.~\ref{fig:EFT_vs_FUVC} for three cases of the UV coupling $\lambda_i =  1/3, 1, 3$, which are denoted by blue dashed-dotted, dotted and dashed lines, respectively.

Besides the bounds obtained from the EFT and the full calculation for both DD (brown) and ID (blue and magenta), we have indicated in Fig.~\ref{fig:EFT_vs_FUVC} the (grey) regions of the parameter space that are excluded for stable DM, $M_{\rm DM} > M_S$ (which depend on $\lambda_i$ due to the choice of our y-axis). It is clear that close to the border of this region the EFT starts to break down, which is illustrated by the discrepancy between the solid and the various dashed lines. In fact, close to the physical boundary the bound in the full theory starts to grow, which is due to the fact that the amplitude in Eq.~\eqref{FDMauv} acquires an infrared divergence in this limit that is regulated by the charged lepton mass. Indeed in this region the DD bound is dominated by the dipole coefficient $C_\sigma$ in Eq.~\eqref{FDMauv}, which for $q^2 \ll m_k^2 \ll M_S^2 \approx M_{\rm DM}^2$ becomes\footnote{This enhancement was already noticed in Ref.~\cite{Kopp:2009et}.}
\begin{align}
C_\sigma \to \frac{\lambda_i^2  M_{\rm DM}}{4 M_S^2} \left( \log \frac{M_S^2}{m_{\ell_i}^2 }- 2\right) \, . 
\label{deglimitFDM}
\end{align} 
The solid curves in the EFT calculation are therefore an excellent approximation of the full calculation, and allow us to infer an approximate bound on the UV model without doing the full calculation. It turns out that a tree-level calculation is enough to estimate the correct bound (using the results in Sec.~\ref{sec:Bounds}) in almost the entire parameter space, as the constraints from DD and ID are largely dominated by the tree-level operators $C_{\psi \psi \ell \ell}$ (Dirac) and $C_{\chi 5 \chi  \ell \ell}$ (Majorana). In the case of ID, this is true in all points of the parameter space, since they are dominated by the tree-level annihilation into leptons (and annihilation into photons at one-loop gives subleading bounds).  For Majorana DM, ID gives the most stringent bound only for coupling with the $\tau$, where the cross section is $s$-wave and proportional to $m_\tau^2$. In this case, DD bounds are very weak, because of an approximate cancellation between the contributions of the anapole and the four-fermion operator (via RG running), see Eq.~\eqref{matching}. For Dirac DM, it is DD that sets the strongest bounds (except for very light DM masses), which at DM masses below roughly 300 GeV arise from the tree-level operator $C_{\psi \psi \ell \ell}$. Above this mass they are surpassed by the constraints on the magnetic dipole $C_{\psi \psi F}$, since this coefficient grows with $M_{\rm DM}$, cf. Eq.~\eqref{FDMmatchingD}.  This behaviour is somewhat surprising; typically the dipole operator is assumed to give the strongest constraints, while in this case we must include the higher-dimensional $C_{\psi \psi \ell \ell}$ to fully capture the behaviour of the DM-photon interactions.

\subsection{Scalar DM: DD and ID Constraints}
\label{sec:Scalar}
We now consider a second UV model, in this case for scalar DM. The logic follows closely that of the Fermion DM Lepton portal discussed in Sec.~\ref{sec:FermionUV}; here we summarize more briefly the scalar DM case and highlight where key differences arise.

In this model, we add to the SM a singlet scalar $s$ that is either complex or real, and a Dirac fermion $F$ with hypercharge $1$. The relevant part of the Lagrangian is given by 
\begin{align}
{\cal L} & = \left( \lambda_i \, s  \overline{F} P_R  \ell_i  + {\rm h.c.} \right) - M_{\rm DM}^2 |s|^2  - M_F \overline{F} F \, , 
\end{align} 
where $\ell_i$ denotes a SM lepton. 
We take $M_{\rm DM} \ll M_F$, so that $s$ is a DM candidate stabilized by a $Z_2$ symmetry. 

\begin{figure}[!t]
 \centering
\includegraphics[width= 0.315 \textwidth]{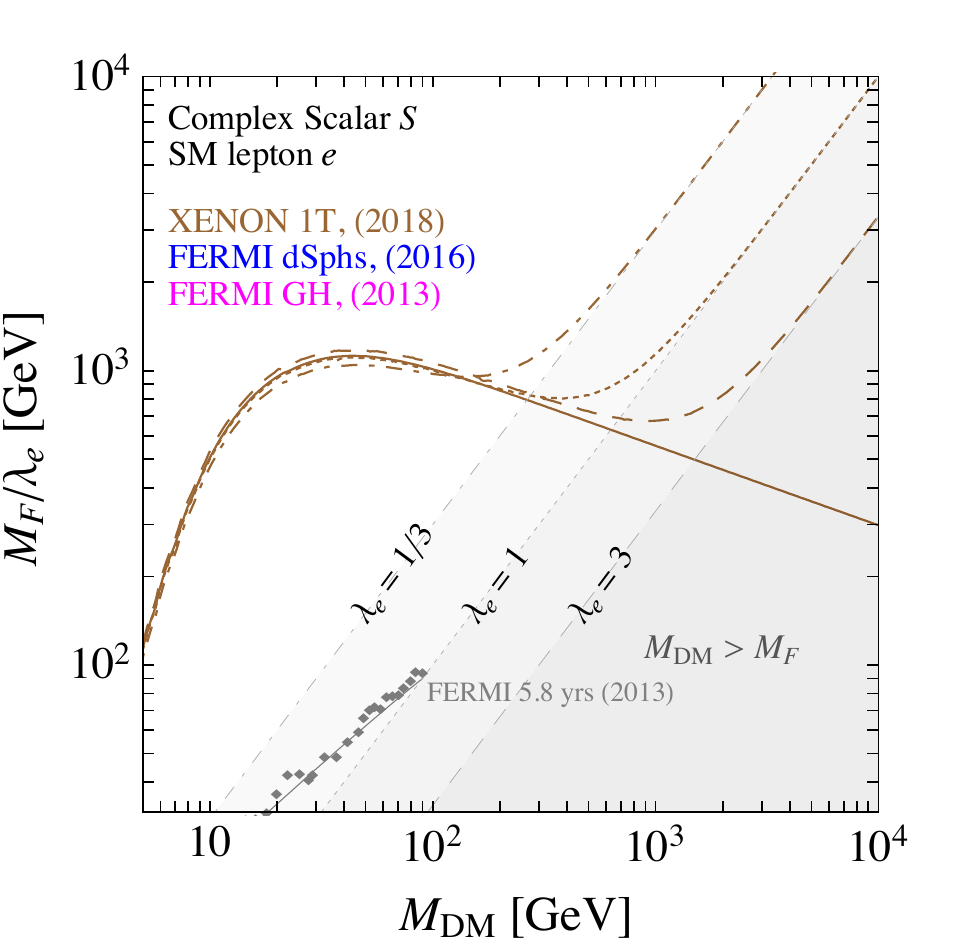} \
\includegraphics[width= 0.315 \textwidth]{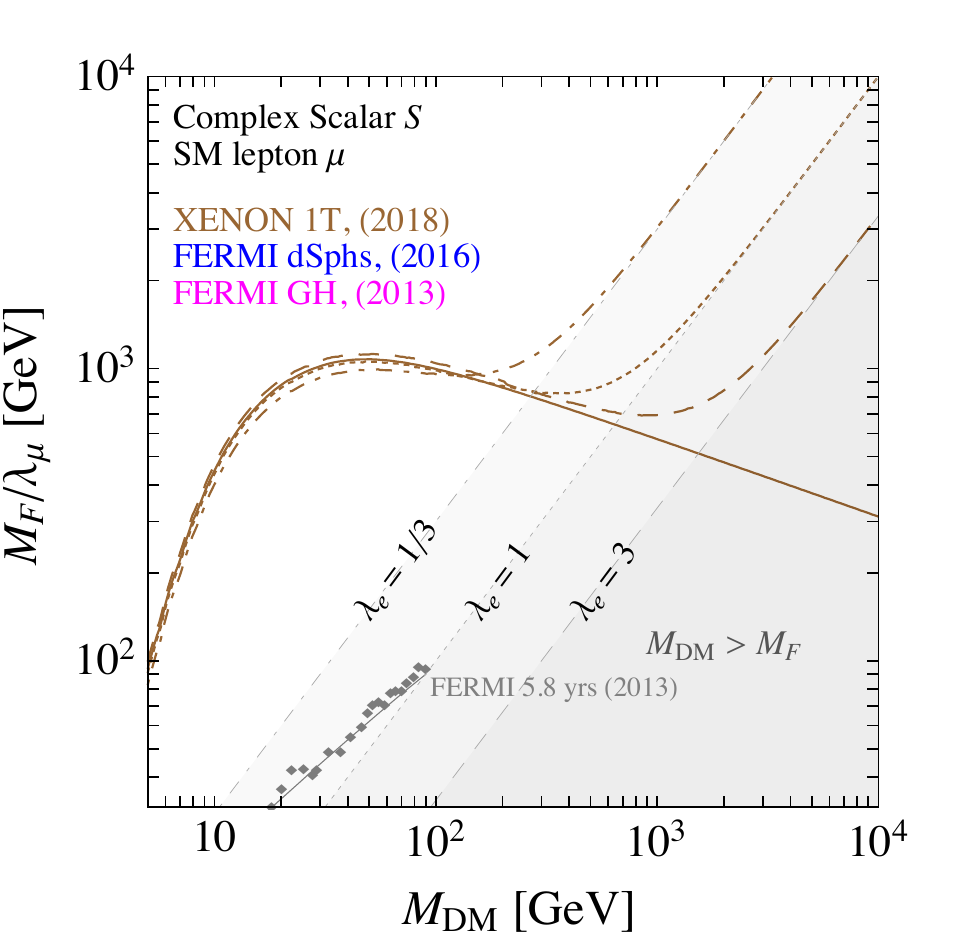} \
\includegraphics[width= 0.315 \textwidth]{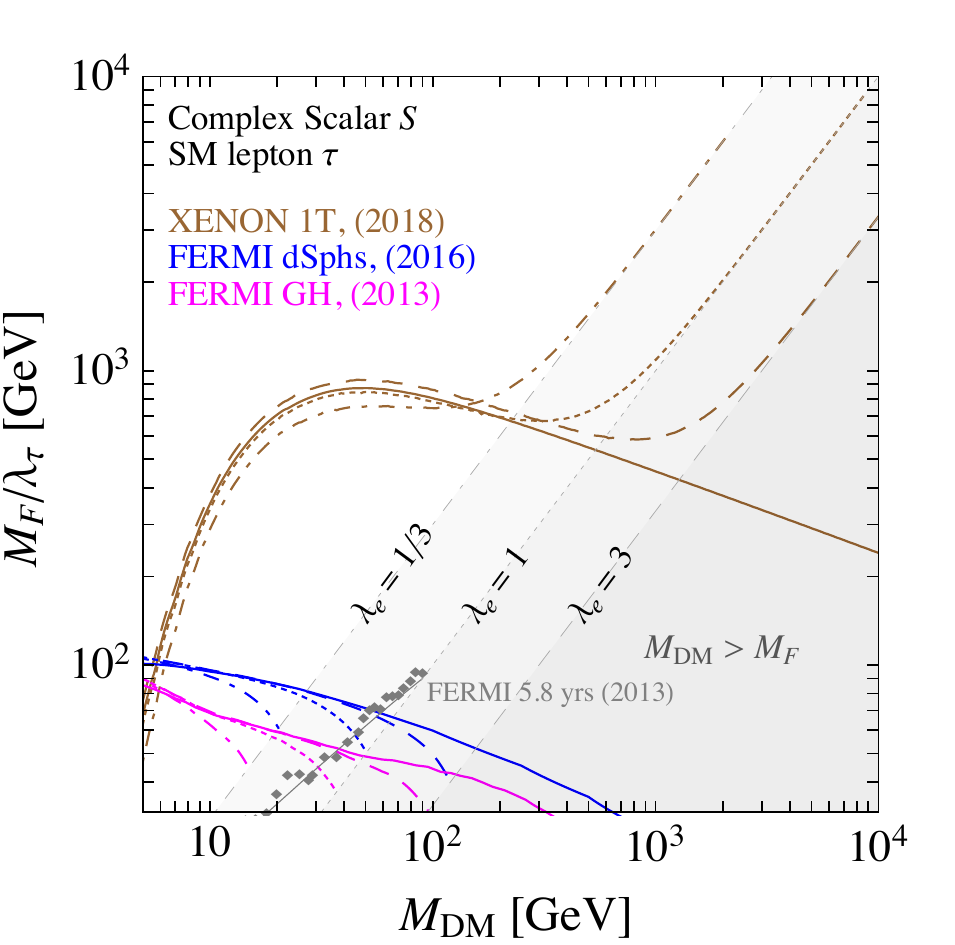} \\

\includegraphics[width= 0.315 \textwidth]{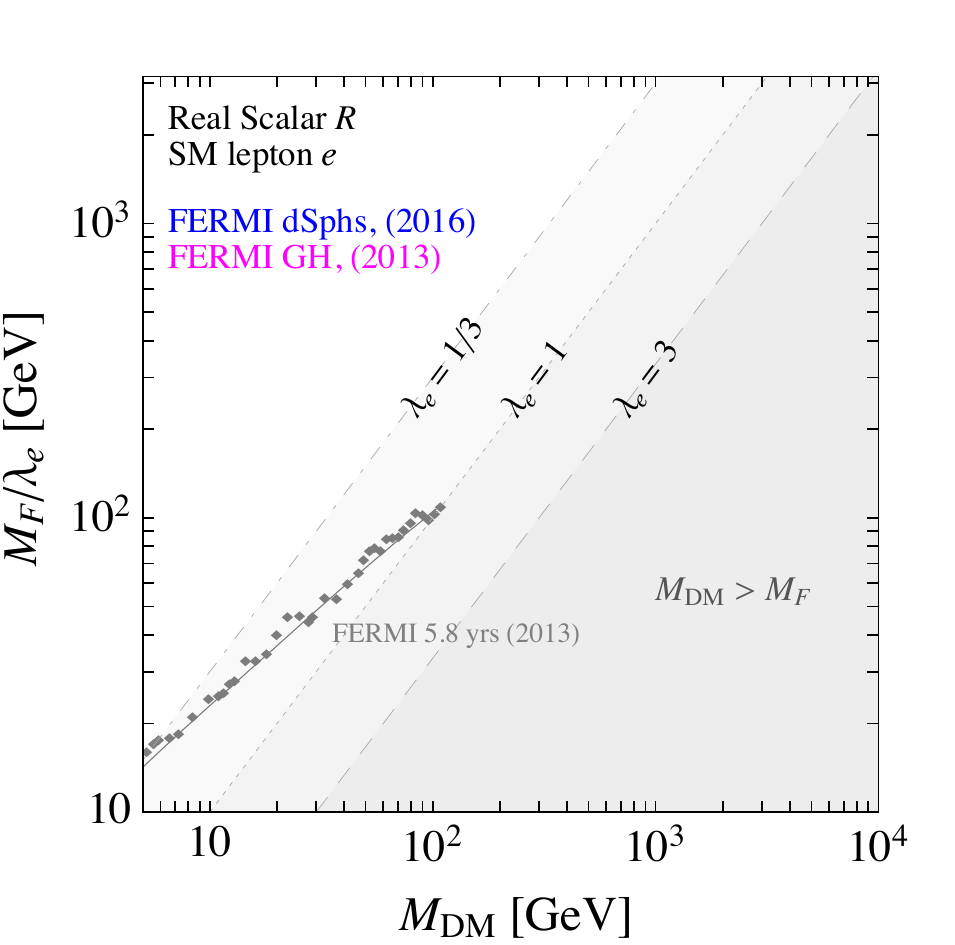} \
\includegraphics[width= 0.315 \textwidth]{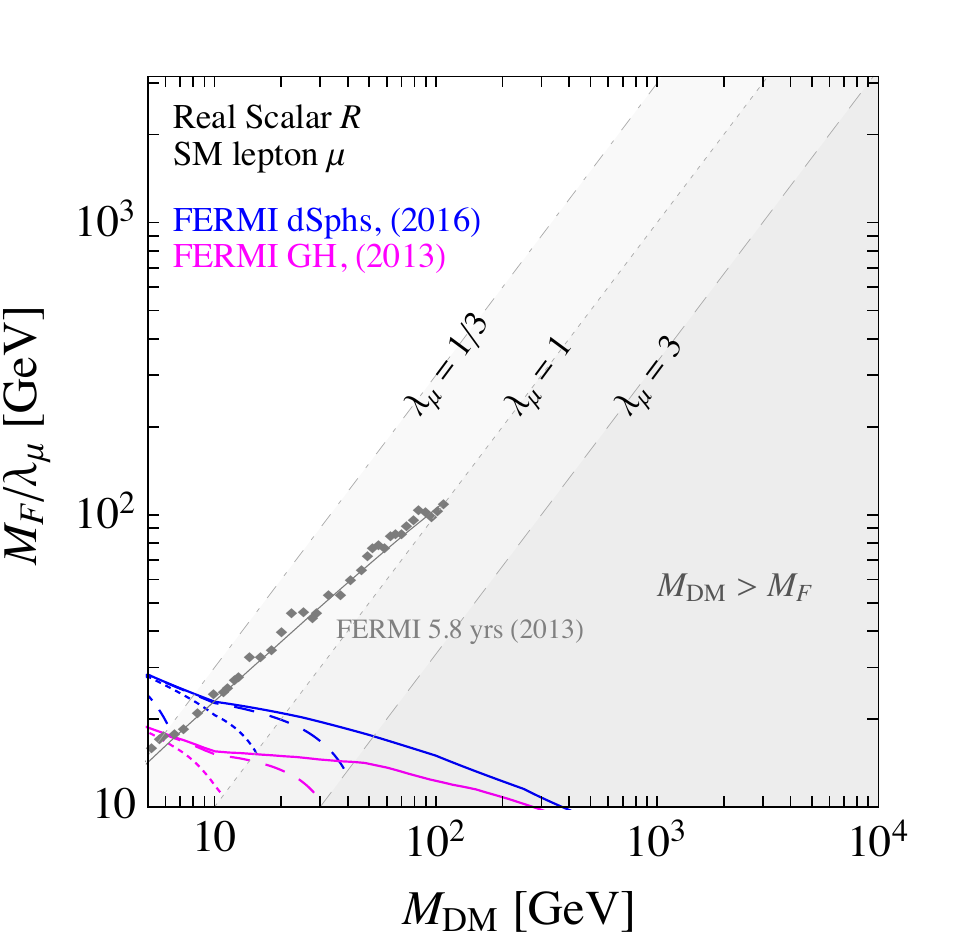} \
\includegraphics[width= 0.315 \textwidth]{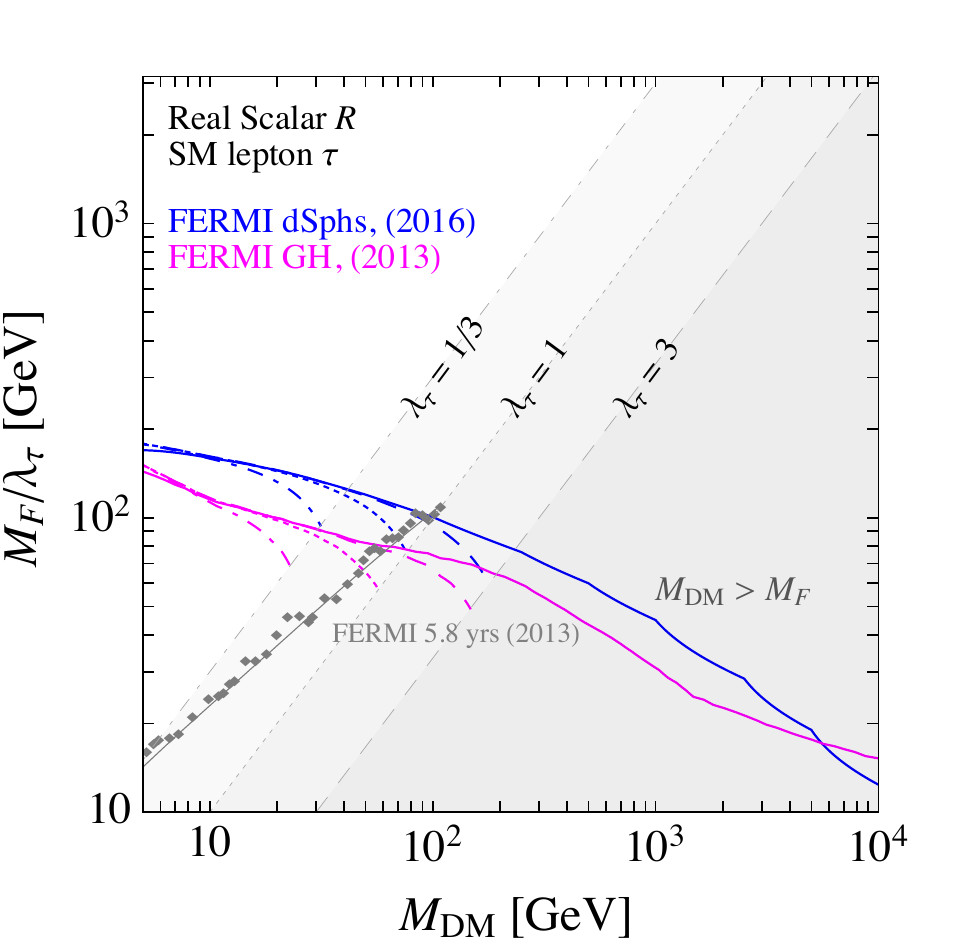} 

\caption{ \label{fig:EFT_vs_SUVC} Same as Fig.~\ref{fig:EFT_vs_FUVC}, but for complex scalar (upper panel) and real scalar DM (lower panel). We also show the limit from gamma ray line searches with FERMI (grey diamonds) and H.E.S.S. (red circles). Note that the bounds from line searches do not strictly apply whenever the continuum limits dominate.} 
\end{figure}

We now match this theory to the dominant effective operators in Table~\ref{Table:Operators}, as we did in the Fermion DM case. As a result of CP invariance, we have $C_{SS F \tilde{F} } = C_{SS\ell 5 \partial \ell}= 0$.  For the complex scalar case, one obtains the following matchings: 
\begin{equation}
\begin{split}
&\frac{C_{\partial S \partial S F}}{ \Lambda^2}   = \frac{\lambda_i^2}{4 M_F^2} \, ,  \qquad\quad \,\,\,
\frac{C_{ S  S F \tilde{F}}}{ \Lambda^2}   = \frac{C_{S  S \ell  5 \partial \ell}}{\Lambda^2} = 0 \, ,  \qquad\quad \,\,\,
\frac{C_{S S FF}}{\Lambda^2}  =   \frac{\lambda_i^2}{3 M_F^2}   \, ,  \\
&\frac{C_{S \partial S \ell \ell}}{\Lambda^2}   =  - \frac{\lambda_i^2}{4 M_F^2}   \, ,  \qquad\quad \,
\frac{C_{S \partial S \ell 5 \ell}}{\Lambda^2}   = -    \frac{\lambda_i^2}{4 M_F^2} \, ,   \qquad\qquad\qquad \,
\frac{C_{S  S \ell  \partial \ell}}{\Lambda^2}   =   \frac{\lambda_i^2}{2 M_F^2}  \, ,
\end{split}
\label{SDMmatching}
\end{equation}
while in the real scalar case one obtains 
\begin{align}
\label{eq:RealMatching}
\frac{C_{R  R \ell \partial \ell}}{\Lambda^2} & =  \frac{\lambda_i^2}{ M_F^2}   \, , &
\frac{C_{R R FF}}{\Lambda^2} & =   \frac{\lambda_i^2}{3 M_F^2}  \, , & 
 \frac{C_{R  R \ell 5 \partial \ell}}{\Lambda^2} & =   \frac{C_{R  R F \tilde{F}}}{\Lambda^2}  = 0 \, . 
\end{align}
The allowed regions in the UV parameter space $M_{\rm DM} - M_F/\lambda_i$ are shown as solid lines in the upper (complex scalar) and lower (real scalar) panels of Fig.~\ref{fig:EFT_vs_SUVC}.  As in the fermion DM case, we present the bounds for three cases of the UV coupling $\lambda_i =  \{ 1/3, 1, 3 \}$.

Notice that there are no constraints from gamma ray lines at zeroth order in $M_R/M_F$, since the amplitude for annihilation into two photons vanishes. Indeed, the tree-level contributions from $C_{R R FF}$ and the 1-loop contribution from $C_{RR \ell \partial \ell}$ cancel each other\footnote{Analogously in the complex scalar case where  $C_{S S FF}$ cancel against the 1-loop contributions from $C_{S \partial S \ell \ell}$, $C_{S \partial S \ell 5 \ell}$ and $C_{SS \ell \partial \ell}$.}. This can be understood by using low-energy theorems~\cite{Shifman:1979eb}. In the limit of $M_R \to 0$, integrating out the leptons yields the (exact) Lagrangian 
\begin{align}
{\cal L} =  \frac{4}{3} \frac{\alpha}{16 \pi} F_{\mu \nu} F^{\mu \nu}  \log  \det M_i^2(R) \, , 
\end{align}
where $\Lambda$ denotes the UV cutoff and $M_i(R)$ is the $R$-dependent ($2\times2$) fermion mass matrix. The couplings of $R$ to photons can then be obtained by expanding the above result in $R$. As a result of the chiral coupling structure, the determinant of the fermion mass matrix is not $R$ dependent, and therefore the amplitude of $RR \to \gamma \gamma$ vanishes at leading order in $M_R$. This result follows from a cancellation of the heavy and light fermions: the contribution of the heavy fermions  alone yields the Wilson coefficient $C_{R R FF}$ in Eq.~\eqref{eq:RealMatching}, while the contribution of the SM leptons gives the same result with opposite sign. At next order in $M_R/M_F$, we find $C_{RR FF} = 7/3\lambda_i^2 \, M_R^2/M_F^4$ (analogous conclusions hold in the complex scalar case).  The resulting bounds from FERMI are depicted in Fig.~\ref{fig:EFT_vs_SUVC} as grey diamonds. We only show the case with $\lambda_i = 1$ because the cross section  into gamma-ray lines is proportional to $(\lambda_i^2/M_F^2)^2 (M_R^2/M_F^2)^2$, and therefore does not allow us to draw limits in the plane ($M_{\rm DM} - M_F/\lambda_i$) without fixing $\lambda_i$, in contrast to the other EFT constraints. As one can see, the limits from gamma-ray lines on $M_F$ are always slightly below $M_{\rm DM}$. In this regime the results are only indicative, because the EFT validity breaks down. A full fixed-order calculation  is therefore needed in order to compute the annihilation cross section into  gamma-ray lines in this case.

\smallskip
We now compare these bounds  to the constraints obtained from an fixed order calculation in the full UV theory. The result for the DM-quark amplitude  is (for a complex scalar)
\begin{align}
 {\cal A} =   \frac{e^2  Q_q}{8 \pi^2} C_S (p+p^\prime)^\mu  \cdot  \overline{u}_q (k^\prime) \gamma_\mu u_q (k) \, ,
  \label{SDMauv}
\end{align}
where the momenta $p$, $p^\prime$ refer to the DM particle and $k$, $k^\prime$ to the quark, as in the Fermion DM case. The function $C_S (\lambda_i, m_{\ell_i}, q, M_{\rm DM}, M_F)$ can be found in Appendix~\ref{appfull}. In the case of real scalar DM, the above amplitude vanishes, $C_S \to 0$. From this expression one can recover the EFT coefficient ${\cal C}_{\partial S \partial S F}$ by taking the limit $m_{\ell_i}^2, M_{\rm DM}^2, q^2 \ll M_F^2$. 
 For $m_{\ell_i}^2 > q^2$ one finds at leading order in $M_{\rm DM}^2/M_F^2$ 
\begin{align}
C_S  & =  \frac{\lambda_i^2}{12 M_F^2} \left( 3 + 2 \log \frac{m_{\ell_i}^2}{M_F^2} \right) \, .
\end{align}
 Again this expression yields the EFT result in Eq.~\eqref{FDMmatchingD}, apart from the logarithm that is reproduced in the EFT by running down the tree-level operator,  recovering Eq.~\eqref{eq:LoopMatch}.
 
Finally, we also need to consider the bound from ID  due to direct annihiliation into leptons in the UV theory. The annihilation cross-sections into leptons $\ell_i$ are given by:
\begin{align}
\sigma(s^* s \to \ell_i \overline{\ell}_i) \, v & = \frac{\lambda_i^4}{16 \pi M_F^2} \frac{1}{(1+r^2)^2}   \left( \frac{m_{\ell_i}^2}{M_F^2} + \frac{v^2 r^2}{3} \right)  \, , \\
\sigma(ss \to \ell_i \overline{\ell}_i) \, v  & = \frac{ \lambda_i^4 }{4 \pi M_F^2} \frac{1}{(1+r^2)^2}  \left( \frac{m_{\ell_i}^2}{M_F^2} + \frac{v^4 r^6}{15 (1+r^2)^2} \right)  \, , 
\end{align}
for the case of real/complex scalar, where $v$ is the relative velocity and $r = M_{\rm DM}/M_F < 1$.

Again, we find that close to the DM stability limit ($M_\mathrm{DM} > M_\mathrm{F}$, grey regions in Fig.~\ref{fig:EFT_vs_SUVC}), the bound in the full theory grows rapidly, due to the fact that the amplitude in Eq.~\eqref{SDMauv} acquires an infrared divergence in this limit that is regulated by the charged lepton mass. Indeed in this region, where $q^2 \ll m_k^2 \ll M_{\rm DM}^2 \approx M_F^2$,  the coefficient $C_S$ in Eq.~\eqref{SDMauv}  becomes 
\begin{align}
C_S \to \frac{\lambda_i^2  }{12 M_F^2} \left( 5 + 3 \log \frac{M_F^2}{m_{\ell_i}^2 } - \frac{3 \pi M_F}{2 m_{\ell_i}}  \right) \, . 
\end{align}

As in the fermion DM case, the solid curves in the EFT calculation are an excellent approximation of the full calculation. In the complex scalar case, the bounds are dominated by DD, with ID playing no role for any DM mass. In contrast to the fermion DM case, however, these DD bounds come predominantly from a single tree-level operator $\mathcal{O}_{S \partial S \ell \ell}$ as expected from the EFT analysis of Sec.~\ref{sec:ScalarEFTBounds}. For real scalar DM, the Rayleigh operator $\mathcal{O}_{R R FF}$ appears with a coefficient of $1/3$, suppressing the (already weak) DD bounds and making them negligible. 
For coupling to $e$ and $\mu$, the Rayleigh operator still gives rise to substantial annihilation into gamma ray lines, so line searches dominate. Instead, for coupling to the $\tau$, the most relevant bounds at low DM mass come from FERMI diffuse searches (blue and magenta) in which case the $ss \rightarrow \ell\overline{\ell}$ cross section scales with $m_{\ell}^2/M_F^2$. The H.E.S.S. line search may still be relevant for strongly coupled theories at high DM mass, though we note that the H.E.S.S. analysis makes optimistic assumptions about the DM density profile in the Milky Way, as discussed at the start of Sec.~\ref{sec:Bounds}.

\subsection{Fermion and Scalar DM: Collider Constraints}
Finally we discuss also the constraints from colliders on the UV models in the previous sections. We consider constraints from direct searches at the LHC and LEP and indirect constraints from Bhabha scattering at LEP-2. The bounds are summarized in Fig.~\ref{fig:collider} and discussed in the following. 
 Note that in other UV scenarios, such as $Z'$-mediated UV completions, the constraints from collider searches can be significantly stronger \cite{DEramo:2016gos,DEramo:2017zqw}. 
\subsubsection{Direct Searches at LHC and LEP}
As discussed in e.g.~Ref.~\cite{Bai:2014osa}, the dominant collider constraints on these UV completions come from scenarios where the charged mediator is pair produced, each sub-sequently decaying into a charged lepton and a DM particle. The signature is therefore $\ell^+ \ell^-$ plus missing energy. 

\paragraph{Fermion DM:} In the case of fermionic DM (scalar mediators), this signature has been considered in searches for sleptons in the MSSM at LHC energies of 8 TeV \cite{TheATLAScollaboration:2013hha,Khachatryan:2014qwa,Aad:2014vma,Aad:2014yka} and 13 TeV \cite{Aaboud:2018jiw,Sirunyan:2018nwe,CMS:2017rio}. We use the 95\% CL bounds from the lower right panel of Fig.~5 (Fig.~6) of Ref.~\cite{Sirunyan:2018nwe} for selectrons (smuons) decaying with a 100\% branching fraction into right-handed electrons (muons) and a neutralino. We show the resulting bounds on $M_S$ as green regions in the upper panel of Fig.~\ref{fig:collider} (note that they do not depend on the coupling $\lambda_i$ as long as the decay is prompt). In the case of staus, LHC searches, e.g.~in Ref.~\cite{Sirunyan:2018vig}, are not yet sufficiently sensitive to derive any constraints.

\smallskip

Direct searches for final state leptons with missing energy have also been carried out at LEP, yielding constraints on heavy charged  scalars of about 80-90~GeV. For the 95\% CL limits we use constraints on sleptons provided by the PDG  in Ref.~\cite{Tanabashi:2018oca}. For different values of the slepton-DM mass difference $\Delta m$, these bounds are $m_{\tilde{e}} > 73 \, \GeV$  for any $\Delta m$, and $m_{\tilde{\mu}} > 88 \, \GeV$, $m_{\tilde{\tau}} > 79 \, \GeV$   for $\Delta m > 15 \,  \GeV$, and shown as blues regions in the upper panel of Fig.~\ref{fig:collider}.

\paragraph{Scalar DM:}  In the case of scalar DM (fermionic mediators), one can recast the LHC slepton searches. In the case of  couplings to right-handed muons, this has been done in e.g.~Ref.~\cite{Calibbi:2018rzv} (see Fig.~5 therein). The case of couplings to right-handed electrons should be similar (as one can see from the slepton case), so for simplicity we use the  bounds from the muon case also for electrons, shown as green regions in the lower panel of Fig.~\ref{fig:collider}. For fermions coupling to right-handed taus, one can recast the stau searches at CMS~\cite{Sirunyan:2018vig}, which yield upper bounds on the stau pair production cross section. These analyses can be potentially relevant, because the production cross section of a heavy fermion is larger than that of heavy scalars (with the same mass and gauge quantum numbers). For the precise values of the cross section we use Fig.~2.1 in Ref.~\cite{Kumar:2015tna}, which is about a factor 10 larger than stau pair production, and allows us to derive bounds on this scenario, shown in green in the lower right panel of Fig.~\ref{fig:collider}.

\smallskip

In this case LEP provides bounds for any mass splitting and all three flavors. We use the pure higgsino bound yielding $m_F > 92.4 \GeV$ at 95\% CL~\cite{LEPhiggsino}, shown as blue regions in the lower panel of Fig.~\ref{fig:collider}.

\subsubsection{LEP Indirect  Searches}
As in section 3, we use the EFT analysis of Refs.~\cite{Falkowski:2015krw, Falkowski:2017pss} to constrain the UV model with couplings to electrons. Integrating out the heavy fermion and scalar and one-loop, one can match to the low-energy 4-electron operator
\begin{align}
{\cal L}_{\rm eff} & =   \frac{c_{RR}}{2v^2} \, (\overline{e} \gamma^\mu P_R e) \cdot (\overline{e} \gamma_\mu P_R e)  \, ,
\end{align}
with a coefficient given in terms of UV parameters as 
\begin{align}
c_{RR} & = - \frac{\lambda_e^4 v^2}{64 \pi^2 M_S^2} \, f_{D,M,S} \left( \frac{M_F^2}{M_S^2} \right) \, .
\end{align}
Here $f_{D,M,S}$ denote the loop functions for the case of Dirac/Majorana fermion and complex scalar DM
\begin{align}
f_D(x) & = f_S(x) = \frac{1-x^2 + 2 x \log{x}}{(1-x)^3} \, , & f_M(x) & = \frac{1 + 4 x - 5 x^2 + ( 4 x + 2 x^2 ) \log{x}}{(1-x)^3} \, ,
\end{align}
while for real scalar DM $c_{RR} = 0$. The EFT coefficient $c_{RR}$ is constrained by Bhabha scattering at LEP-2 according to 
\begin{align}
c_{RR} = \left( 3.8 \pm 2.8 \right) \cdot 10^{-3} \, ,
\end{align} 
and we use this bound at 95\% CL to derive bounds on the UV parameters, shown as red contours in Fig.~\ref{fig:collider} for different values of $\lambda_e = \{ 1,3 \}$, distinguishing Dirac  (light red) and Majorana (dark red) DM. 

It is instructive to compare this result with the EFT estimate in Section 3, see Eq.~\eqref{RRmatching}. Using the tree-level matching conditions in Eqs.~\eqref{FDMmatchingD} and \eqref{SDMmatching}, one can check that the EFT result exactly reproduces the results of the full calculation in the EFT limit $M_{F,S}^2 \to 0$ for all four cases, upon the identification $\Lambda = M_{S,F}$ in the EFT.

\subsubsection{Comparison with DD and ID Constraints}
 \begin{figure}[!t]
 \centering
\includegraphics[width= 0.315 \textwidth]{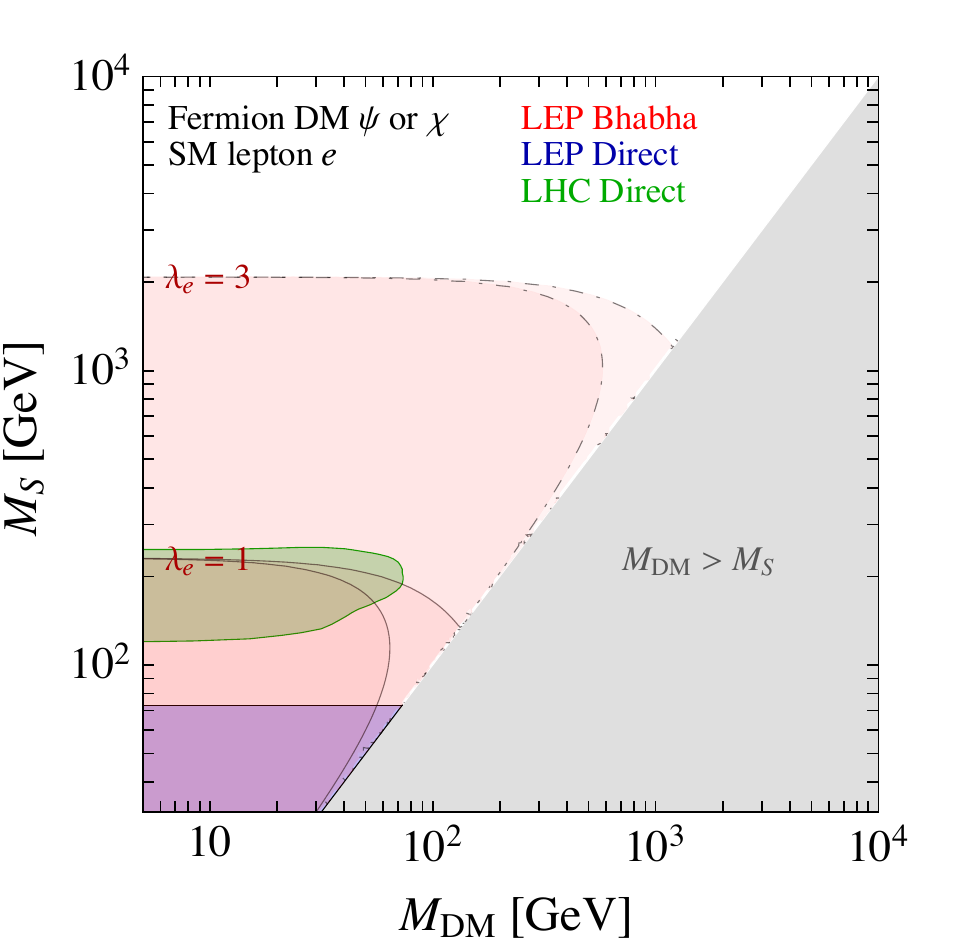} \
\includegraphics[width= 0.315 \textwidth]{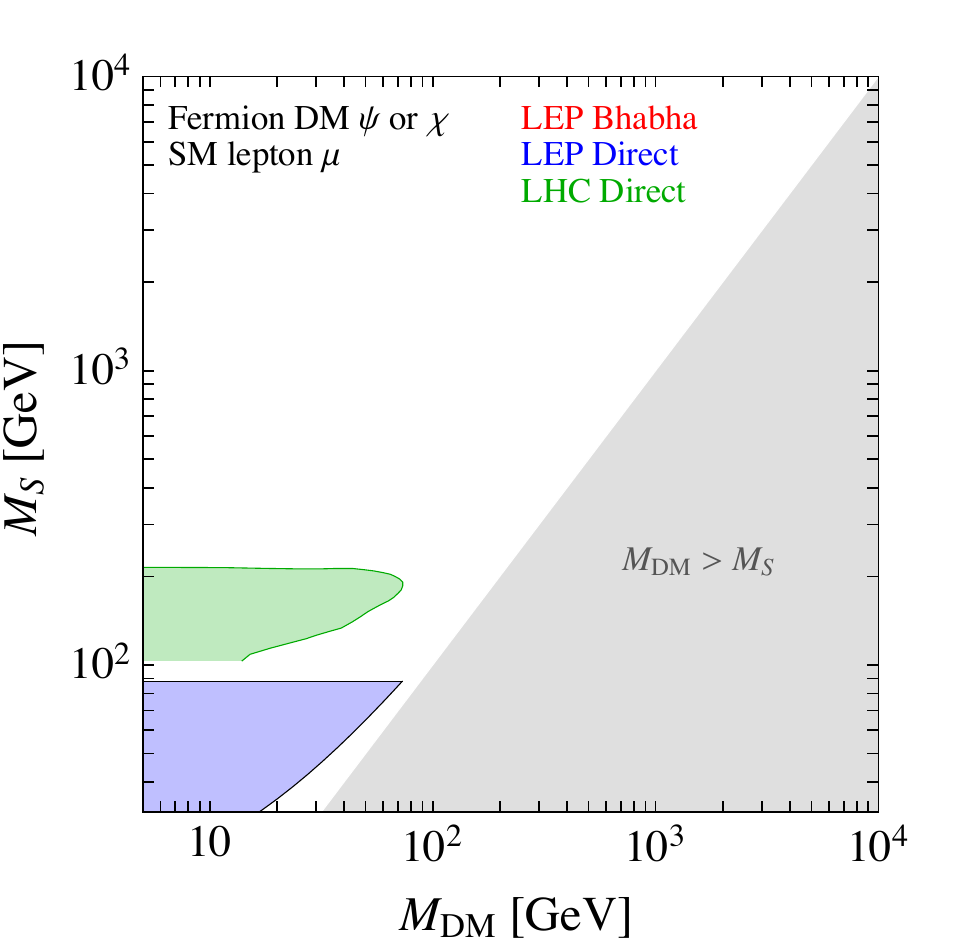} \
\includegraphics[width= 0.315 \textwidth]{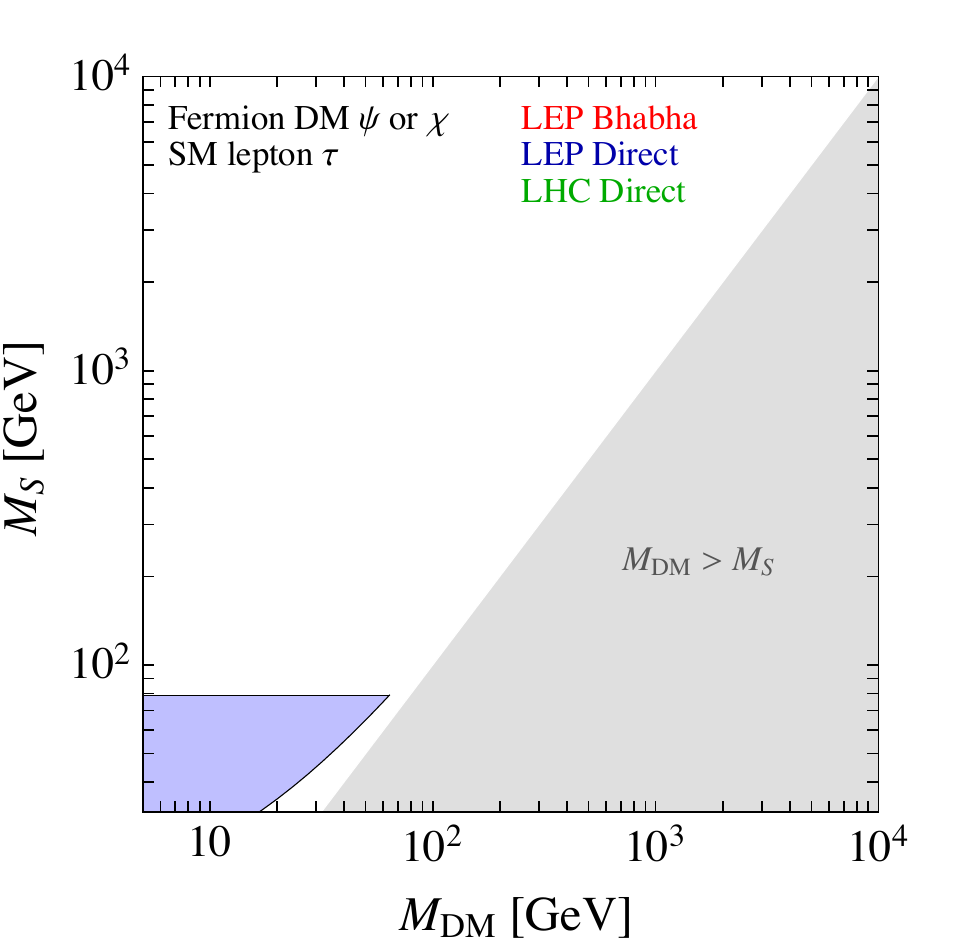} \\

\includegraphics[width= 0.315 \textwidth]{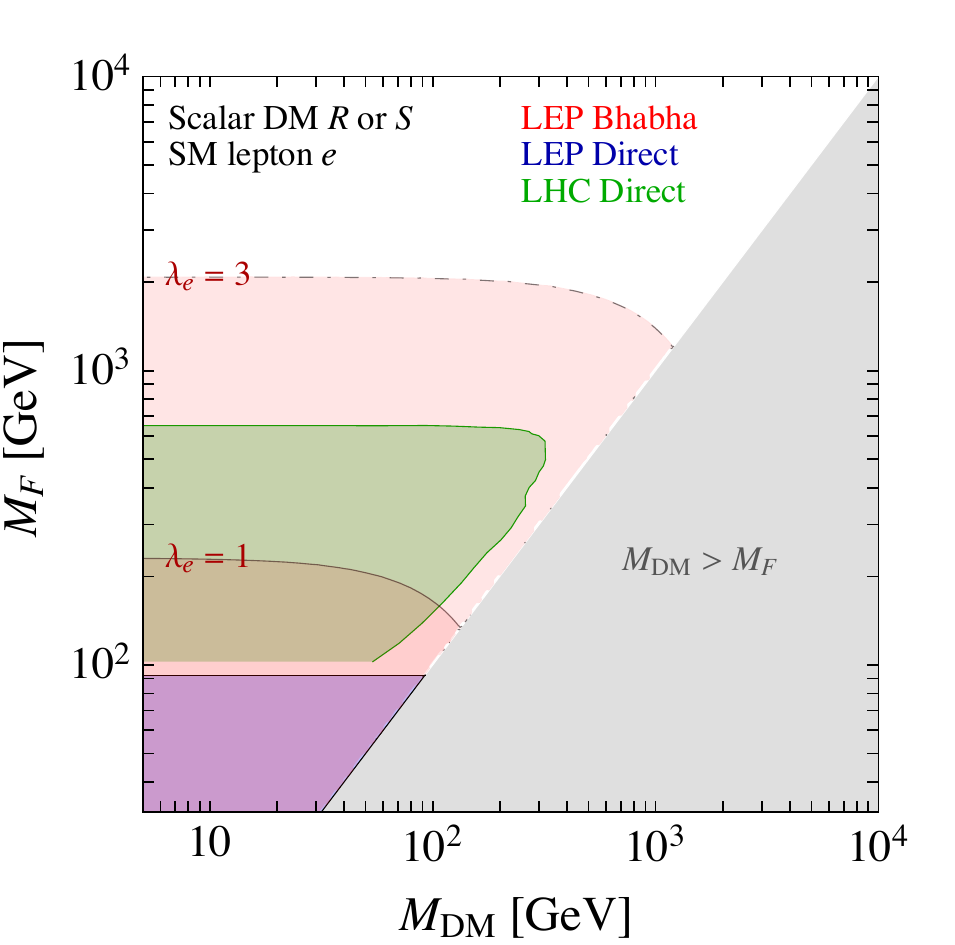} \
\includegraphics[width= 0.315 \textwidth]{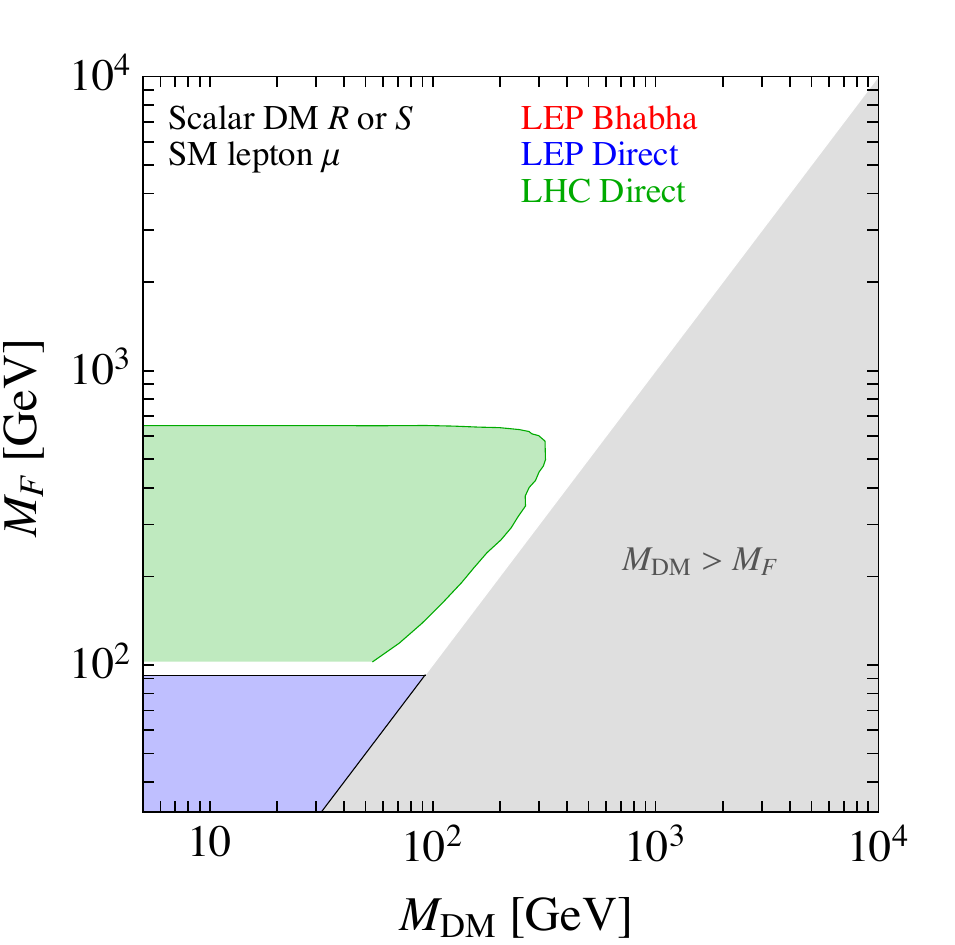} \
\includegraphics[width= 0.315 \textwidth]{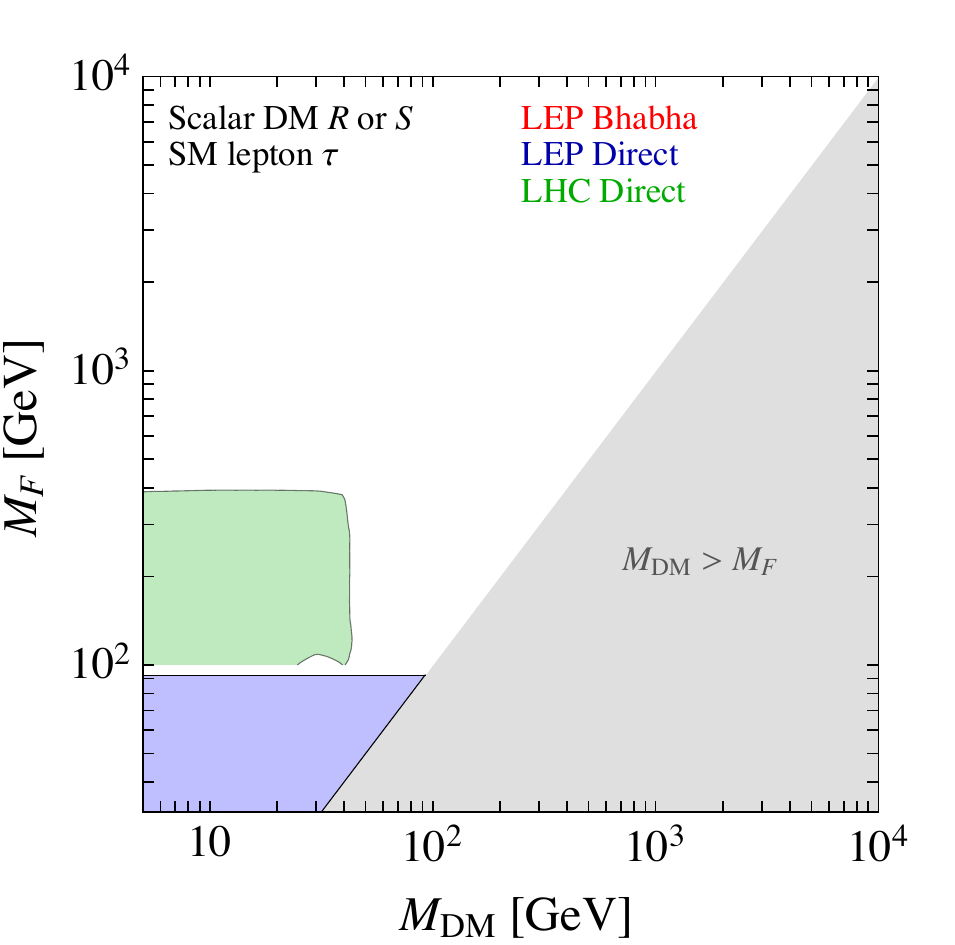} 

\caption{\label{fig:collider} Collider constraints for fermion (upper panel) and scalar DM (lower panel) for UV models couplings to right-handed electrons, muon and taus. The grey area in the $M_{\rm DM}-M_{S,F}$ plane is excluded by DM stability, while the colored regions are excluded by direct searches at LHC (green) and LEP (blue). For DM coupling to electrons one has also constraints from Bhabha scattering at LEP-2 (red), which we display for different values of the coupling $\lambda_e =1$ (solid) and $\lambda_e =3$ (dashed). These constraints do not apply to real scalar DM, and for fermions are different for Dirac (light red) and Majorana (dark red) DM.}
\end{figure}
In this section we compare the bounds from colliders in Fig.~\ref{fig:collider} to those from direct and indirect DM searches in Figs.~\ref{fig:EFT_vs_FUVC} and \ref{fig:EFT_vs_SUVC}. 
\paragraph{Dirac DM:}
For Dirac fermions, bounds from DD prevail for $\lambda_i = 1$ for all SM leptons and for $M_{\rm DM} > 10 \, \GeV$ (see upper panels in Fig.~\ref{fig:EFT_vs_FUVC}). This holds even more for $\lambda_{\mu, \tau} >1$ in the cases of $\mu$ and $\tau$, while for $e$ the LEP-2 indirect searches give stronger constraints than DD for $\lambda_e \gtrsim 4$. Instead for $\lambda_i <1$ colliders typically provide stronger constraints than DD. For $e, \mu$ LHC limits prevail for $\lambda_{e , \mu} \lesssim 0.2$, while LEP starts to become more relevant than DD for all leptons at around $\lambda_i \lesssim 0.1$. Note that close to the degenerate limit DD bounds always dominate, since in this case collider searches loose sensitivity while DD cross sections are enhanced, cf. Eq.~\eqref{deglimitFDM}.

\paragraph{Majorana DM:}
For Majorana fermions, bounds from colliders prevail for $\lambda_i < 1$ for all SM leptons (see lower panels in Fig.~\ref{fig:EFT_vs_FUVC}).  ID bounds are only important for taus with $\lambda_\tau \ge1$, while DD bounds only play a role close to the degenerate limit (as discussed above). Note in particular that in the electron case, Bhabha scattering provides stringent bounds (in the TeV regime) for $\lambda_e >1$.

\paragraph{Complex Scalar DM:}
The complex scalar case is very similar to the Dirac fermion case, apart from stronger limits from colliders. In particular, bounds from DD prevail for $\lambda_i = 1$ for all SM leptons and for $M_{\rm DM} > 10 \, \GeV$ (see upper panels in Fig.~\ref{fig:EFT_vs_SUVC}). This holds even more for $\lambda_{\mu, \tau} >1$ in the cases of $\mu$ and $\tau$, while for $e$ the LEP-2 indirect searches give stronger constraints than DD for $\lambda_e \gtrsim 4$. Instead for $\lambda_i <1$ colliders typically provide stronger constraints than DD. LHC limits prevail for $\lambda_{e , \mu} \lesssim 0.5$ and $\lambda_{\tau} \lesssim 0.4$ , while LEP starts to become more relevant than DD for all leptons at around $\lambda_i \lesssim 0.1$.  Close to the degenerate limit DD always dominate, since in this case collider searches loose sensitivity while DD cross sections are enhanced, cf. Eq.~\eqref{deglimitFDM} (although in contrast to fermion DM  LEP completely closes the kinematic gap).

\paragraph{Real Scalar DM:}
The real scalar case is very similar to the Majorana fermion case, apart from stronger limits from colliders (especially the presence of LHC bounds in the $\tau$ case), and no bounds from DD and Bhabha scattering, In particular,  bounds from colliders prevail for $\lambda_i < 1$ for all SM leptons (see lower panel in Fig.~\ref{fig:EFT_vs_SUVC}).  ID bounds are only important for taus with $\lambda_\tau \gtrsim 2$. In the degenerate limit, LHC (but not LEP) bounds can be avoided, but a full fixed-order calculation is needed in this case in order to assess the limits from annihilation into gamma ray lines, as we have discussed in Section~\ref{sec:Scalar}. 

\section{Summary and Conclusions}
\label{sec:summary}

In this work, we have studied the scenario where DM couples to quarks only via photon exchange. To this end, we have considered the most general effective interactions of fermion and scalar Dark Matter with the photon, distinguishing between Dirac/Majorana and complex/real scalar DM, see Eqs.~\eqref{MajoranaL}--\eqref{CScalarL}. We have systematically classified these operators according to their CP properties and dimension, as we summarize in Table~\ref{Table:Operators}. Since DM is neutral under electromagnetism, these operators can only arise at loop level at the cutoff scale. This implies that for direct detection (DD) and indirect detection (ID), there are other relevant operators, which arise at tree-level at the cutoff scale and mix into DM-photon operators via RG evolution. According to our philosophy, these can only be four-fermion operators coupling DM to vector currents of SM leptons, which we therefore include in the classification and list in Eq.~\eqref{D4l}. Moreover, such operators are usually accompanied by other DM-lepton currents due to SM gauge invariance. We also include these operators in the classification, referring to them as ``UV-related operators", see Eqs.~\eqref{D4l5}--\eqref{dim8}.  

\medskip

We have then studied the model-independent constraints from direct and indirect searches in Section~\ref{sec:Bounds}, considering one operator at a time. The results are shown for Dirac DM in Fig.~\ref{fig:EFT_Constraints_Dirac}, for Majorana DM in Fig.~\ref{fig:EFT_Constraints_Majorana} and for complex and real scalar DM in Fig.~\ref{fig:EFT_Constraints_Scalar}. For the operators coupling directly to photons, like the dipole operators, we update existing bounds in the literature, in particular showing bounds from the latest Xenon1T (2018) results. We find that tree-level operators are strongly constrained by both DD and ID, depending on the Lorentz structure. In the case of DD the resulting bounds provide stringent constraints, due to large 
RG mixing from the SM vector current. For example, the 4-fermion operator $\mathcal{O}_{\psi\psi\ell\ell} = \overline{\psi} \gamma^\mu \psi \cdot \overline{\ell} \gamma_\mu \ell$ mixes into the charge radius operator $\mathcal{O}_{\psi\psi\partial F} = \overline{\psi} \gamma^\mu \psi \cdot \partial^\nu F_{\mu\nu}$ and bounds from the former interaction are stronger than those from the latter. In the Dirac DM case, these bounds (on dimension-6 operators) can even compete with those on the electromagnetic dipole (of dimension-5), which has the same CP quantum numbers. We also note the bounds on the UV-related operators can be important in the case of Majorana or real scalar DM coupling to tau leptons (see Figs.~\ref{fig:UVrelated_Constraints_Majorana} and \ref{fig:UVrelated_Constraints_Scalar}), which in general are more weakly constrained. Finally, we also derive bounds on four-fermion operators involving DM and electrons from Bhabha scattering at LEP-2, which can be quite stringent due to the fixed sign of this new contribution relative to the SM. 

\medskip

In Section \ref{sec:UVmodels} we have studied two examples of explicit UV completions, in which DM couples to SM leptons and a new heavy fermion or scalar. The main purpose of this section is to demonstrate the usefulness of the EFT analysis, which allows us to  gain further insight into the resulting constraints from DD and ID, shown in Figs.~\ref{fig:EFT_vs_FUVC} and \ref{fig:EFT_vs_SUVC}. Here we compare the bounds obtained in the EFT prediction to those from the full fixed-order calculation, finding excellent agreement except in the degenerate limit for DD.  
In particular, in most of the parameter space the dominant bounds come from 4-fermion operators (as expected from the results of Section~\ref{sec:Bounds}), which can be obtained by a simple tree-level calculation. 
Finally, we have discussed the constraints from colliders (see Fig.~\ref{fig:collider}) on these UV completions, which can be more important that DD and ID for Majorana fermion and real scalar DM. In particular, we also studied new constraints from LEP-2 on the electron portal DM scenario, which give important constraints for sizable couplings.  

As an interesting result, we found that the annihilation cross section of scalar DM into gamma ray lines vanishes at leading order in the DM mass, in contrast to previous results in the literature~\cite{Giacchino:2014moa}. We have checked our calculation by applying low-energy Higgs theorems, which allow us to elegantly derive the result in the limit of vanishing scalar mass.   

\medskip

In conclusion, we have provided model-independent constraints on a wide class of effective operators coupling DM to photons or leptons. This compendium will allow the reader to quickly assess the constraints  on any DM model that interacts with quarks mainly via photon exchange. We have illustrated this approach in two explicit UV models. Updating the bounds on these UV models, we find that DM-photon interactions are typically constrained by multiple complementary probes. The EFT analysis provides a simple way of taking into account all relevant constraints and shedding light on DM-photon interactions without exhaustive calculations.

\acknowledgments

We thank Eugenio Del Nobile and Andrew Cheek for helpful discussions concerning the anapole interaction, Adam Falkowski for providing details on Refs.~\cite{Falkowski:2015krw, Falkowski:2017pss}, and Lorenzo Calibbi, Jose Ramon Espinosa and  Rodrigo Alonso for useful discussions on collider constraints and technical aspects of EFT calculations. We are especially grateful to Emmanuel Stamou for cross-checking some of the EFT results for the Rayleigh operator in the scalar DM case. BJK acknowledges funding from the Netherlands Organization for Scientific Research (NWO) through the VIDI research program ``Probing the Genesis of Dark Matter" (680-47-532).


\appendix
 \numberwithin{equation}{section}
  \section{Direct Detection}
  \label{app:DirectDetection}
  In this appendix we collect the relevant expressions that have been used to derive the bounds from direct detection. We begin by reviewing the non-relativistic effective field theory (NREFT) formalism, which gives the differential scattering cross section in terms of non-relativistic operator coefficients.  Then these coefficients are matched to the relativistic operators below the weak scale, which allows us to provide the contribution of a single operator to the differential scattering cross section. Finally we summarize the approximate bounds from XENON1T. 
  
 \subsection{NREFT Formalism}
 \label{app:NREFT}
Here we list the NREFT operators \cite{Dobrescu:2006au,Fan:2010gt,Fitzpatrick:2012ix,Anand:2013yka,Fitzpatrick:2012ib,Panci:2014gga,Dent:2015zpa,DelNobile:2018dfg} we consider in this work, with normalisations chosen to match that of Ref.~\cite{DelNobile:2013sia} (this is the same normalisation as in the public tool \href{http://www.marcocirelli.net/NRopsDD.html}{\textsc{NRopsDD} tools}, which we have used to obtain the bounds on relativistic operators in Section~\ref{sec:Bounds}). In the context of DM-photon interactions, the most relevant operators

\footnote{For a more complete list of NREFT operators see, for example, Ref.~\cite{DelNobile:2018dfg}, which constructs the basis of 16 Galilean-invariant `building blocks' for NREFT interactions with spin-0 and spin-1/2 DM. For spin-1 DM, see also Refs.~\cite{Dent:2015zpa,Fan:2010gt}, though in that case it is not clear whether the full operator basis has been constructed yet.}

 describing DM interactions with nucleons $N = p,\,n$ are:
 \begin{align}
 \label{eq:operators}
\begin{split}
\mathcal{O}^{\rm NR}_1 &= 1 \, , \\
\mathcal{O}^{\rm NR}_4 &= \vec{S}_\textrm{DM}\cdot\vec{S}_N\, ,  \\
\mathcal{O}^{\rm NR}_5 &= i  \vec{S}_\textrm{DM} \cdot (\vec{q} \times \vec{v}^{\perp})\, , \\
\mathcal{O}^{\rm NR}_6 &= (\vec{S}_\textrm{DM} \cdot \vec{q})(\vec{S}_N \cdot \vec{q})\, , \\
\end{split}
\qquad
\begin{split}
\mathcal{O}^{\rm NR}_8 &= \vec{S}_\textrm{DM} \cdot \vec{v}^{\perp}\, , \\
\mathcal{O}^{\rm NR}_9 &= i \vec{S}_\textrm{DM} \cdot (\vec{S}_N \times \vec{q})\, ,  \\
\mathcal{O}^{\rm NR}_{11} &= i \vec{S}_\textrm{DM} \cdot \vec{q} \, . \\
\end{split}
\end{align}
Here $\vec{q}$ denotes the momentum transfer, $\vec{v}^{\perp}$ the transverse WIMP-nucleon velocity, $\vec{S}_\textrm{DM}$ the DM spin and $\vec{S}_N$ the nucleon spin. The transverse velocity is given by 
\begin{equation}
\vec{v}^\perp = \vec{v} + \frac{\vec{q}}{2 \mu_{\textrm{DM}, N}}\,,
\end{equation}
where $\mu_{\textrm{DM}, N} = m_\textrm{DM} m_N/(m_\textrm{DM} + m_N)$ is the reduced mass of the WIMP-nucleon system and $m_N$ is the nucleon mass. 
Note that $\ONR{1}$ is the standard spin-independent (SI) operator, which is coherently enhanced and therefore typically dominates. The spin-dependent (SD) operator corresponds to $\ONR{4}$. 
 
 The DM-nucleon matrix element can then be expressed as a sum over operators,
 \begin{equation}
 \label{eq:DMnucleon}
 \mathcal{M}^{(N)} = \sum_{i} c_i^{(N)} \mathcal{O}^{\mathrm{NR}}_{i\,(N)}\,,
 \end{equation}
 where the coefficients $c_i^{N=p,n}$ can be calculated from a given relativistic Lagrangian (see e.g.~Ref.~\cite{DelNobile:2018dfg}). The DM-\textit{nucleus} matrix element is then obtained by summing over nucleons in a given target nucleus $T$.
The differential scattering cross section can then be written in terms of the appropriate nuclear response functions $F$, which encode the nuclear structure:
\begin{align}
\begin{split}
\label{XSc}
\frac{\mathrm{d}\sigma_T}{\mathrm{d}E_R} &= \frac{m_T}{32 \pi m_\mathrm{DM}^2 m_N^2} \frac{1}{v^2} \sum_{i, j = 1}^{15} \sum_{N, N' = p, n} c_i^N c_j^{N'} F_{i,j}^{(N,N')}(q^2)\,.
\end{split}
\end{align}
The nuclear recoil energy $E_R$ and recoil momentum are related by $q^2 = 2 m_T E_R$.  The nuclear response functions $F_{i,j}$ can be written in terms of nuclear form factors $F_{X}$, where $ X = M, \Sigma^\prime, \Sigma^{\prime\prime}, \Delta, \tilde{\Phi}^\prime, \Phi^{\prime \prime}, \Sigma^\prime \Delta$ correspond to sums over different nucleon properties in the target nucleus $T$.  We employ the following relations (suppressing the $N=p,n$ indices and $q^2$ dependence in all but the first expression):
 \begin{align}
\begin{split}
\label{eq:FormFactors}
F_{1,1}^{(N,N^\prime)}&= F_{M}^{(N,N^\prime)}(q^2)\,, \\
F_{4,4} &= \frac{C(j_\mathrm{DM})}{16}\left( F_{\Sigma'} + F_{\Sigma''}\right)\,, \\
F_{5,5} &= \frac{C(j_\mathrm{DM}) }{4}  q^2  \left( v_\perp^2 F_{M} +  \frac{q^2}{m_N^2} F_{\Delta} \right)\,, \\
F_{6,6} &= \frac{C(j_\mathrm{DM})}{16} q^4 F_{\Sigma''}\,,\\
F_{8,8} &=  \frac{C(j_\mathrm{DM})}{4} \left(  v_\perp^2 F_{M} +  \frac{q^2}{m_N^2} F_{\Delta} \right) \,,
\end{split}
\qquad
\begin{split}
F_{9,9} &=  \frac{C(j_\mathrm{DM})}{16} q^2 F_{\Sigma'}\,,\\
F_{11,11} &= \frac{1}{4} q^2 F_M\,,\\
F _ { 4,5 }  &= - C \left( j _ { \chi } \right) \frac { q ^ { 2 } } { 8 m _ { N } } F _ { \Sigma ^ { \prime }  \Delta } \,,\\
F _ { 4,6 } &= C \left( j _ { \chi } \right) \frac { q ^ { 2 } } { 16 } F _ { \Sigma ^ { \prime \prime } } \,,\\
F _ { 8,9 }  &= C \left( j _ { \chi } \right) \frac { q ^ { 2 } } { 8 m _ { N } } F _ { \Sigma^\prime \Delta } \,.
\end{split}
\end{align}
Here, $C(j_{\rm DM}) = 4 j_{\rm DM} (j_{\rm DM} + 1)/3$, where $j_{\rm DM}$ is the DM spin. The transverse velocity appearing here is the WIMP-\textit{nucleus} velocity:
\begin{equation}
\vec{v}_\perp = \vec{v} + \frac{\vec{q}}{2 \mu_{{\rm DM}, T}}\, ; \qquad v_\perp^2 = v^2 - \frac{m_T E_R}{2\mu_{{\rm DM}, T}^2}\,,
\end{equation}
where $\mu_{\textrm{DM}, T} = m_\textrm{DM} m_T/(m_\textrm{DM} + m_T)$ is the reduced mass of the WIMP-(target nucleus) system with $m_T$ denoting the target nucleus mass. When the operator $\mathcal{O}_i^\mathrm{NR}$ appears with additional powers of $q^n$, the corresponding response function is $q^{2n} F_{i,i}$.

The response $F_M$ is the standard SI form factor which is coherently enhanced and therefore scales as $A^2$ (for equal couplings to protons and neutrons). The standard SD form factor is a combination of $F_{\Sigma'}$ and $F_{\Sigma''}$. Nuclear form factors have been calculated in Refs.~\cite{Fitzpatrick:2012ix,Anand:2013yka,Catena:2015uha}. For Xenon, we use the form factors listed in Appendix A.3 of Ref.~\cite{Fitzpatrick:2012ix}, apart from the spin-dependent form factors, which we take from Ref.~\cite{Menendez:2012tm} which includes two-body currents.
  
Finally, the differential recoil rate with a given target $T$ is obtained from the differential cross section as:
\begin{align}
\frac{\mathrm{d}R_T}{\mathrm{d}E_R} = \frac{\rho_\chi}{m_{\rm DM} m_T} \int_{v > v_\mathrm{min}} v f(v) \,\frac{\mathrm{d}\sigma_T}{\mathrm{d}E_R}\,\mathrm{d}v\,,
\end{align}
where $\rho_\chi$ is the DM density, $f(v)$ the DM velocity distribution and $v_\mathrm{min} = \sqrt{m_T E_R/(2 \mu_{\chi T}^2)}$. Further details can be found in e.g.~Ref.~\cite{Cerdeno:2010jj}. 
  
\subsection{Non-relativistic Operator Matching and Cross Sections}
\label{sec:DD_cross_sections}

Full details of the matching of the photon interactions with quark-level and nucleon-level interactions can be found in a number of references \cite{Fitzpatrick:2012ib,DelNobile:2013sia,Gresham:2014vja,Bishara:2017pfq}. Here we highlight a subtlety in the matching which is not always considered.

We can use the equations of motions in order to rewrite operators involving $\partial^\nu F_{\mu \nu}$ in terms of currents of Standard Model fermions $f$. We are considering interactions with nucleons, so we restrict ourself to interactions with quark currents $q$:
\begin{align}
\partial^\nu F_{\mu \nu} = - e \sum_{q} Q_q \overline{q} \gamma_\mu q\,.
 \end{align}
Note that this expression describes the interaction with \textit{free} fermions, while the nucleons we are interested in consist of bounds states of quarks. The expectation value of the quark vector current inside the nucleon can be written as (see e.g.~Eq.~(22) of Ref.~\cite{Bishara:2017pfq}):
\begin{align}
\label{eq:q_to_N}
\left\langle N \left| \overline { q } \gamma ^ { \mu } q \right| N \right\rangle = \overline { u } _ { N }  \left[ F _ { 1 } ^ { (q,N) } \left( q ^ { 2 } \right) \gamma ^ { \mu } + \frac { i } { 2 m _ { N } } F _ { 2 } ^ {( q , N) } \left( q ^ { 2 } \right) \sigma ^ { \mu \nu } q _ { \nu } \right] u _ { N }\,,
\end{align}
where $u_N$ are  nucleon spinors. The two form factors $F _ { 1 } ^ { (q,N) }(q^2)$ (`Dirac') and $F _ { 2} ^ { (q,N) }(q^2)$ (`Pauli') encode the internal nucleon structure (as a function of the momentum transfer $q^2$). The first describes the contribution of quarks $q$ to the charge of nucleon $N$ while the second describes the contribution to the nucleon magnetic moment. Galactic Dark Matter is not energetic enough to probe the internal $nucleon$ structure, so we can safely set $q^2 \rightarrow 0$, in which case we obtain (see Appendix A of Ref.~\cite{Bishara:2017pfq}):\footnote{The corresponding expression for neutrons is obtained by isospin symmetry: $n \leftrightarrow p$, $u \leftrightarrow d$, $s \leftrightarrow s$.}
\begin{align}
F _ { 1 } ^ { (u, p) } ( 0 ) &= 2\,,& F _ { 1 } ^ { (d, p) } ( 0 ) =& 1 \,,& F _ { 1 } ^ { (s, p) } ( 0 ) &= 0\,, \nonumber\\
F _ { 2 } ^ { (u, p) } ( 0 ) &= 1.609 \,,& F _ { 2 } ^ { (d , p) } ( 0 )  =& - 2.097\,,& F _ { 2 } ^ { (s, p) } ( 0 ) &= - 0.064\,.
\end{align}
Now, using Eq.~\eqref{eq:q_to_N}, we see that the quark electromagnetic current embedded in the nucleon is equivalent to:
\begin{align}
\label{eq:EMmapping}
e\sum_{q} Q_q \overline { q } \gamma ^ { \mu } q \rightarrow e \sum_{N=p,n} Q_N \overline{N} \gamma^\mu N + \frac{e}{2m_N} \sum_{N=p,n} a_N \overline{N} i\sigma^{\mu\nu}q_\nu N\,,
\end{align}
where $a_p = 1.793$ and $a_n = -1.913$ are calculated from the Pauli form factors given above (these are in fact the anomalous magnetic moments of the nucleons). Typically when coupling to the DM vector current, the first term in Eq.~\eqref{eq:EMmapping} leads to the standard spin-independent interaction, in which case the second term is sub-dominant and may be neglected \cite{DelNobile:2013sia}. However, in general (and in particular in the case of the anapole interaction ${\cal O}_{\chi 5 \chi \partial F}$), both terms must be included.

The relativistic interactions at the DM-\textit{nucleon} level can then be matched onto the NREFT operators described in Sec.~\ref{app:NREFT} using standard `dictionaries' (e.g.~Ref.~\cite{DelNobile:2018dfg}). Below we list the non-relativistic matrix elements for DM scattering with nucleons $N = p,\,n$ (as in Eq.~\eqref{eq:DMnucleon}) which are induced by the EMSM$_\chi$ operators after this matching has been performed: 
 
 \begin{align}
\label{eq:DMnucleonlist}
 {\mathcal{M}}_{\psi \psi F} & = 4 e \,\frac{\mathcal{C}}{\Lambda} \left[ Q_N m_N \ONR{1} + 4 Q_N \frac{m_\mathrm{DM} m_N}{q^2} \ONR{5} + 2 g_N m_\mathrm{DM} (\ONR{4} - \frac{1}{q^2} \ONR{6})\right]\,,  \nonumber \\
  {\mathcal{M}}_{\psi 5 \psi F} & =  16e  \,\frac{\mathcal{C}}{\Lambda}Q_N   \frac{m_\mathrm{DM} m_N}{q^2} \mathcal{O}^{\rm NR}_{11} \, ,  \nonumber \\
   {\mathcal{M}}_{\psi \psi \partial F} & =  -4 e   \,\frac{\mathcal{C}}{\Lambda^2}Q_N m_\mathrm{DM} m_N \ONR{1} \, ,  \nonumber \\
{\mathcal{M}}_{\chi 5 \chi \partial F} & = {\mathcal{M}}_{\psi 5 \psi \partial F} =    -4e   \,\frac{\mathcal{C}}{\Lambda^2}m_\mathrm{DM} (2 Q_N m_N \ONR{8} + g_N \ONR{9}) \, , \nonumber  \\
 {\mathcal{M}}_{\partial S \partial S F} & = - 2 e   \,\frac{\mathcal{C}}{\Lambda^2}Q_N m_\mathrm{DM} m_N \ONR{1} \,.
 \end{align}
 Here $Q_p = 1$, $Q_n = 0$ are the nucleon electric charges and $g_p = 5.59$, $g_n = -3.83$ are the nucleon $g$-factors  \cite{DelNobile:2013sia}. The factor $\mathcal{C}$ denotes the dimensionless numerical coefficient of each of the operators appearing in the relativistic Lagrangian. The precise definitions of the operators and their numerical coefficients (loop factors and $\mathcal{O}(1)$ factors) are given in Sec.~\ref{sec:photops}. As an example, in the case of the magnetic dipole interaction, we have $\mathcal{C} \equiv e \,\mathcal { C } _ { \psi \psi F }/(32 \pi^2)$. The operators in Sec.~\ref{sec:running} may induce some of the interactions appearing in Eq.~\eqref{eq:DMnucleonlist}, with the appropriate coefficients given in Eq.~\eqref{eq:LoopMatch}. 
 
 Using Eq.~\eqref{XSc}, a single EMSM$_\chi$ operator leads to a contribution to the differential cross-section for a target nucleus $T$ given by:
 \begin{align}
\frac{\mathrm{d}\sigma^T_{\psi \psi F}}{\mathrm{d}E_R} &=    \frac{\mathcal{C}^2 }{\Lambda^2} \frac{4 \alpha }{E_R}  \left[ 1 - \frac{  m_\mathrm{DM}  + 2  m_T  }{2 m_\mathrm{DM} m_T v^2} E_R   \right] F_M^{pp}    \nonumber \\
& +\frac{ \mathcal{C}^2 }{\Lambda^2} \frac{\alpha m_T}{2  m_N^2} \frac{1}{v^2} \left[ 16 F_{\Delta}^{pp} + \sum_{N, N' = p, n} g_N g_{N^\prime} \left(    F_{\Sigma^\prime }^{(N,N')} - 2 Q_N g_{N^\prime}   F_{\Sigma^{\prime }, \Delta }^{(N,N')} \right) \right]  \, , \nonumber \\
\frac{\mathrm{d}\sigma^T_{\psi 5 \psi F}}{\mathrm{d}E_R} &= \frac{\mathcal{C}^2}{\Lambda^2} \frac{4 \alpha}{E_R v^2}   F_{M}^{pp} \,, \nonumber \\
   \frac{\mathrm{d}\sigma^T_{\psi \psi \partial F}}{\mathrm{d}E_R} &= \frac{\mathcal{C}^2}{\Lambda^4} \frac{2 \alpha m_T}{v^2} F_M^{pp}\,, \nonumber \\
   \frac{\mathrm{d}\sigma^T_{\chi 5 \chi \partial F}}{\mathrm{d}E_R} &=    \frac{\mathrm{d}\sigma^T_{\psi 5 \psi \partial F}}{\mathrm{d}E_R} = \frac{\mathcal{C}^2}{\Lambda^4} \frac{4 \alpha m_T^2 E_R}{v^2 m_N^2} \left[ F_\Delta^{pp} + \frac{1}{16} \sum_{N, N^\prime = p, n} g_N g_{N^\prime} F_{\Sigma^\prime}^{NN^\prime}  + \frac{1}{2} \sum_{N = p,n} g_N F_{\Sigma^\prime, \Delta}^{pN} \right] \nonumber \\
   & \qquad\quad\quad\quad\quad+ \frac{\mathcal{C}^2}{\Lambda^4}  2 \alpha m_T \left[ 1- \frac{E_R}{2 m_\mathrm{DM}^2 m_T v^2} \left( m_\mathrm{DM} + m_T \right)^2 \right] F^{pp}_M \,,  \nonumber \\
   \frac{\mathrm{d}\sigma^T_{\partial S \partial S F}}{\mathrm{d}E_R} &= \frac{\mathcal{C}^2}{\Lambda^4} \frac{ \alpha m_T}{2 v^2} F_M^{pp} \,,
\end{align}
where $\alpha = e^2/(4\pi)$.

Following Ref.~\cite{Weiner:2012cb}, the differential cross section for the Majorana and Dirac Rayleigh operators ${\cal O}_{\chi \chi FF}$ and ${\cal O}_{\psi \psi FF}$ is given by:
\begin{equation}
\frac{\mathrm{d}\sigma^T_{\chi \chi FF}}{\mathrm{d}E_R} = \frac{\mathrm{d}\sigma^T_{\psi \psi FF}}{\mathrm{d}E_R}  = \frac{ m_T}{\pi^2 v^2}\left|  \alpha Z^2 Q_0 \frac{\mathcal{C}}{\Lambda^3} \mathcal{F} \left(\frac{2 m_T E_R}{Q_0^2}  \right)     \right|^2\,,
\end{equation}
where the nuclear coherence scale is $Q_0 = \sqrt{6}R_N$ and the nuclear radius is given by~\cite{Jungman:1995df} 
\begin{equation}
R_N = (0.3 + 0.91 A^{1/3}) \,\,\mathrm{fm}\,.
\end{equation}
We normalize the form factor $\mathcal{F}$ associated with the Rayleigh operator such that $\mathcal{F}(0) = 1$. An explicit expression is given in Appendix A of Ref.~\cite{Frandsen:2012db} (which corrects a number of mistakes in the derivation of Ref.~\cite{Weiner:2012cb}). As detailed in Ref.~\cite{Frandsen:2012db, DEramo:2016aee}, the Rayleigh operator mixes into the operators $\overline{\chi} \chi \overline{q} q$ and $\overline{\chi} \chi G^{\mu\nu} G_{\mu\nu}$, which in turn lead to standard SI scattering with nucleons. These different operators may interfere and we include this effect in our analysis\footnote{We do not include threshold corrections due to the top quark, because the energy scale $\Lambda$ probed by direct detection experiments is small compared to the top mass.}. However, the impact on the limits turns out to be small. The energy scale $\Lambda$ probed by direct detection experiments is rather low, so the mixing effects are not substantial.

Reference~\cite{Ovanesyan:2014fha} pointed out that for Rayleigh scattering, the effects of 2-nucleon scattering should not be ignored in calculating the direct detection rate. While $\mathcal{O}(1)$ corrections are expected, the exact form of the proton-proton form factor (which accounts for proton-proton correlations in the nucleus) is not known, so we neglect these effects here.

The remaining fermion Rayleigh operators give a negligible contribution to direct detection cross sections. The operator $\mathcal{O}_{\chi \chi F \tilde{F}}$ gives a vanishing DM-nucleon matrix element, while $\mathcal{O}_{\chi 5 \chi FF}$ and $\mathcal{O}_{\chi \chi F \tilde{F}}$ give rise to momentum-suppressed scattering \cite{Frandsen:2012db}. Given the already weak constraints on $\Lambda$ which arise from the unsuppressed operator $\mathcal{O}_{\chi\chi FF}$, we therefore neglect direct detection limits coming from the remaining operators. Note that the operators $\mathcal{O}_{\chi 5 \chi FF}$ and $\mathcal{O}_{\chi \chi F \tilde{F}}$ are still constrained by indirect detection, where they give rise to an $s$-wave annihilation cross section.

For scalar DM, the Rayleigh operators ${\cal O}_{RR FF}$ and ${\cal O}_{SS FF}$ lead to a similar cross section as in the fermion case:
\begin{equation}
\frac{\mathrm{d}\sigma^T_{RRFF}}{\mathrm{d}E_R} =  \frac{\mathrm{d}\sigma^T_{SSFF}}{\mathrm{d}E_R} = \frac{ m_T}{ \pi^2 m_\mathrm{DM}^2 v^2}\left|  \alpha Z^2 Q_0 \frac{\mathcal{C}}{\Lambda^2} \mathcal{F} \left(\frac{2 m_T E_R}{Q_0^2}  \right)     \right|^2\,.
\end{equation}
The other scalar Rayleigh operators $\mathcal{O}_{RRF\tilde{F}}$ and $\mathcal{O}_{SSF\tilde{F}}$ gives a vanishing contribution to the direct detection cross section (due to the anti-symmetry properties of $\tilde{F}_{\mu\nu}$) \cite{Frandsen:2012db}.

\subsection{Approximate XENON1T Bounds}
\label{app:Xenon1T}

The XENON1T \cite{Aprile:2017iyp,Aprile:2018dbl} detector is a liquid Xenon time projection chamber, operating with a total Xenon mass of 3.2t and a fiducial target mass of 1.30t. Table I of Ref.~\cite{Aprile:2018dbl} reports the number of observed events and the number of expected background events for a number of different cuts. We use here the 0.9t `reference mass' and events only in the `reference' signal region, between the median and $-2\sigma$ quantile in (cS2$_\mathrm{b}$, cS1) space. We therefore calculate the number of expected WIMP signal events for a total exposure of 278.8 days $\times$ 0.9t, corrected by an overall factor of $0.475$ to account for the nuclear recoil acceptance of the `reference' region in  (cS2$_\mathrm{b}$, cS1). The nuclear recoil detection efficiency is taken from Fig.~1 of Ref.~\cite{Aprile:2018dbl}.

For this exposure, the number of observed events is $N_\mathrm{obs} = 2$, while the best fit number of expected background events is $N_\mathrm{BG} = 1.62$. We use this to set a single-bin Poisson upper limit on the number of signal events $N_\mathrm{sig}^{90\%}$. Neglecting background uncertainties, this limit is calculated by solving \cite{Feldman:1997qc}:
\begin{equation}\label{eq:Nsig90}
\sum_{k \geqslant N_\mathrm{obs} + 1} P(k | N_\mathrm{BG} + N_\mathrm{sig}^{90\%}) = 90\%\,,
\end{equation}
where $P(k | N)$ is the Poisson probability of observing $k$ events, given $N$ expected events. This limit on the number of signal events is then converted into a limit on the relevant operator coupling. Python code for calculating this limit is available online \cite{XENON1T-code}.

\begin{figure}[!t]
 \centering
 
\includegraphics[width= 0.60 \textwidth]{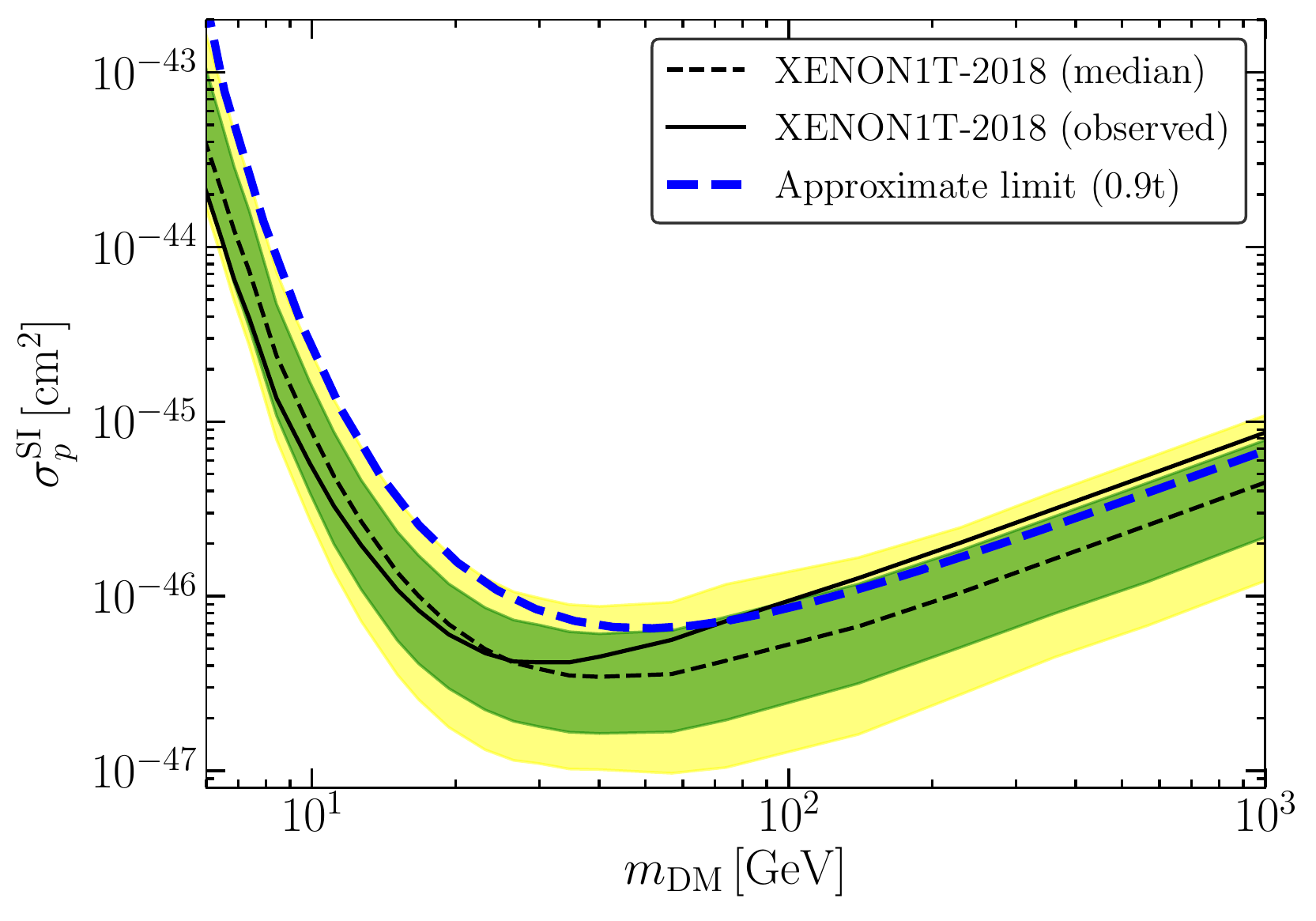} 

\caption{Approximate XENON1T bound (dashed blue) calculated from a single-bin Poisson upper limit. We show for comparison the median expected sensitivity (dashed black) and observed limit (solid black) of the XENON1T 2018 exposure \cite{Aprile:2018dbl}. Shaded bands show the $1\sigma$ and $2\sigma$ bands on the sensitivity. Our limit matches the official limit within a factor of a few for all DM masses above $10\,\,\mathrm{GeV}$. \label{fig:Xe1T}}
\end{figure}

We show in Fig.~\ref{fig:Xe1T} the resulting approximate limit (dashed blue) for the standard spin-independent interaction, as well as the median expected sensitivity and observed limit as reported by the XENON1T collaboration (dashed and solid black, respectively). Our approximate limit is within a factor of a few of the official XENON1T limit for DM masses heavier than $10 \,\,\mathrm{GeV}$ and is consistent with other approximate limits presented in the literature \cite{Workgroup:2017lvb,Fowlie:2018svr,Athron:2018hpc,DDCalc:url}. The XENON1T 2018 exposure observed an upward fluctuation at high recoil energies, leading the observed limit to be weaker than the expected sensitivity at high DM mass. Our limit matches the official limit closely at high DM mass but deviates for low masses, as we do not include any recoil energy information about the individual events. A more detailed treatment would require more information about the background and signal distributions in (cS2$_\mathrm{b}$, cS1) and is beyond the scope of this work.

\newpage 
  \section{Indirect Detection}
 \label{app:IndirectDetection}
  In this Appendix we summarize the annihilation cross sections for all relevant operators in Table~\ref{XS}. Note that below $\psi$ denotes a Dirac fermion, $\chi$ a Majorana fermion, $S$ a complex scalar and $R$ a real scalar.  Where applicable, we have checked that the cross sections match those given in Ref.~\cite{DelNobile:2012tx}.
\begin{table}[h!]
\centering 
\begin{tabular}{|c||cc|c|}
\hline
& & $\sigma  (s) $& $\sigma v_{\rm rel}$  \\
\hline
${\cal O}_{\psi \psi F}$ & $\psi \overline{\psi} \to f \overline{f}:$
  & $ \frac{\alpha N_c^f Q_f^2}{3 } \frac{\beta_f}{\beta_\psi}  \left(  3 - 2 \beta_\psi^2 \right)  \left(  3 - \beta_f^2 \right) $ & $ 4  \alpha Q_f^2 N_c^f $ \\
 & $\psi \overline{\psi} \to \gamma \gamma:$ & $ \frac{s}{ 6 \pi \beta_\psi } \left( 9- 7 \beta_\psi^2  - \frac{6 (1-\beta_\psi^2)^2}{\beta_\psi}  {\rm arctanh} \, \beta_\psi \right)$  & $  \frac{4 m_\psi^2}{ \pi }$\\
  \hline
${\cal O}_{\psi 5 \psi F}$  & $\psi \overline{\psi} \to f \overline{f}:$ & $\frac{\alpha N_c^f Q_f^2}{3  } \beta_f \beta_\psi \left( 3 - \beta_f^2\right)$  & $\frac{\alpha Q_f^2 N_c^f}{3 } v_{\rm rel}^2$\\
& $\psi \overline{\psi} \to \gamma \gamma:$ & $\frac{s }{ 6 \pi \beta_\psi  } \left(9 - 7 \beta_\psi^2 - \frac{6 (1-\beta_\psi^2)^2  }{\beta_\psi}{\rm arctanh} \, \beta_\psi \right)$& $ \frac{4 m_\psi^2}{ \pi} $ \\ 
\hline
\hline
${\cal O}_{\psi  \psi \partial F}$  & $\psi \overline{\psi} \to f \overline{f}:$ & $\frac{\alpha N_c^f Q_f^2 s}{12 } \frac{\beta_f}{\beta_\psi } \left( 3- \beta_\psi^2 \right) \left( 3- \beta_f^2 \right)$ & $ 4 \alpha m_\psi^2 Q_f^2 N_c^f $\\
\hline
${\cal O}_{\psi 5 \psi \partial F}$  & $\psi \overline{\psi} \to f \overline{f}:$ & $\frac{ \alpha N_c^f Q_f^2 s}{6  } \beta_f \beta_\psi  \left(3 -\beta_f^2 \right)$ & $ \frac{ 2 \alpha m_\psi^2 Q_f^2 N_c^f }{ 3  } v_{\rm rel}^2$\\
\hline
${\cal O}_{ \partial S  \partial S  F}$  & $S S^* \to f \overline{f}:$ & $\frac{\alpha N_c^f Q_f^2 s}{24  }   \beta_f \beta_S  \left(3 -\beta_f^2 \right)$ & $ \frac{\alpha m_S^2 Q_f^2 N_c^f  }{6  } v_{\rm rel}^2$ \\
\hline
\hline
${\cal O}_{ \psi \psi  F F}, {\cal O}_{ \psi \psi  F \tilde{F}}$  & $\psi \overline{\psi} \to \gamma \gamma:$ & $\frac{s^2  \beta_\psi}{8 \pi  }$ & $\frac{ m_\psi^4}{ \pi } v_{\rm rel}^2$ \\
\hline
${\cal O}_{ \psi 5 \psi  F F},{\cal O}_{ \psi 5 \psi  F \tilde{F}}$  & $\psi \overline{\psi}  \to \gamma \gamma:$ & $\frac{s^2}{8 \pi \beta_\psi }$ & $\frac{4 m_\psi^4}{ \pi  } $ \\
\hline
${\cal O}_{ SS F F},{\cal O}_{ SS  F \tilde{F}}$  & $S S^* \to \gamma \gamma:$ & $ \frac{s}{4 \pi \beta_S  } $ & $\frac{2 m_S^2}{ \pi  } $ \\
\hline
\hline
${\cal O}_{ \psi \psi  \ell \ell }$  & $\psi \overline{\psi}  \to \ell \overline{\ell}:$ & $\frac{s }{48 \pi  }  \frac{\beta_\ell }{\beta_\psi}  \left( 3 - \beta_\psi^2 \right)  \left( 3 - \beta_\ell^2 \right) $ & $\frac{m_\psi^2}{\pi }$ \\
\hline
${\cal O}_{ \psi 5 \psi  \ell \ell }$  & $\psi \overline{\psi}  \to \ell \overline{\ell}:$ & $ \frac{s  }{24 \pi  }  \beta_\ell \beta_\psi  \left( 3 - \beta_\ell^2 \right)  $ & $  \frac{m_\psi^2}{6 \pi } v_{\rm rel}^2$ \\
\hline
${\cal O}_{ S \partial S \ell \ell }$  & $S S^* \to \ell \overline{\ell}:$ & $\frac{s }{24 \pi  }  \beta_\ell \beta_S \left( 3 - \beta_\ell^2 \right)$ & $ \frac{ m_S^2}{6 \pi  }  v_{\rm rel}^2$ \\
\hline
\hline
${\cal O}_{ \psi \psi  \ell 5 \ell }$  & $\psi \overline{\psi}  \to \ell \overline{\ell}:$ & $\frac{s }{24 \pi  }  \frac{\beta_\ell^3 }{\beta_\psi}  \left( 3 - \beta_\psi^2 \right)   $ & $\frac{m_\psi^2}{\pi }$ \\
\hline
${\cal O}_{ \psi 5 \psi  \ell 5 \ell }$  & $\psi \overline{\psi}  \to \ell \overline{\ell}:$ & $ \frac{s  }{48 \pi  }  \frac{\beta_\ell}{ \beta_\psi}  \left( 7 \beta_\ell^2 \beta_\psi^2 - 3 \beta_\ell^2 - 3 \beta_\psi^2 + 3  \right)  $ & $  \frac{m_\ell^2}{2 \pi} +  \frac{ m_\psi^2}{6 \pi  }  v_{\rm rel}^2$ \\
\hline
${\cal O}_{ S \partial S \ell 5 \ell }$  & $S S^* \to \ell \overline{\ell}:$ & $\frac{s }{12 \pi  }  \beta_\ell^3 \beta_S$ & $ \frac{ m_S^2}{6 \pi  }  v_{\rm rel}^2$ \\
\hline
\hline
${\cal O}_{ \psi \psi  \ell  \ell -8}, {\cal O}_{ \psi \psi  \ell 5  \ell -8}$  & $\psi \overline{\psi}  \to \ell \overline{\ell}:$ & $m_\psi^2 m_\ell^2 \frac{s }{16 \pi  }  \beta_\ell^3 \beta_\psi  \, , \, m_\psi^2 m_\ell^2 \frac{s }{16 \pi  }  \beta_\ell \beta_\psi  $ & $m_\psi^2 m_\ell^2 \frac{m_\psi^2}{8 \pi } v_{\rm rel}^2 $ \\
\hline
${\cal O}_{ \psi 5 \psi  \ell  \ell -8}, {\cal O}_{ \psi 5 \psi  \ell  5 \ell -8}$  & $\psi \overline{\psi}  \to \ell \overline{\ell}:$ & $m_\psi^2 m_\ell^2 \frac{s }{16 \pi  }  \frac{\beta_\ell^3 }{\beta_\psi} \, , \, m_\psi^2 m_\ell^2 \frac{s }{16 \pi  }  \frac{\beta_\ell}{\beta_\psi}     $ & $m_\psi^2 m_\ell^2 \frac{m_\psi^2} {2 \pi }$ \\
\hline
${\cal O}_{ S  S \ell 5 \partial \ell }, {\cal O}_{ S  S \ell  \partial \ell }$  & $S S^* \to \ell \overline{\ell}:$ & $m_\ell^2 \frac{1}{8 \pi  }  \frac{\beta_\ell}{ \beta_S} \,,\, m_\ell^2 \frac{1 }{8 \pi  }  \frac{\beta_\ell^3}{ \beta_S}$ & $  m_\ell^2 \frac{1}{4 \pi  }  $ \\
\hline
\end{tabular}
\caption{\label{XS} Annihilation cross sections for operators defined in Section 2 (with no numerical prefactors except 1/2 for operators with Majorana fermions and real scalars). The scaling due to an operator coefficient ${\cal C}$ is $\propto {\cal C}^2$ in each cross section, except for the annihilation into photons due to ${\cal O}_{\psi \psi F}$ and ${\cal O}_{\psi 5 \psi F}$, which scale as $\propto {\cal C}^4$.  Here $\beta_i \equiv (1 - 4 m^2_i/s)^{1/2} $, $Q_f$ is the fermion charge and $N_c^f$ the fermion color factor. For the non-relativistic annihilation cross-sections to fermions we only keep the leading terms in $m_f^2/m^2_{\rm DM}$. Note that operators involving $\partial F$ do not contribute to annihilation into photons at tree-level. }
\end{table}

 \section{Full DM-Quark Scattering Amplitudes}
 \label{appfull}
\setcounter{equation}{0}
For the general Lagrangian
\begin{align}
{\cal L} & = \left( \lambda_i \, S  \overline{F} P_R  \ell_i  + {\rm h.c.} \right) - M_{S}^2 |S|^2  - M_F \overline{F} F \, , 
\end{align} 
the loop functions for the DM-quark scattering amplitude in Eq.~\eqref{FDMauv} and \eqref{SDMauv} are given by (in the limit $m_{\ell_i}^2 , q^2 \ll  M_S^2, M_F^2$)
\begin{align}
C_\gamma & = - \frac{\lambda_i^2}{2} \int_0^1 \int_0^1 dx \, dy  \left[ \frac{3x -1}{\Delta_e} + 2 M_{F}^2 \frac{x (1-x)^2 (2x-1) }{\Delta_e^2}  \right] y (1-y) \, ,  \nonumber \\
C_{\gamma 5} & =  \frac{\lambda_i^2}{2} \int_0^1 \int_0^1 dx \, dy    \left[ \frac{3x -1}{\Delta_e}  y (1-y) \right] \, ,  \\
C_\sigma & =  \frac{\lambda_i^2}{2} M_{F} \int_0^1 \int_0^1 dx \, dy    \left[ \frac{x  (1-x)}{\Delta_e}  \right] \, , \nonumber \\
C_S & =\frac{\lambda_i^2}{2} \int_0^1 \int_0^1 dx \, dy  \left[  \frac{ 6 x^3 - 6 x^2 - 5 x +3 }{ \Delta_S} - M_S^2  \frac{x (1-x)^2  (1+x) (2x-1)}{ \Delta_S^2} \right]  y (1-y) \, , \nonumber
\end{align}
where 
\begin{align}
\Delta_e & = - x (1-x) M_{F}^2 + (1-x) M_S^2 + x \, m_{\ell_i}^2 - x^2 y (1-y) q^2 \, , \nonumber \\
\Delta_S & = -  x (1-x)  M_S^2    + (1-x) M_F^2 + x \, m_{\ell_i}^2 - x^2 y (1-y) q^2 \, .
\end{align}

 \bibliography{biblio}

\providecommand{\href}[2]{#2}\begingroup\raggedright\begin{thebibliography}{100}

\bibitem{Jungman:1995df}
G.~Jungman, M.~Kamionkowski and K.~Griest, \emph{{Supersymmetric dark matter}},
  \href{http://dx.doi.org/10.1016/0370-1573(95)00058-5}{\emph{Phys. Rept.} {\bf
  267} (1996) 195--373}, [\href{http://arxiv.org/abs/hep-ph/9506380}{{\tt
  hep-ph/9506380}}].

\bibitem{Bertone:2004pz}
G.~Bertone, D.~Hooper and J.~Silk, \emph{{Particle dark matter: Evidence,
  candidates and constraints}},
  \href{http://dx.doi.org/10.1016/j.physrep.2004.08.031}{\emph{Phys. Rept.}
  {\bf 405} (2005) 279--390}, [\href{http://arxiv.org/abs/hep-ph/0404175}{{\tt
  hep-ph/0404175}}].

\bibitem{Feng:2010gw}
J.~L. Feng, \emph{{Dark Matter Candidates from Particle Physics and Methods of
  Detection}},
  \href{http://dx.doi.org/10.1146/annurev-astro-082708-101659}{\emph{Ann. Rev.
  Astron. Astrophys.} {\bf 48} (2010) 495--545},
  [\href{http://arxiv.org/abs/1003.0904}{{\tt 1003.0904}}].

\bibitem{Undagoitia:2015gya}
T.~MarrodÃ¡n~Undagoitia and L.~Rauch, \emph{{Dark matter direct-detection
  experiments}},
  \href{http://dx.doi.org/10.1088/0954-3899/43/1/013001}{\emph{J. Phys.} {\bf
  G43} (2016) 013001}, [\href{http://arxiv.org/abs/1509.08767}{{\tt
  1509.08767}}].

\bibitem{Gaskins:2016cha}
J.~M. Gaskins, \emph{{A review of indirect searches for particle dark matter}},
  \href{http://dx.doi.org/10.1080/00107514.2016.1175160}{\emph{Contemp. Phys.}
  {\bf 57} (2016) 496--525}, [\href{http://arxiv.org/abs/1604.00014}{{\tt
  1604.00014}}].

\bibitem{Kahlhoefer:2017dnp}
F.~Kahlhoefer, \emph{{Review of LHC Dark Matter Searches}},
  \href{http://dx.doi.org/10.1142/S0217751X1730006X}{\emph{Int. J. Mod. Phys.}
  {\bf A32} (2017) 1730006}, [\href{http://arxiv.org/abs/1702.02430}{{\tt
  1702.02430}}].

\bibitem{DeRujula:1989fe}
A.~De~Rujula, S.~L. Glashow and U.~Sarid, \emph{{CHARGED DARK MATTER}},
  \href{http://dx.doi.org/10.1016/0550-3213(90)90227-5}{\emph{Nucl. Phys.} {\bf
  B333} (1990) 173--194}.

\bibitem{Barkana:2018lgd}
R.~Barkana, \emph{{Possible interaction between baryons and dark-matter
  particles revealed by the first stars}},
  \href{http://dx.doi.org/10.1038/nature25791}{\emph{Nature} {\bf 555} (2018)
  71--74}, [\href{http://arxiv.org/abs/1803.06698}{{\tt 1803.06698}}].

\bibitem{Munoz:2018pzp}
J.~B. Munoz and A.~Loeb, \emph{{A small amount of mini-charged dark matter
  could cool the baryons in the early Universe}},
  \href{http://dx.doi.org/10.1038/s41586-018-0151-x}{\emph{Nature} {\bf 557}
  (2018) 684}, [\href{http://arxiv.org/abs/1802.10094}{{\tt 1802.10094}}].

\bibitem{Fraser:2018acy}
S.~Fraser et~al., \emph{{The EDGES 21 cm Anomaly and Properties of Dark
  Matter}}, \href{http://dx.doi.org/10.1016/j.physletb.2018.08.035}{\emph{Phys.
  Lett.} {\bf B785} (2018) 159--164},
  [\href{http://arxiv.org/abs/1803.03245}{{\tt 1803.03245}}].

\bibitem{Berlin:2018sjs}
A.~Berlin, D.~Hooper, G.~Krnjaic and S.~D. McDermott, \emph{{Severely
  Constraining Dark Matter Interpretations of the 21-cm Anomaly}},
  \href{http://dx.doi.org/10.1103/PhysRevLett.121.011102}{\emph{Phys. Rev.
  Lett.} {\bf 121} (2018) 011102}, [\href{http://arxiv.org/abs/1803.02804}{{\tt
  1803.02804}}].

\bibitem{Holdom:1985ag}
B.~Holdom, \emph{{Two U(1)'s and Epsilon Charge Shifts}},
  \href{http://dx.doi.org/10.1016/0370-2693(86)91377-8}{\emph{Phys. Lett.} {\bf
  166B} (1986) 196--198}.

\bibitem{Holdom:1986eq}
B.~Holdom, \emph{{Searching for $\epsilon$ Charges and a New U(1)}},
  \href{http://dx.doi.org/10.1016/0370-2693(86)90470-3}{\emph{Phys. Lett.} {\bf
  B178} (1986) 65--70}.

\bibitem{Abel:2003ue}
S.~A. Abel and B.~W. Schofield, \emph{{Brane anti-brane kinetic mixing,
  millicharged particles and SUSY breaking}},
  \href{http://dx.doi.org/10.1016/j.nuclphysb.2004.02.037}{\emph{Nucl. Phys.}
  {\bf B685} (2004) 150--170}, [\href{http://arxiv.org/abs/hep-th/0311051}{{\tt
  hep-th/0311051}}].

\bibitem{Batell:2005wa}
B.~Batell and T.~Gherghetta, \emph{{Localized U(1) gauge fields, millicharged
  particles, and holography}},
  \href{http://dx.doi.org/10.1103/PhysRevD.73.045016}{\emph{Phys. Rev.} {\bf
  D73} (2006) 045016}, [\href{http://arxiv.org/abs/hep-ph/0512356}{{\tt
  hep-ph/0512356}}].

\bibitem{DelNobile:2015bqo}
E.~Del~Nobile, M.~Nardecchia and P.~Panci, \emph{{Millicharge or Decay: A
  Critical Take on Minimal Dark Matter}},
  \href{http://dx.doi.org/10.1088/1475-7516/2016/04/048}{\emph{JCAP} {\bf 1604}
  (2016) 048}, [\href{http://arxiv.org/abs/1512.05353}{{\tt 1512.05353}}].

\bibitem{Davidson:2000hf}
S.~Davidson, S.~Hannestad and G.~Raffelt, \emph{{Updated bounds on millicharged
  particles}},
  \href{http://dx.doi.org/10.1088/1126-6708/2000/05/003}{\emph{JHEP} {\bf 05}
  (2000) 003}, [\href{http://arxiv.org/abs/hep-ph/0001179}{{\tt
  hep-ph/0001179}}].

\bibitem{McDermott:2010pa}
S.~D. McDermott, H.-B. Yu and K.~M. Zurek, \emph{{Turning off the Lights: How
  Dark is Dark Matter?}},
  \href{http://dx.doi.org/10.1103/PhysRevD.83.063509}{\emph{Phys. Rev.} {\bf
  D83} (2011) 063509}, [\href{http://arxiv.org/abs/1011.2907}{{\tt
  1011.2907}}].

\bibitem{Pospelov:2000bq}
M.~Pospelov and T.~ter Veldhuis, \emph{{Direct and indirect limits on the
  electromagnetic form-factors of WIMPs}},
  \href{http://dx.doi.org/10.1016/S0370-2693(00)00358-0}{\emph{Phys. Lett.}
  {\bf B480} (2000) 181--186}, [\href{http://arxiv.org/abs/hep-ph/0003010}{{\tt
  hep-ph/0003010}}].

\bibitem{Sigurdson:2004zp}
K.~Sigurdson, M.~Doran, A.~Kurylov, R.~R. Caldwell and M.~Kamionkowski,
  \emph{{Dark-matter electric and magnetic dipole moments}},
  \href{http://dx.doi.org/10.1103/PhysRevD.70.083501,
  10.1103/PhysRevD.73.089903}{\emph{Phys. Rev.} {\bf D70} (2004) 083501},
  [\href{http://arxiv.org/abs/astro-ph/0406355}{{\tt astro-ph/0406355}}].

\bibitem{Barger:2010gv}
V.~Barger, W.-Y. Keung and D.~Marfatia, \emph{{Electromagnetic properties of
  dark matter: Dipole moments and charge form factor}},
  \href{http://dx.doi.org/10.1016/j.physletb.2010.12.008}{\emph{Phys. Lett.}
  {\bf B696} (2011) 74--78}, [\href{http://arxiv.org/abs/1007.4345}{{\tt
  1007.4345}}].

\bibitem{Banks:2010eh}
T.~Banks, J.-F. Fortin and S.~Thomas, \emph{{Direct Detection of Dark Matter
  Electromagnetic Dipole Moments}},  \href{http://arxiv.org/abs/1007.5515}{{\tt
  1007.5515}}.

\bibitem{DelNobile:2017fzy}
E.~Del~Nobile, \emph{{Direct detection signals of dark matter with magnetic
  dipole moment}}, \href{http://dx.doi.org/10.22323/1.314.0626}{\emph{PoS} {\bf
  EPS-HEP2017} (2017) 626}, [\href{http://arxiv.org/abs/1709.08700}{{\tt
  1709.08700}}].

\bibitem{Ho:2012bg}
C.~M. Ho and R.~J. Scherrer, \emph{{Anapole Dark Matter}},
  \href{http://dx.doi.org/10.1016/j.physletb.2013.04.039}{\emph{Phys. Lett.}
  {\bf B722} (2013) 341--346}, [\href{http://arxiv.org/abs/1211.0503}{{\tt
  1211.0503}}].

\bibitem{Gao:2013vfa}
Y.~Gao, C.~M. Ho and R.~J. Scherrer, \emph{{Anapole Dark Matter at the LHC}},
  \href{http://dx.doi.org/10.1103/PhysRevD.89.045006}{\emph{Phys. Rev.} {\bf
  D89} (2014) 045006}, [\href{http://arxiv.org/abs/1311.5630}{{\tt
  1311.5630}}].

\bibitem{DelNobile:2014eta}
E.~Del~Nobile, G.~B. Gelmini, P.~Gondolo and J.-H. Huh, \emph{{Direct detection
  of Light Anapole and Magnetic Dipole DM}},
  \href{http://dx.doi.org/10.1088/1475-7516/2014/06/002}{\emph{JCAP} {\bf 1406}
  (2014) 002}, [\href{http://arxiv.org/abs/1401.4508}{{\tt 1401.4508}}].

\bibitem{Alves:2017uls}
A.~Alves, A.~C.~O. Santos and K.~Sinha, \emph{{Collider Detection of Dark
  Matter Electromagnetic Anapole Moments}},
  \href{http://dx.doi.org/10.1103/PhysRevD.97.055023}{\emph{Phys. Rev.} {\bf
  D97} (2018) 055023}, [\href{http://arxiv.org/abs/1710.11290}{{\tt
  1710.11290}}].

\bibitem{Weiner:2012cb}
N.~Weiner and I.~Yavin, \emph{{How Dark Are Majorana WIMPs? Signals from MiDM
  and Rayleigh Dark Matter}},
  \href{http://dx.doi.org/10.1103/PhysRevD.86.075021}{\emph{Phys. Rev.} {\bf
  D86} (2012) 075021}, [\href{http://arxiv.org/abs/1206.2910}{{\tt
  1206.2910}}].

\bibitem{Weiner:2012gm}
N.~Weiner and I.~Yavin, \emph{{UV completions of magnetic inelastic and
  Rayleigh dark matter for the Fermi Line(s)}},
  \href{http://dx.doi.org/10.1103/PhysRevD.87.023523}{\emph{Phys. Rev.} {\bf
  D87} (2013) 023523}, [\href{http://arxiv.org/abs/1209.1093}{{\tt
  1209.1093}}].

\bibitem{Frandsen:2012db}
M.~T. Frandsen, U.~Haisch, F.~Kahlhoefer, P.~Mertsch and K.~Schmidt-Hoberg,
  \emph{{Loop-induced dark matter direct detection signals from gamma-ray
  lines}}, \href{http://dx.doi.org/10.1088/1475-7516/2012/10/033}{\emph{JCAP}
  {\bf 1210} (2012) 033}, [\href{http://arxiv.org/abs/1207.3971}{{\tt
  1207.3971}}].

\bibitem{Latimer:2016kdg}
D.~C. Latimer, \emph{{Two-photon interactions with Majorana fermions}},
  \href{http://dx.doi.org/10.1103/PhysRevD.94.093010}{\emph{Phys. Rev.} {\bf
  D94} (2016) 093010}, [\href{http://arxiv.org/abs/1706.05071}{{\tt
  1706.05071}}].

\bibitem{Bai:2014osa}
Y.~Bai and J.~Berger, \emph{{Lepton Portal Dark Matter}},
  \href{http://dx.doi.org/10.1007/JHEP08(2014)153}{\emph{JHEP} {\bf 08} (2014)
  153}, [\href{http://arxiv.org/abs/1402.6696}{{\tt 1402.6696}}].

\bibitem{DEramo:2014nmf}
F.~D'Eramo and M.~Procura, \emph{{Connecting Dark Matter UV Complete Models to
  Direct Detection Rates via Effective Field Theory}},
  \href{http://dx.doi.org/10.1007/JHEP04(2015)054}{\emph{JHEP} {\bf 04} (2015)
  054}, [\href{http://arxiv.org/abs/1411.3342}{{\tt 1411.3342}}].

\bibitem{Crivellin:2014qxa}
A.~Crivellin, F.~D'Eramo and M.~Procura, \emph{{New Constraints on Dark Matter
  Effective Theories from Standard Model Loops}},
  \href{http://dx.doi.org/10.1103/PhysRevLett.112.191304}{\emph{Phys. Rev.
  Lett.} {\bf 112} (2014) 191304}, [\href{http://arxiv.org/abs/1402.1173}{{\tt
  1402.1173}}].

\bibitem{DEramo:2016gos}
F.~D'Eramo, B.~J. Kavanagh and P.~Panci, \emph{{You can hide but you have to
  run: direct detection with vector mediators}},
  \href{http://dx.doi.org/10.1007/JHEP08(2016)111}{\emph{JHEP} {\bf 08} (2016)
  111}, [\href{http://arxiv.org/abs/1605.04917}{{\tt 1605.04917}}].

\bibitem{Hisano:2015bma}
J.~Hisano, R.~Nagai and N.~Nagata, \emph{{Effective Theories for Dark Matter
  Nucleon Scattering}},
  \href{http://dx.doi.org/10.1007/JHEP05(2015)037}{\emph{JHEP} {\bf 05} (2015)
  037}, [\href{http://arxiv.org/abs/1502.02244}{{\tt 1502.02244}}].

\bibitem{runDM}
B.~J. Kavanagh, F.~D'Eramo and P.~Panci, \emph{{runDM v1.0 [Computer
  Software]}},  2016.
\newblock 10.5281/zenodo.823249.

\bibitem{Haisch:2013uaa}
U.~Haisch and F.~Kahlhoefer, \emph{{On the importance of loop-induced
  spin-independent interactions for dark matter direct detection}},
  \href{http://dx.doi.org/10.1088/1475-7516/2013/04/050}{\emph{JCAP} {\bf 1304}
  (2013) 050}, [\href{http://arxiv.org/abs/1302.4454}{{\tt 1302.4454}}].

\bibitem{Akerib:2016vxi}
{\scshape LUX} collaboration, D.~S. Akerib et~al., \emph{{Results from a search
  for dark matter in the complete LUX exposure}},
  \href{http://dx.doi.org/10.1103/PhysRevLett.118.021303}{\emph{Phys. Rev.
  Lett.} {\bf 118} (2017) 021303}, [\href{http://arxiv.org/abs/1608.07648}{{\tt
  1608.07648}}].

\bibitem{Aprile:2018dbl}
{\scshape XENON} collaboration, E.~Aprile et~al., \emph{{Dark Matter Search
  Results from a One Tonne$\times$Year Exposure of XENON1T}},
  \href{http://arxiv.org/abs/1805.12562}{{\tt 1805.12562}}.

\bibitem{Akerib:2015cja}
{\scshape LZ} collaboration, D.~S. Akerib et~al., \emph{{LUX-ZEPLIN (LZ)
  Conceptual Design Report}},  \href{http://arxiv.org/abs/1509.02910}{{\tt
  1509.02910}}.

\bibitem{Mount:2017qzi}
B.~J. Mount et~al., \emph{{LUX-ZEPLIN (LZ) Technical Design Report}},
  \href{http://arxiv.org/abs/1703.09144}{{\tt 1703.09144}}.

\bibitem{Cui:2017nnn}
{\scshape PandaX-II} collaboration, X.~Cui et~al., \emph{{Dark Matter Results
  From 54-Ton-Day Exposure of PandaX-II Experiment}},
  \href{http://dx.doi.org/10.1103/PhysRevLett.119.181302}{\emph{Phys. Rev.
  Lett.} {\bf 119} (2017) 181302}, [\href{http://arxiv.org/abs/1708.06917}{{\tt
  1708.06917}}].

\bibitem{Hehn:2016nll}
{\scshape EDELWEISS} collaboration, L.~Hehn et~al., \emph{{Improved
  EDELWEISS-III sensitivity for low-mass WIMPs using a profile likelihood
  approach}},
  \href{http://dx.doi.org/10.1140/epjc/s10052-016-4388-y}{\emph{Eur. Phys. J.}
  {\bf C76} (2016) 548}, [\href{http://arxiv.org/abs/1607.03367}{{\tt
  1607.03367}}].

\bibitem{Agnese:2017jvy}
{\scshape SuperCDMS} collaboration, R.~Agnese et~al., \emph{{Low-mass dark
  matter search with CDMSlite}},
  \href{http://dx.doi.org/10.1103/PhysRevD.97.022002}{\emph{Phys. Rev.} {\bf
  D97} (2018) 022002}, [\href{http://arxiv.org/abs/1707.01632}{{\tt
  1707.01632}}].

\bibitem{Petricca:2017zdp}
{\scshape CRESST} collaboration, F.~Petricca et~al., \emph{{First results on
  low-mass dark matter from the CRESST-III experiment}},  in \emph{{15th
  International Conference on Topics in Astroparticle and Underground Physics
  (TAUP 2017) Sudbury, Ontario, Canada, July 24-28, 2017}}, 2017.
\newblock \href{http://arxiv.org/abs/1711.07692}{{\tt 1711.07692}}.

\bibitem{DelNobile:2013sia}
M.~Cirelli, E.~Del~Nobile and P.~Panci, \emph{{Tools for model-independent
  bounds in direct dark matter searches}},
  \href{http://dx.doi.org/10.1088/1475-7516/2013/10/019}{\emph{JCAP} {\bf 1310}
  (2013) 019}, [\href{http://arxiv.org/abs/1307.5955}{{\tt 1307.5955}}].

\bibitem{Kavanagh:2016pyr}
B.~J. Kavanagh, R.~Catena and C.~Kouvaris, \emph{{Signatures of
  Earth-scattering in the direct detection of Dark Matter}},
  \href{http://dx.doi.org/10.1088/1475-7516/2017/01/012}{\emph{JCAP} {\bf 1701}
  (2017) 012}, [\href{http://arxiv.org/abs/1611.05453}{{\tt 1611.05453}}].

\bibitem{Ackermann:2015lka}
{\scshape Fermi-LAT} collaboration, M.~Ackermann et~al., \emph{{Updated search
  for spectral lines from Galactic dark matter interactions with pass 8 data
  from the Fermi Large Area Telescope}},
  \href{http://dx.doi.org/10.1103/PhysRevD.91.122002}{\emph{Phys. Rev.} {\bf
  D91} (2015) 122002}, [\href{http://arxiv.org/abs/1506.00013}{{\tt
  1506.00013}}].

\bibitem{Lefranc:2016fgn}
V.~Lefranc, E.~Moulin, P.~Panci, F.~Sala and J.~Silk, \emph{{Dark Matter in
  $\gamma$ lines: Galactic Center vs dwarf galaxies}},
  \href{http://dx.doi.org/10.1088/1475-7516/2016/09/043}{\emph{JCAP} {\bf 1609}
  (2016) 043}, [\href{http://arxiv.org/abs/1608.00786}{{\tt 1608.00786}}].

\bibitem{Abramowski:2013ax}
{\scshape H.E.S.S.} collaboration, A.~Abramowski et~al., \emph{{Search for
  Photon-Linelike Signatures from Dark Matter Annihilations with H.E.S.S.}},
  \href{http://dx.doi.org/10.1103/PhysRevLett.110.041301}{\emph{Phys. Rev.
  Lett.} {\bf 110} (2013) 041301}, [\href{http://arxiv.org/abs/1301.1173}{{\tt
  1301.1173}}].

\bibitem{Ackermann:2015zua}
{\scshape Fermi-LAT} collaboration, M.~Ackermann et~al., \emph{{Searching for
  Dark Matter Annihilation from Milky Way Dwarf Spheroidal Galaxies with Six
  Years of Fermi Large Area Telescope Data}},
  \href{http://dx.doi.org/10.1103/PhysRevLett.115.231301}{\emph{Phys. Rev.
  Lett.} {\bf 115} (2015) 231301}, [\href{http://arxiv.org/abs/1503.02641}{{\tt
  1503.02641}}].

\bibitem{DEramo:2017zqw}
F.~D'Eramo, B.~J. Kavanagh and P.~Panci, \emph{{Probing Leptophilic Dark
  Sectors with Hadronic Processes}},
  \href{http://dx.doi.org/10.1016/j.physletb.2017.05.063}{\emph{Phys. Lett.}
  {\bf B771} (2017) 339--348}, [\href{http://arxiv.org/abs/1702.00016}{{\tt
  1702.00016}}].

\bibitem{Ackermann:2012rg}
{\scshape Fermi-LAT} collaboration, M.~Ackermann et~al., \emph{{Constraints on
  the Galactic Halo Dark Matter from Fermi-LAT Diffuse Measurements}},
  \href{http://dx.doi.org/10.1088/0004-637X/761/2/91}{\emph{Astrophys. J.} {\bf
  761} (2012) 91}, [\href{http://arxiv.org/abs/1205.6474}{{\tt 1205.6474}}].

\bibitem{Cirelli:2009vg}
M.~Cirelli and P.~Panci, \emph{{Inverse Compton constraints on the Dark Matter
  e+e- excesses}},
  \href{http://dx.doi.org/10.1016/j.nuclphysb.2009.06.034}{\emph{Nucl. Phys.}
  {\bf B821} (2009) 399--416}, [\href{http://arxiv.org/abs/0904.3830}{{\tt
  0904.3830}}].

\bibitem{Cirelli:2009dv}
M.~Cirelli, P.~Panci and P.~D. Serpico, \emph{{Diffuse gamma ray constraints on
  annihilating or decaying Dark Matter after Fermi}},
  \href{http://dx.doi.org/10.1016/j.nuclphysb.2010.07.010}{\emph{Nucl. Phys.}
  {\bf B840} (2010) 284--303}, [\href{http://arxiv.org/abs/0912.0663}{{\tt
  0912.0663}}].

\bibitem{Fichet:2016clq}
S.~Fichet, \emph{{Shining Light on Polarizable Dark Particles}},
  \href{http://dx.doi.org/10.1007/JHEP04(2017)088}{\emph{JHEP} {\bf 04} (2017)
  088}, [\href{http://arxiv.org/abs/1609.01762}{{\tt 1609.01762}}].

\bibitem{Falkowski:2015krw}
A.~Falkowski and K.~Mimouni, \emph{{Model independent constraints on
  four-lepton operators}},
  \href{http://dx.doi.org/10.1007/JHEP02(2016)086}{\emph{JHEP} {\bf 02} (2016)
  086}, [\href{http://arxiv.org/abs/1511.07434}{{\tt 1511.07434}}].

\bibitem{Falkowski:2017pss}
A.~Falkowski, M.~Gonz{\'a}lez-Alonso and K.~Mimouni, \emph{{Compilation of
  low-energy constraints on 4-fermion operators in the SMEFT}},
  \href{http://dx.doi.org/10.1007/JHEP08(2017)123}{\emph{JHEP} {\bf 08} (2017)
  123}, [\href{http://arxiv.org/abs/1706.03783}{{\tt 1706.03783}}].

\bibitem{LEP:2003aa}
{\scshape SLD Electroweak Group, SLD Heavy Flavor Group, DELPHI, LEP, ALEPH,
  OPAL, LEP Electroweak Working Group, L3} collaboration, t.~S. Electroweak,
  \emph{{A Combination of preliminary electroweak measurements and constraints
  on the standard model}},  \href{http://arxiv.org/abs/hep-ex/0312023}{{\tt
  hep-ex/0312023}}.

\bibitem{Aghanim:2018eyx}
{\scshape Planck} collaboration, N.~Aghanim et~al., \emph{{Planck 2018 results.
  VI. Cosmological parameters}},  \href{http://arxiv.org/abs/1807.06209}{{\tt
  1807.06209}}.

\bibitem{DelNobile:2012tx}
E.~Del~Nobile, C.~Kouvaris, P.~Panci, F.~Sannino and J.~Virkajarvi,
  \emph{{Light Magnetic Dark Matter in Direct Detection Searches}},
  \href{http://dx.doi.org/10.1088/1475-7516/2012/08/010}{\emph{JCAP} {\bf 1208}
  (2012) 010}, [\href{http://arxiv.org/abs/1203.6652}{{\tt 1203.6652}}].

\bibitem{Chung:1998rq}
D.~J.~H. Chung, E.~W. Kolb and A.~Riotto, \emph{{Production of massive
  particles during reheating}},
  \href{http://dx.doi.org/10.1103/PhysRevD.60.063504}{\emph{Phys. Rev.} {\bf
  D60} (1999) 063504}, [\href{http://arxiv.org/abs/hep-ph/9809453}{{\tt
  hep-ph/9809453}}].

\bibitem{Giudice:2000ex}
G.~F. Giudice, E.~W. Kolb and A.~Riotto, \emph{{Largest temperature of the
  radiation era and its cosmological implications}},
  \href{http://dx.doi.org/10.1103/PhysRevD.64.023508}{\emph{Phys. Rev.} {\bf
  D64} (2001) 023508}, [\href{http://arxiv.org/abs/hep-ph/0005123}{{\tt
  hep-ph/0005123}}].

\bibitem{DEramo:2017gpl}
F.~D'Eramo, N.~Fernandez and S.~Profumo, \emph{{When the Universe Expands Too
  Fast: Relentless Dark Matter}},
  \href{http://dx.doi.org/10.1088/1475-7516/2017/05/012}{\emph{JCAP} {\bf 1705}
  (2017) 012}, [\href{http://arxiv.org/abs/1703.04793}{{\tt 1703.04793}}].

\bibitem{Hamdan:2017psw}
S.~Hamdan and J.~Unwin, \emph{{Dark Matter Freeze-out During Matter
  Domination}}, \href{http://dx.doi.org/10.1142/S021773231850181X}{\emph{Mod.
  Phys. Lett.} {\bf A33} (2018) 1850181},
  [\href{http://arxiv.org/abs/1710.03758}{{\tt 1710.03758}}].

\bibitem{Binder:2017rgn}
T.~Binder, T.~Bringmann, M.~Gustafsson and A.~Hryczuk, \emph{{Early kinetic
  decoupling of dark matter: when the standard way of calculating the thermal
  relic density fails}},
  \href{http://dx.doi.org/10.1103/PhysRevD.96.115010}{\emph{Phys. Rev.} {\bf
  D96} (2017) 115010}, [\href{http://arxiv.org/abs/1706.07433}{{\tt
  1706.07433}}].

\bibitem{Hall:2009bx}
L.~J. Hall, K.~Jedamzik, J.~March-Russell and S.~M. West, \emph{{Freeze-In
  Production of FIMP Dark Matter}},
  \href{http://dx.doi.org/10.1007/JHEP03(2010)080}{\emph{JHEP} {\bf 03} (2010)
  080}, [\href{http://arxiv.org/abs/0911.1120}{{\tt 0911.1120}}].

\bibitem{Petraki:2013wwa}
K.~Petraki and R.~R. Volkas, \emph{{Review of asymmetric dark matter}},
  \href{http://dx.doi.org/10.1142/S0217751X13300287}{\emph{Int. J. Mod. Phys.}
  {\bf A28} (2013) 1330028}, [\href{http://arxiv.org/abs/1305.4939}{{\tt
  1305.4939}}].

\bibitem{DAgnolo:2015ujb}
R.~T. D'Agnolo and J.~T. Ruderman, \emph{{Light Dark Matter from Forbidden
  Channels}},
  \href{http://dx.doi.org/10.1103/PhysRevLett.115.061301}{\emph{Phys. Rev.
  Lett.} {\bf 115} (2015) 061301}, [\href{http://arxiv.org/abs/1505.07107}{{\tt
  1505.07107}}].

\bibitem{Agrawal:2011ze}
P.~Agrawal, S.~Blanchet, Z.~Chacko and C.~Kilic, \emph{{Flavored Dark Matter,
  and Its Implications for Direct Detection and Colliders}},
  \href{http://dx.doi.org/10.1103/PhysRevD.86.055002}{\emph{Phys. Rev.} {\bf
  D86} (2012) 055002}, [\href{http://arxiv.org/abs/1109.3516}{{\tt
  1109.3516}}].

\bibitem{Giacchino:2013bta}
F.~Giacchino, L.~Lopez-Honorez and M.~H.~G. Tytgat, \emph{{Scalar Dark Matter
  Models with Significant Internal Bremsstrahlung}},
  \href{http://dx.doi.org/10.1088/1475-7516/2013/10/025}{\emph{JCAP} {\bf 1310}
  (2013) 025}, [\href{http://arxiv.org/abs/1307.6480}{{\tt 1307.6480}}].

\bibitem{Garny:2015wea}
M.~Garny, A.~Ibarra and S.~Vogl, \emph{{Signatures of Majorana dark matter with
  t-channel mediators}},
  \href{http://dx.doi.org/10.1142/S0218271815300190}{\emph{Int. J. Mod. Phys.}
  {\bf D24} (2015) 1530019}, [\href{http://arxiv.org/abs/1503.01500}{{\tt
  1503.01500}}].

\bibitem{Baker:2018uox}
M.~J. Baker and A.~Thamm, \emph{{Leptonic WIMP Coannihilation and the Current
  Dark Matter Search Strategy}},  \href{http://arxiv.org/abs/1806.07896}{{\tt
  1806.07896}}.

\bibitem{Pierce:2014spa}
A.~Pierce and Z.~Zhang, \emph{{Hidden Dipole Dark Matter}},
  \href{http://dx.doi.org/10.1103/PhysRevD.90.015026}{\emph{Phys. Rev.} {\bf
  D90} (2014) 015026}, [\href{http://arxiv.org/abs/1405.1937}{{\tt
  1405.1937}}].

\bibitem{Herrero-Garcia:2018koq}
J.~Herrero-Garcia, E.~Molinaro and M.~A. Schmidt, \emph{{Dark matter direct
  detection of a fermionic singlet at one loop}},
  \href{http://dx.doi.org/10.1140/epjc/s10052-018-5935-5}{\emph{Eur. Phys. J.}
  {\bf C78} (2018) 471}, [\href{http://arxiv.org/abs/1803.05660}{{\tt
  1803.05660}}].

\bibitem{Kopp:2009et}
J.~Kopp, V.~Niro, T.~Schwetz and J.~Zupan, \emph{{DAMA/LIBRA and leptonically
  interacting Dark Matter}},
  \href{http://dx.doi.org/10.1103/PhysRevD.80.083502}{\emph{Phys. Rev.} {\bf
  D80} (2009) 083502}, [\href{http://arxiv.org/abs/0907.3159}{{\tt
  0907.3159}}].

\bibitem{Shifman:1979eb}
M.~A. Shifman, A.~I. Vainshtein, M.~B. Voloshin and V.~I. Zakharov,
  \emph{{Low-Energy Theorems for Higgs Boson Couplings to Photons}},
  {\emph{Sov. J. Nucl. Phys.} {\bf 30} (1979) 711--716}.

\bibitem{TheATLAScollaboration:2013hha}
{\scshape ATLAS} collaboration, T.~A. collaboration, \emph{{Search for
  direct-slepton and direct-chargino production in final states with two
  opposite-sign leptons, missing transverse momentum and no jets in 20/fb of pp
  collisions at sqrt(s) = 8 TeV with the ATLAS detector}}, .

\bibitem{Khachatryan:2014qwa}
{\scshape CMS} collaboration, V.~Khachatryan et~al., \emph{{Searches for
  electroweak production of charginos, neutralinos, and sleptons decaying to
  leptons and W, Z, and Higgs bosons in pp collisions at 8 TeV}},
  \href{http://dx.doi.org/10.1140/epjc/s10052-014-3036-7}{\emph{Eur. Phys. J.}
  {\bf C74} (2014) 3036}, [\href{http://arxiv.org/abs/1405.7570}{{\tt
  1405.7570}}].

\bibitem{Aad:2014vma}
{\scshape ATLAS} collaboration, G.~Aad et~al., \emph{{Search for direct
  production of charginos, neutralinos and sleptons in final states with two
  leptons and missing transverse momentum in $pp$ collisions at $\sqrt{s} =$ 8
  TeV with the ATLAS detector}},
  \href{http://dx.doi.org/10.1007/JHEP05(2014)071}{\emph{JHEP} {\bf 05} (2014)
  071}, [\href{http://arxiv.org/abs/1403.5294}{{\tt 1403.5294}}].

\bibitem{Aad:2014yka}
{\scshape ATLAS} collaboration, G.~Aad et~al., \emph{{Search for the direct
  production of charginos, neutralinos and staus in final states with at least
  two hadronically decaying taus and missing transverse momentum in $pp$
  collisions at $\sqrt{s}$ = 8 TeV with the ATLAS detector}},
  \href{http://dx.doi.org/10.1007/JHEP10(2014)096}{\emph{JHEP} {\bf 10} (2014)
  096}, [\href{http://arxiv.org/abs/1407.0350}{{\tt 1407.0350}}].

\bibitem{Aaboud:2018jiw}
{\scshape ATLAS} collaboration, M.~Aaboud et~al., \emph{{Search for electroweak
  production of supersymmetric particles in final states with two or three
  leptons at $\sqrt{s}=13\,$TeV with the ATLAS detector}},
  \href{http://arxiv.org/abs/1803.02762}{{\tt 1803.02762}}.

\bibitem{Sirunyan:2018nwe}
{\scshape CMS} collaboration, A.~M. Sirunyan et~al., \emph{{Search for
  supersymmetric partners of electrons and muons in proton-proton collisions at
  $\sqrt{s}=$ 13 TeV}}, {\emph{Submitted to: Phys. Lett.} (2018) },
  [\href{http://arxiv.org/abs/1806.05264}{{\tt 1806.05264}}].

\bibitem{CMS:2017rio}
{\scshape CMS} collaboration, C.~Collaboration, \emph{{Search for pair
  production of tau sleptons in $\sqrt{s}=13~\mathrm{TeV}$ pp collisions in the
  all-hadronic final state}}, .

\bibitem{Sirunyan:2018vig}
{\scshape CMS} collaboration, A.~M. Sirunyan et~al., \emph{{Search for
  supersymmetry in events with a $\tau$ lepton pair and missing transverse
  momentum in proton-proton collisions at $\sqrt{s}=$ 13 TeV}},
  \href{http://arxiv.org/abs/1807.02048}{{\tt 1807.02048}}.

\bibitem{Tanabashi:2018oca}
{\scshape Particle Data Group} collaboration, M.~Tanabashi et~al.,
  \emph{{Review of Particle Physics}},
  \href{http://dx.doi.org/10.1103/PhysRevD.98.030001}{\emph{Phys. Rev.} {\bf
  D98} (2018) 030001}.

\bibitem{Calibbi:2018rzv}
L.~Calibbi, R.~Ziegler and J.~Zupan, \emph{{Minimal models for dark matter and
  the muon g-2 anomaly}},
  \href{http://dx.doi.org/10.1007/JHEP07(2018)046}{\emph{JHEP} {\bf 07} (2018)
  046}, [\href{http://arxiv.org/abs/1804.00009}{{\tt 1804.00009}}].

\bibitem{Kumar:2015tna}
N.~Kumar and S.~P. Martin, \emph{{Vectorlike Leptons at the Large Hadron
  Collider}}, \href{http://dx.doi.org/10.1103/PhysRevD.92.115018}{\emph{Phys.
  Rev.} {\bf D92} (2015) 115018}, [\href{http://arxiv.org/abs/1510.03456}{{\tt
  1510.03456}}].

\bibitem{LEPhiggsino}
D.~L. LEPSUSYWG, ALEPH and O.~collaboration, ``Combined lep chargino results,
  up to 208 gev for low dm.''

\bibitem{Giacchino:2014moa}
F.~Giacchino, L.~Lopez-Honorez and M.~H.~G. Tytgat, \emph{{Bremsstrahlung and
  Gamma Ray Lines in 3 Scenarios of Dark Matter Annihilation}},
  \href{http://dx.doi.org/10.1088/1475-7516/2014/08/046}{\emph{JCAP} {\bf 1408}
  (2014) 046}, [\href{http://arxiv.org/abs/1405.6921}{{\tt 1405.6921}}].

\bibitem{Dobrescu:2006au}
B.~A. Dobrescu and I.~Mocioiu, \emph{{Spin-dependent macroscopic forces from
  new particle exchange}},
  \href{http://dx.doi.org/10.1088/1126-6708/2006/11/005}{\emph{JHEP} {\bf 11}
  (2006) 005}, [\href{http://arxiv.org/abs/hep-ph/0605342}{{\tt
  hep-ph/0605342}}].

\bibitem{Fan:2010gt}
J.~Fan, M.~Reece and L.-T. Wang, \emph{{Non-relativistic effective theory of
  dark matter direct detection}},
  \href{http://dx.doi.org/10.1088/1475-7516/2010/11/042}{\emph{JCAP} {\bf 1011}
  (2010) 042}, [\href{http://arxiv.org/abs/1008.1591}{{\tt 1008.1591}}].

\bibitem{Fitzpatrick:2012ix}
A.~L. Fitzpatrick, W.~Haxton, E.~Katz, N.~Lubbers and Y.~Xu, \emph{{The
  Effective Field Theory of Dark Matter Direct Detection}},
  \href{http://dx.doi.org/10.1088/1475-7516/2013/02/004}{\emph{JCAP} {\bf 1302}
  (2013) 004}, [\href{http://arxiv.org/abs/1203.3542}{{\tt 1203.3542}}].

\bibitem{Anand:2013yka}
N.~Anand, A.~L. Fitzpatrick and W.~C. Haxton, \emph{{Weakly interacting massive
  particle-nucleus elastic scattering response}},
  \href{http://dx.doi.org/10.1103/PhysRevC.89.065501}{\emph{Phys. Rev.} {\bf
  C89} (2014) 065501}, [\href{http://arxiv.org/abs/1308.6288}{{\tt
  1308.6288}}].

\bibitem{Fitzpatrick:2012ib}
A.~L. Fitzpatrick, W.~Haxton, E.~Katz, N.~Lubbers and Y.~Xu, \emph{{Model
  Independent Direct Detection Analyses}},
  \href{http://arxiv.org/abs/1211.2818}{{\tt 1211.2818}}.

\bibitem{Panci:2014gga}
P.~Panci, \emph{{New Directions in Direct Dark Matter Searches}},
  \href{http://dx.doi.org/10.1155/2014/681312}{\emph{Adv. High Energy Phys.}
  {\bf 2014} (2014) 681312}, [\href{http://arxiv.org/abs/1402.1507}{{\tt
  1402.1507}}].

\bibitem{Dent:2015zpa}
J.~B. Dent, L.~M. Krauss, J.~L. Newstead and S.~Sabharwal, \emph{{General
  analysis of direct dark matter detection: From microphysics to observational
  signatures}}, \href{http://dx.doi.org/10.1103/PhysRevD.92.063515}{\emph{Phys.
  Rev.} {\bf D92} (2015) 063515}, [\href{http://arxiv.org/abs/1505.03117}{{\tt
  1505.03117}}].

\bibitem{DelNobile:2018dfg}
E.~Del~Nobile, \emph{{A complete Lorentz-to-Galileo dictionary for direct Dark
  Matter detection}},  \href{http://arxiv.org/abs/1806.01291}{{\tt
  1806.01291}}.

\bibitem{Catena:2015uha}
R.~Catena and B.~Schwabe, \emph{{Form factors for dark matter capture by the
  Sun in effective theories}},
  \href{http://dx.doi.org/10.1088/1475-7516/2015/04/042}{\emph{JCAP} {\bf 1504}
  (2015) 042}, [\href{http://arxiv.org/abs/1501.03729}{{\tt 1501.03729}}].

\bibitem{Menendez:2012tm}
J.~Menendez, D.~Gazit and A.~Schwenk, \emph{{Spin-dependent WIMP scattering off
  nuclei}}, \href{http://dx.doi.org/10.1103/PhysRevD.86.103511}{\emph{Phys.
  Rev.} {\bf D86} (2012) 103511}, [\href{http://arxiv.org/abs/1208.1094}{{\tt
  1208.1094}}].

\bibitem{Cerdeno:2010jj}
D.~G. Cerdeno and A.~M. Green, \emph{{Direct detection of WIMPs}},
  \href{http://arxiv.org/abs/1002.1912}{{\tt 1002.1912}}.

\bibitem{Gresham:2014vja}
M.~I. Gresham and K.~M. Zurek, \emph{{Effect of nuclear response functions in
  dark matter direct detection}},
  \href{http://dx.doi.org/10.1103/PhysRevD.89.123521}{\emph{Phys. Rev.} {\bf
  D89} (2014) 123521}, [\href{http://arxiv.org/abs/1401.3739}{{\tt
  1401.3739}}].

\bibitem{Bishara:2017pfq}
F.~Bishara, J.~Brod, B.~Grinstein and J.~Zupan, \emph{{From quarks to nucleons
  in dark matter direct detection}},
  \href{http://dx.doi.org/10.1007/JHEP11(2017)059}{\emph{JHEP} {\bf 11} (2017)
  059}, [\href{http://arxiv.org/abs/1707.06998}{{\tt 1707.06998}}].

\bibitem{DEramo:2016aee}
F.~D'Eramo, J.~de~Vries and P.~Panci, \emph{{A 750 GeV Portal: LHC
  Phenomenology and Dark Matter Candidates}},
  \href{http://dx.doi.org/10.1007/JHEP05(2016)089}{\emph{JHEP} {\bf 05} (2016)
  089}, [\href{http://arxiv.org/abs/1601.01571}{{\tt 1601.01571}}].

\bibitem{Ovanesyan:2014fha}
G.~Ovanesyan and L.~Vecchi, \emph{{Direct detection of dark matter
  polarizability}},
  \href{http://dx.doi.org/10.1007/JHEP07(2015)128}{\emph{JHEP} {\bf 07} (2015)
  128}, [\href{http://arxiv.org/abs/1410.0601}{{\tt 1410.0601}}].

\bibitem{Aprile:2017iyp}
{\scshape XENON} collaboration, E.~Aprile et~al., \emph{{First Dark Matter
  Search Results from the XENON1T Experiment}},
  \href{http://dx.doi.org/10.1103/PhysRevLett.119.181301}{\emph{Phys. Rev.
  Lett.} {\bf 119} (2017) 181301}, [\href{http://arxiv.org/abs/1705.06655}{{\tt
  1705.06655}}].

\bibitem{Feldman:1997qc}
G.~J. Feldman and R.~D. Cousins, \emph{{A Unified approach to the classical
  statistical analysis of small signals}},
  \href{http://dx.doi.org/10.1103/PhysRevD.57.3873}{\emph{Phys. Rev.} {\bf D57}
  (1998) 3873--3889}, [\href{http://arxiv.org/abs/physics/9711021}{{\tt
  physics/9711021}}].

\bibitem{XENON1T-code}
B.~J. Kavanagh, ``\textnormal{Xenon1T-2018 v1.0 [Computer Software]}.''
  \href{https://doi.org/10.5281/zenodo.1888420}{\textnormal{doi:10.5281/zenodo.1888420}}\textnormal{.
  Available at }\url{https://github.com/bradkav/Xenon1T-2018}, 2018.

\bibitem{Workgroup:2017lvb}
{\scshape The GAMBIT Dark Matter Workgroup} collaboration, T.~Bringmann et~al.,
  \emph{{DarkBit: A GAMBIT module for computing dark matter observables and
  likelihoods}},
  \href{http://dx.doi.org/10.1140/epjc/s10052-017-5155-4}{\emph{Eur. Phys. J.}
  {\bf C77} (2017) 831}, [\href{http://arxiv.org/abs/1705.07920}{{\tt
  1705.07920}}].

\bibitem{Fowlie:2018svr}
A.~Fowlie, \emph{{Non-parametric uncertainties in the dark matter velocity
  distribution}},
  \href{http://dx.doi.org/10.1088/1475-7516/2019/01/006}{\emph{JCAP} {\bf 1901}
  (2019) 006}, [\href{http://arxiv.org/abs/1809.02323}{{\tt 1809.02323}}].

\bibitem{Athron:2018hpc}
{\scshape GAMBIT} collaboration, P.~Athron et~al., \emph{{Global analyses of
  Higgs portal singlet dark matter models using GAMBIT}},
  \href{http://dx.doi.org/10.1140/epjc/s10052-018-6513-6}{\emph{Eur. Phys. J.}
  {\bf C79} (2019) 38}, [\href{http://arxiv.org/abs/1808.10465}{{\tt
  1808.10465}}].

\bibitem{DDCalc:url}
{\scshape The GAMBIT Dark Matter Workgroup} collaboration, ``{DDcalc v2.0
  [Software]}.'' \url{http://ddcalc.hepforge.org/}.

\end{thebibliography}\endgroup


\providecommand{\href}[2]{#2}\begingroup\raggedright\endgroup

\end{document}